\documentclass[12pt]{article}
\usepackage{bbm}
\usepackage{bm}
\usepackage{textgreek}
\usepackage{color}
\usepackage{authblk}
\usepackage{latexsym}
\usepackage{epsfig,amssymb,euscript, mathrsfs,cite}
\usepackage{amsmath}
\usepackage{tikz}
\usetikzlibrary{decorations.pathmorphing}
\usepackage{slashed}
\usepackage{mathrsfs} 
\usepackage{enumerate}
\usepackage{graphicx}
\usepackage{caption}
\usepackage{subcaption}
\definecolor{MyDarkBlue}{rgb}{0.15,0.15,0.45}
\usepackage[linktocpage=true]{hyperref}
\hypersetup{
colorlinks=true,
citecolor=MyDarkBlue,
linkcolor=MyDarkBlue,
urlcolor=MyDarkBlue,
pdfauthor={},
pdftitle={},
pdfsubject={hep-th}
}
\newcommand{\stoptocwriting}{%
  \addtocontents{toc}{\protect\setcounter{tocdepth}{-5}}}
  \newcommand{\resumetocwriting}{%
  \addtocontents{toc}{\protect\setcounter{tocdepth}{\arabic{tocdepth}}}}

\newsavebox{\ns}
\newsavebox{\dbrane}
\newsavebox{\dbshort}

\def\be{\begin{equation}}
\def\ee{\end{equation}}
\def\bea{\begin{eqnarray}}
\def\eea{\end{eqnarray}}

\newcommand{\nn}{\nonumber\\}

\newcommand\R{\mathbb{R}}
\newcommand\Z{\mathbb{Z}}

\newcommand\D{\mathcal{D}}

\newcommand\C{\mathbb{C}}

\newcommand\diff{\mathrm{d}}

\newcommand{\ii}{\mathrm{i}}

\newcommand{\ex}{\mathrm{e}}
\newcommand{\vol}{\mathrm{vol}}

\newcommand{\Omegach}{H}
\newcommand{\Hch}{Z}
\newcommand{\hch}{w}

\newcommand{\jphi}{\varphi}

\newcommand{\zetaJ}{\zeta}

\newlength{\sswidth}
\newcommand{\sla}[1]{
   \settowidth{\sswidth}{$#1$}
   \mbox{$\not{\hspace*{-0.3\sswidth}#1}$}}

\numberwithin{equation}{section}       

\newcommand{\PP}{\mathcal{P}}
\newcommand{\QQ}{\mathcal{Q}}

\newcommand{\psip}{\psi}
\newcommand{\sigmap}{z}
\newcommand{\qp}{\mathtt{j}}
\newcommand{\Qp}{\mathtt{a}}

\newcommand{\kp}{\mathtt{p}}
\newcommand{\pp}{\mathtt{q}}
\newcommand{\gf}{h}

\newcommand{\pJ}{\mathtt{p}}
\newcommand{\qJ}{\mathtt{q}}

\newcommand{\kapp}{\mathtt{r}}
\newcommand{\anotherangle}{\nu}

\newcommand{\gammathree}{\mbox{\textgamma}}

\begin{document}

\begin{titlepage}

\begin{flushright}
Imperial/TP/2020/JG/05\\
\end{flushright}

\vskip 1cm

\begin{center}


{\Large \bf Accelerating Black Holes and Spinning Spindles}

\vskip 1cm
{Pietro Ferrero$^{\mathrm{a}}$, Jerome P. Gauntlett$^{\mathrm{b}}$, Juan Manuel P\'erez Ipi\~na$^{\mathrm{a}}$,\\[2mm]
Dario Martelli$^{\mathrm{c,d,e}}$
and James Sparks$^{\mathrm{a}}$}

\vskip 1cm

${}^{\,\mathrm{a}}$\textit{Mathematical Institute, University of Oxford,\\
Andrew Wiles Building, Radcliffe Observatory Quarter,\\
Woodstock Road, Oxford, OX2 6GG, U.K.\\}
\vskip 0.2 cm

${}^{\mathrm{b}}$\textit{Blackett Laboratory, Imperial College, \\
Prince Consort Rd., London, SW7 2AZ, U.K.\\}

\vskip 0.2cm

${}^{\mathrm{c}}$\textit{Dipartimento di Matematica ``Giuseppe Peano'', Universit\`a di Torino,\\
Via Carlo Alberto 10, 10123 Torino, Italy}

\vskip 0.2cm

${}^{\mathrm{d}}$\textit{INFN, Sezione di Torino \&}   ${}^{\mathrm{e}}$\textit{Arnold--Regge Center,\\
 Via Pietro Giuria 1, 10125 Torino, Italy}

\end{center}

\vskip 0.5 cm

\begin{abstract}
\noindent
We study solutions in the 
Pleba\'nski--Demia\'nski family which describe an accelerating, rotating and dyonically charged
black hole in $AdS_4$. These are solutions of $D=4$ Einstein-Maxwell theory with a negative cosmological constant and hence minimal $D=4$
gauged supergravity. It is well known that when the acceleration is non-vanishing the $D=4$ black hole metrics 
have conical singularities. 
By uplifting the solutions to $D=11$ supergravity using a regular Sasaki-Einstein $7$-manifold, $SE_7$,
we show how the free parameters can be chosen to eliminate the conical singularities.
Topologically, the $D=11$ solutions incorporate an $SE_7$ fibration over a two-dimensional
weighted projective space, $\mathbb{WCP}^1_{[n_-,n_+]}$, also known as a spindle, which is labelled by two integers that determine the conical singularities of the $D=4$ metrics.  
We also discuss the supersymmetric and extremal limit and show that the near horizon limit gives rise to a new family of regular supersymmetric $AdS_2\times Y_9$ solutions of $D=11$ supergravity, which generalise a known family by the addition of a rotation parameter. We calculate the entropy of these black holes and argue that it should be possible to derive this from certain ${\cal N}=2$, $d=3$ quiver gauge theories compactified on a spinning spindle with appropriate magnetic flux.

\end{abstract}

\end{titlepage}

\pagestyle{plain}
\setcounter{page}{1}
\newcounter{bean}
\baselineskip18pt

\tableofcontents

\newpage

\section{Introduction}

The $C$-metrics of $D=4$ Einstein-Maxwell theory \cite{PhysRevD.2.1359} describe two charged black holes
undergoing uniform acceleration. The force for the acceleration is provided by a conical angle excess (a strut) between the two black holes, 
a conical angle deficit (a string) attached to the black holes and extending out to infinity, or a combination of the two. 
By extending the model to include additional matter fields that allow for cosmic strings, the conical singularity can be smoothed out by
having two cosmic strings pull the black holes apart \cite{Achucarro:1995nu,Eardley:1995au}. Another approach for removing the conical singularities, in the case of electrically or magnetically charged black holes, is provided by the Ernst metric in which the black holes are being
accelerated by a background electric or magnetic field, respectively \cite{doi:10.1063/1.522935}. 

Generalizations of the Ernst metric were found for Einstein-Maxwell-dilaton gravity in \cite{Dowker:1993bt}. For the particular case associated with
$D=5$ Kaluza-Klein theory it was shown in \cite{Dowker:1995gb} that the accelerating $D=4$ extremal, magnetically charged black hole solutions can be obtained by a dimensional reduction of a double Wick rotation of the $D=5$ generalization of the Kerr solution \cite{Myers:1986un}. 
Recall the extremal magnetically charged black holes in Kaluza-Klein theory 
are Kaluza-Klein monopoles \cite{Sorkin:1983ns,Gross:1983hb}, which are in fact naked singularities from the four-dimensional point of view. A key ingredient in the construction of \cite{Dowker:1995gb} is that the Kaluza-Klein circle action associated with the reduction has two fixed points which precisely correspond to a Kaluza-Klein monopole, anti-monopole pair.

In this paper we provide a new way of desingularising the conical deficits for a specific class of accelerating, 
rotating and dyonically charged
black holes in $AdS_4$, by embedding them in $D=11$ supergravity. We study the Pleba\'nski--Demia\'nski (PD) solutions \cite{Plebanski:1976gy} of $D=4$ Einstein-Maxwell theory, generalising
the $C$-metrics of \cite{PhysRevD.2.1359}, which in particular allow for a cosmological constant, which we take to be negative. 
Such accelerating $AdS_4$ black holes have been extensively studied (e.g. \cite{Podolsky:2002nk,Dias:2002mi,Anabalon:2018qfv}) and here we will only be interested in the case of 
small acceleration for which there is just a single black hole. 
From a $D=4$ perspective we will consider a single accelerating black hole with a two-dimensional horizon given by a ``spindle", 
a weighted projective space, $\mathbb{WCP}^1_{[n_-,n_+]}$, which is topologically a sphere with conical deficits at both poles specified by 
$n_\pm \in\mathbb{N}$,
which stretch out to the $AdS_4$ boundary. The net acceleration of the black hole is due to the mismatch of the conical deficits on either side of the black hole. 
Interestingly, the same mismatch also leads to a non-zero magnetic flux though the horizon, given by $G_{(4)}Q_m=\tfrac{n_--n_+}{4 n_+n_-}$.

We can embed these solutions in $D=11$ supergravity, locally, using an arbitrary seven-dimensional Sasaki-Einstein manifold $SE_7$.
Indeed it has been shown \cite{Gauntlett:2007ma} that there is a consistent Kaluza-Klein truncation of $D=11$ supergravity on any $SE_7$ down to
minimal gauged supergravity in $D=4$, whose bosonic content is Einstein-Maxwell theory (with a negative cosmological constant). 
By definition, this means that any solution of $D=4$ Einstein-Maxwell theory can be uplifted on an arbitrary $SE_7$ manifold to
obtain a solution of $D=11$ supergravity. Furthermore, if the $D=4$ solution preserves supersymmetry then so does the uplifted solution. For example
the vacuum $AdS_4$ solution, which preserves all of the $D=4$ supersymmetry, uplifts to the $AdS_4\times SE_7$ solution which, in general,
preserves 1/4 of the supersymmetry and is dual to an $\mathcal{N}=2$ supersymmetric conformal field theory (SCFT) in $d=3$. 

In our construction it is only the {\it regular class} of $SE_7$ that
play a role in obtaining smooth $D=11$ solutions. These have the property that the $SE_7$ manifold is a circle bundle over $KE_6$, a six-dimensional K\"ahler-Einstein manifold with positive curvature. It will also be important to recall that this includes the generic class where the circle bundle is the canonical bundle of the K\"ahler-Einstein manifold, but it is also possible, depending on the choice of $KE_6$, to enlarge the period of the circle to
get other Sasaki-Einstein manifolds. The simplest example is for the case of $\mathbb{CP}^3$: for this case
$S^7/\mathbb{Z}_4$ is the canonical circle bundle over $\mathbb{CP}^3$ and is a smooth Sasaki-Einstein manifold, but so too
are $S^7/\mathbb{Z}_2$ and $S^7$. 

After uplifting the PD black hole solutions with horizon $\mathbb{WCP}^1_{[n_-,n_+]}$ on a regular $SE_7$ we will show, somewhat
miraculously, that by choosing the parameters in the PD metric appropriately,
the $D=11$ solution is free from any conical deficit singularities.\footnote{Of course the 
black hole curvature singularity remains, behind the black hole horizon.} Importantly, precisely which regular Sasaki-Einstein manifold one can uplift upon
depends on the integers $n_\pm \in\mathbb{N}$. For example, for the case of $\mathbb{CP}^3$, for given $n_\pm \in\mathbb{N}$ 
we can uplift on precisely one of the three cases $S^7$, $S^7/\mathbb{Z}_2$ or $S^7/\mathbb{Z}_4$. 
It is also important to note that for the Kaluza-Klein reduction on the $SE_7$ the fibred circle action 
does not have any fixed points and hence there are no Kaluza-Klein monopoles 
as in the construction of \cite{Dowker:1995gb}. On the other hand, the circle action is not free and this leads
to the conical deficits of the spindle $\mathbb{WCP}^1_{[n_-,n_+]}$. 

Of particular interest is that our construction also includes accelerating black holes that preserve supersymmetry, 
as considered in a general context in \cite{Klemm:2013eca},
and, moreover, have regular extremal horizons. 
From the $D=4$ perspective, the near horizon limit of these supersymmetric, extremal, rotating black holes are of the form 
$AdS_2\times \mathbb{WCP}^1_{[n_-,n_+]}$, 
with non-vanishing magnetic flux through the horizon $\mathbb{WCP}^1_{[n_-,n_+]}$. 
The way that supersymmetry is preserved for these $AdS_2$ solutions is non-standard: 
the magnetic flux is not describing a  topological twist \cite{Witten:1988ze,Bershadsky:1995qy}, and the Killing spinors 
are then correspondingly  not simply constant, even when the rotation is turned off; indeed, they are sections of non-trivial 
bundles over the spindle $\mathbb{WCP}^1_{[n_-,n_+]}$.
The uplifted supersymmetric $D=11$ solutions have a near horizon limit of the form $AdS_2\times Y_9$. In the special case that
we set the rotation parameter of the PD metrics to zero, we find an $AdS_2\times Y_9$ geometry
in the general class of \cite{Kim:2006qu}, where $Y_9$ is a nine-dimensional Gauntlett-Kim (GK) geometry \cite{Gauntlett:2007ts}.  In fact, remarkably, when the rotation parameter vanishes we find exactly the same class
of supersymmetric $AdS_2\times Y_9$ solutions first constructed in \cite{Gauntlett:2006ns} from a quite different perspective. 
When the rotation parameter of the PD metrics 
is non-zero we find a new class of supersymmetric $AdS_2\times Y_9$ solutions that lie outside the class considered in \cite{Kim:2006qu},
and generalize those  of \cite{Gauntlett:2006ns} with an extra rotation parameter.\footnote{ 
These should arise within the recent classification of rotating 
$D=11$ $AdS_2$ solutions in \cite{Couzens:2020jgx}, and it would be interesting 
to investigate this further.}

We calculate the entropy $S_{BH}$ for the PD black hole solutions. In particular, for the 
``supersymmetric spinning spindles'' that in addition have extremal horizons, we find
\begin{align}\label{blackholent}
S_{BH} = \frac{\pi}{G_{(4)}}  \frac{J}{Q_e} = \frac{\pi}{G_{(4)}}  \frac{\sqrt{2} \sqrt{n_-^2+n_+^2 +8 n_-^2 n_+^2(G_{(4)}Q_e)^2}-(n_-+n_+)}{4 n_- n_+}\, ,
\end{align}
where $J$ is the angular momentum of the black hole, and $Q_e$ is its electric charge. Notice this gives a relation between 
$J$ and $Q_e$ for extremal solutions. We also note that with non-zero acceleration, we can still have supersymmetric extremal black holes with
$J=Q_e=0$; for these black holes the second expression in \eqref{blackholent} is the valid expression. We also recall that 
$\frac{\pi}{2G_{(4)}}=\mathcal{F}_{S^3}$ is the free energy of the $d=3$ SCFT dual to the $AdS_4\times SE_7$ solution.

We define the angular momentum $J$ of the black hole, appearing in \eqref{blackholent}, to be a conserved quantity 
that can be equally evaluated either at the conformal boundary or at the black hole horizon (see e.g. \cite{Papadimitriou:2005ii}).  As a consequence it is a type of Page charge that depends on the gauge used for the gauge field. However, there is a different natural angular momentum defined in the near horizon limit which has a gauge field that is invariant under the $AdS_2$ symmetries, 
that we denote $J_{AdS_2}$. We find
\begin{align}\label{angmomsurprise}
J_{AdS_2}-J = \frac{Q_e}{4}\chi(\Sigma)\, ,
\end{align}
where $\chi(\Sigma)=\frac{n_-+n_+}{n_+n_-}$ is the Euler number of the spindle 
horizon $\Sigma=\mathbb{WCP}^1_{[n_-,n_+]}$.
 
 It would be very interesting to be able to derive this entropy from the $d=3$ SCFT
dual to the corresponding $AdS_4\times SE_7$ vacuum solution.
In the context of the topological twist, there has been considerable progress in obtaining such derivations for supersymmetric $AdS_4$ black holes
with magnetic flux through a Riemann surface horizon using the principle of $I$-extremization \cite{Benini:2015eyy,Benini:2016rke,Cabo-Bizet:2017jsl,Azzurli:2017kxo,Hosseini:2019use,Gauntlett:2019roi,Hosseini:2019ddy,Kim:2019umc}. 
The relevant black hole solutions approach
$AdS_4\times SE_7$ in the UV and $AdS_2\times Y_9$ in the near horizon, with $Y_9$ a $SE_7$ fibration over
the Riemann surface, $\Sigma_g$. The black hole entropy can be obtained from the $d=3$ field theory by
extremizing a certain twisted topological index \cite{Benini:2015noa}
associated with the $d=3$ SCFT compactified on $\Sigma_g$. 
This index can be calculated using localization techniques and then evaluated in the large $N$ limit for many different examples \cite{Hosseini:2016tor,Hosseini:2016ume}.
A geometric dual of $I$-extremization was proposed in \cite{Couzens:2018wnk,Gauntlett:2018dpc} and later
shown to agree with the $I$-extremization procedure in field theory 
for infinite classes of examples of such $AdS_2\times Y_9$ solutions in \cite{Hosseini:2019use,Gauntlett:2019roi,Hosseini:2019ddy,Kim:2019umc}. 

The family of black holes that we consider includes the well-known Kerr-Newman-AdS spacetime \cite{Carter:1968ks} in the non-accelerating case, whose Bogomolnyi-Prasad-Somerfeld (BPS), i.e. supersymmetric, and extremal limits have been 
extensively discussed both from the gravitational, as well as from the dual field theory points of view,
see e.g. \cite{Kostelecky:1995ei,AlonsoAlberca:2000cs,Choi:2018fdc,Cassani:2019mms} and \cite{Bobev:2019zmz,Nian:2019pxj}, respectively.
Interestingly, the entropy of the BPS and extremal Kerr-Newman-AdS black hole can be immediately recovered  from (\ref{blackholent}) by 
setting $n_+=n_-=1$.
In addition, the expression \eqref{angmomsurprise} is also valid for the Kerr-Newman-AdS black holes, a point that seems to have been overlooked in the literature. 

The results of this paper lead to the challenge of recovering the entropy of the rotating and accelerating   
black hole, by evaluating a suitable index. This index will be an appropriately defined localized 
partition function of the dual
$d=3$ SCFT compactified on $S^1\times \mathbb{WCP}^1_{[n_-,n_+]}$, with electric flux and magnetic flux through $\mathbb{WCP}^1_{[n_-,n_+]}$, and with additional rotation. 
There are a number of subtleties related to this computation, which will need to be further developed in future work.
Here we will make a number of particularly interesting observations concerning the BPS extremal black holes with acceleration but
without rotation or electric charge i.e. $J=Q_e=0$. One feature is that these solutions have an
acceleration horizon that intersects the conformal boundary,  effectively dividing the spindle in half. Moreover, we
find that the Killing spinor on this conformal boundary is given by a topological 
twist, so that the spinor is constant, but  it is a \emph{different} constant spinor on each half of the space! We also explain how these
unusual features are ``regulated'' by keeping, for example, supersymmetry but relaxing the extremality condition (although this class with $J=0$ then
has a naked singularity in the bulk).

The plan of the rest of this paper is as follows. In section \ref{sec:gensetting} we briefly
review $D=4$ minimal gauged supergravity, and its uplift to $D=11$ supergravity on Sasaki-Einstein 
seven-manifolds. Section \ref{sec:PD} introduces the class of Pleba\'nski--Demia\'nski 
solutions of interest, showing that the parameters can be chosen so that the black hole horizon 
is topologically a spindle, $ \mathbb{WCP}^1_{[n_-,n_+]}$. In section \ref{sec:uplift}
we show that on uplifting to $D=11$ supergravity the PD solutions become completely 
smooth, with properly quantized flux. Supersymmetric and extremal solutions 
are studied in section \ref{sec:BPS}, where we focus on the near horizon 
$AdS_2$ geometries and associated Killing spinors. In section 
\ref{sec:conboundary} we discuss some global properties 
of the accelerating but non-rotating extremal supersymmetric black hole, 
discussing the conformal boundary and the acceleration horizon. 
We conclude with a discussion 
of open problems in section \ref{sec:discussion}. 

We have also included a number of appendices. Appendix A has some general comments on circle bundles over spindles.
Appendix \ref{nonacc} briefly discusses some aspects of the well-known non-accelerating class of 
Kerr-Newman-AdS black holes. Appendix~\ref{appnh} contains some 
technical details concerning the near horizon limit, while appendix \ref{oldsols} shows that the resulting supersymmetric $AdS_2\times Y_9$ solutions 
generalize those of \cite{Gauntlett:2006ns} by the addition of an angular momentum parameter. In appendix \ref{sec:angmom} we 
discuss how we define the angular momentum of the black holes, clarify various aspects of the gauge dependence
and also derive \eqref{angmomsurprise}. Appendix \ref{appeee} discusses conventions for Killing spinors while appendix \ref{bulkks} contains some details of how
to obtain the bulk Killing spinor, as well as how to obtain the associated Killing spinor on the three-dimensional conformal boundary.

\section{General setting}\label{sec:gensetting}
We will be interested in solutions of $D=4$ Einstein-Maxwell theory with action given by
\begin{align}\label{act}
S=\frac{1}{16\pi G_{(4)}}\int d^4x\sqrt{-g}\big(R+6-F^2\big)\,,
\end{align}
with $F=dA$.
This is the action for the bosonic fields of $D=4$, $\mathcal{N}=2$ minimal gauged supergravity. A solution to the equations of motion will preserve supersymmetry provided that it admits appropriate Killing spinors which we define later. Note that we have performed a scaling to
set the cosmological constant as $\Lambda=-3$ so that a unit radius $AdS_4$ vacuum solution, with vanishing gauge field,
solves the equations of motion. This vacuum solution preserves all of the $D=4$ supersymmetry. 

Any (supersymmetric) solution of this theory automatically uplifts to a (supersymmetric)
solution of $D=11$ supergravity on an arbitrary Sasaki-Einstein 7-manifold, $SE_7$ \cite{Gauntlett:2007ma}.  
The general uplifting ansatz is\footnote{Note that in \cite{Gauntlett:2007ma} $A^{\mathrm{there}}=2A^{\mathrm{here}}$.}
\begin{align}\label{lift}
d s^2_{11}& = L^2\left[ \tfrac{1}{4}d s^2_4 + \left(\eta + \tfrac{1}{2}A\right)^2 + d s^2_T\right]\, ,\nn
G &  =  L^3\left[\tfrac{3}{8}\vol_4 - \tfrac{1}{2}*_4 F \wedge J\right]\, ,
\end{align}
where $L>0$ is a constant which is eventually fixed by flux quantization.
Here $\eta$ is the contact one-form of the Sasaki-Einstein manifold, with transverse K\"ahler two-form\footnote{We have used the letter $J$ to denote both the K\"ahler two-form and the angular momentum of the black holes; the context should make it clear which one we are referring to.}
$J$ and associated transverse metric $d s^2_T$, 
with $d\eta=2J$. 
The vacuum $AdS_4$ solution, with $\diff s^2_4$ a unit radius $AdS_4$ and $A=0$, uplifts to the supersymmetric
$AdS_4\times SE_7$ solution, dual to an $\mathcal{N}=2$ SCFT in $d=3$. For later use we note that
\begin{align}\label{dualflux}
*_{11}G=L^6\Big[J^3\wedge (\eta+\tfrac{1}{2}A)+  \frac{1}{2^4}F\wedge J^2\wedge  (\eta+\tfrac{1}{2}A) \Big]\,.
\end{align}

In this paper we will be interested in the case that the Sasaki-Einstein manifold is in the {\it regular class}, for which $d s^2_T=ds^2_{KE_6}$ is a metric on $KE_6$, a K\"ahler-Einstein manifold. Necessarily, $KE_6$ has positive curvature and $d s^2_T$
is normalized so that $Ric(KE_6)=8g(KE_6)$ and hence $\rho=8J$, where $\rho$ is the Ricci form. Introducing local coordinates we 
can write 
\begin{align}\label{contactform}
\eta=\tfrac{1}{4}d\psip+\sigma\, ,\qquad d\sigma=2J\,.
\end{align}
The Sasaki-Einstein manifold admits a Killing spinor which has charge $1/2$ under $\partial_{\psip}$ (see appendix \ref{appeee}). That is, in 
a frame invariant under $\partial_{\psip}$ there is an explicit phase $\ex^{\ii \psip/2}$ in the spinor. This gives rise to a phase 
$\ex^{\ii \psip}$ in the holomorphic $(4,0)$-form 
on the cone over the $SE_7$, which is a bilinear in the Killing spinor.

In the sequel it will be important to recall that the $KE_6$ has a {\it Fano index} $I\in\mathbb{N}$. If $\mathcal{L}$ is the canonical line bundle over the $KE_6$
then $I$ is the largest integer for which there is a line bundle $\mathcal{N}$ with $\mathcal{L}=\mathcal{N}^I$, i.e. we are able to take the $I$th root of the canonical bundle. In general we may then take the period of $\psip$ to be
\begin{align}\label{psipp}
\Delta\psip = \frac{2\pi I}{k}\, ,
\end{align}
where $k$ is a positive integer that divides $I$. The fundamental group is then $\pi_1(SE_7)\cong \Z_k$, 
so that in particular when $k=1$ 
the resulting manifold is simply-connected.
For example, if $KE_6=\mathbb{CP}^3$, then $I=4$ and the associated $SE_7$ manifolds with $k=1, 2$ and $4$ 
are respectively $S^7$, $S^7/\Z_2$ and $S^7/\Z_4$. Note that $I/k\in\mathbb{N}$ ensures that the Killing spinor and 
holomorphic $(4,0)$-form 
 mentioned above are well-defined.
Some other well-known examples of  regular $SE_7$ are summarized in Table \ref{tabone}, with the associated values of $I$ (see e.g.  Theorem 3.1 of \cite{Boyer:1999ms}.)

Another integer quantity associated with the $KE_6$ that will frequently appear is defined by
\begin{align}\label{defM}
M\equiv \int_{KE_6}\left(\frac{\rho}{2\pi}\right)^3=\int_{KE_6}c_1^3\,,
\end{align}
where $\rho$ is the Ricci form for the $KE_6$. Values of $M$ for examples of $SE_7$ can also be found in Table \ref{tabone}.
We note that the volume of the $KE_6$ and the $SE_7$ can be expressed in terms of $M$ as follows
\begin{align}\label{volumes}
\mathrm{vol}(KE_6) 
= \frac{\pi^3 M}{3\times 2^7}~,\qquad
\vol(SE_7)=\frac{\pi^4 M I}{3\times 2^8 k}\,.
\end{align}

Our conventions for $D=11$ supergravity are as in e.g. \cite{Gauntlett:2002fz,Gauntlett:2003di}. We define $N_{SE}>0$ to be the quantized flux 
through the $SE_7$ manifold: 
\begin{align}\label{fluxse7}
N_{SE}\equiv \frac{1}{(2\pi \ell_p)^6}\int_{SE_7}*_{11}G=\frac{3 L^6\vol(SE_7)}{\pi^6 2^5 \ell_p^6}\,.
\end{align}
Here $\ell_p$ is the $D=11$ Planck length with the $D=11$ Newton's constant defined by $1/(16\pi G_{(11)})=1/(2\pi)^8\ell_p^9$. Carrying out the dimensional reduction to $D=4$
we can now express the $D=4$ Newton constant in the following useful form
\begin{align}\label{4dnewt}
\frac{1}{G_{(4)}}=\frac{2^{3/2}\pi^2}{3^{3/2}\vol(SE_7)^{1/2}}{N_{SE_7}^{3/2}}
=\frac{2^{11/2} k^{1/2}}{3 M^{1/2} I^{1/2}}N_{SE_7}^{3/2}\,.
\end{align}
We also recall that the free energy of the $d=3$ SCFT dual to the $AdS_4\times SE_7$ solution is given by
\begin{align}\label{4dnewtfree}
\mathcal{F}_{S^3}=\frac{\pi}{2G_{(4)}}\,.
\end{align}

\begin{table}[h!]
  \begin{center}
    \begin{tabular}{|l|c|c|c|} %
    \hline
      $KE_6$ & $I$ & $M$& $SE_7$  (with $k=1$)  \\
      \hline
      $\mathbb{CP}^3$ & 4 & $2^6$& $S^7$\\
      $\mathbb{CP}^1\times \mathbb{CP}^1\times \mathbb{CP}^1$& 2 & $2^4\times 3$&$Q^{1,1,1}$\\
      $\mathbb{CP}^2\times \mathbb{CP}^1$ & 1 &$2\times 3^3$ &$M^{3,2}$\\
            $SU(3)/T^2$ & 2 & $2^4\times3$&$N^{1,1}$\\
                  $Gr_{5,2}$ & 3 & $2\times 3^3$&$V_{5,2}$\\
                                    $dP_n\times \mathbb{CP}^1$ & 1 & $6\times (9-n)$& $\star$\\
                                    \hline
    \end{tabular}
  \end{center}
      \caption{Some examples of simply-connected ($k=1$) $SE_7$ manifolds obtained as circle bundles over a 
 $KE_6$ manifold. The integer $I$
      is the Fano index for the $KE_6$ and $M$ is defined in \eqref{defM}. Note that $dP_n$ are del-Pezzo surfaces with $n=3,\dots,8$ and the associated $SE_7$ do not have a name.}
          \label{tabone}
\end{table}

\section{$AdS_4$ PD black holes in $D=4$}\label{sec:PD}

We start with a sub-class of the class of Pleba\'nski--Demia\'nski (PD) solutions \cite{Plebanski:1976gy} of Einstein-Maxwell theory,
as presented in \cite{Podolsky:2006px}. 
The metric is given by
  \begin{align} \label{PDmetric}
d s^2 & =\frac{1}{\Omegach^2}\bigg\{
 -\frac{Q}{\Sigma}\big[d t- a\sin^2\theta\,d\phi \big]^2 +\frac{\Sigma}{Q}\,d r^2  
+\frac{\Sigma}{P}d\theta^2 \nn
& \qquad \qquad  \qquad \qquad +\frac{P}{\Sigma}\sin^2\theta \big[ ad t -(r^2+a^2)d\phi \big]^2  
 \bigg\}, 
\end{align}
where
\begin{align}
 \Omegach&=1-\alpha\, r\cos\theta\,,\nn
\Sigma&=r^2+a^2\cos^2\theta \,, \nn
 P&= 1-2\alpha m\cos\theta +\Big(\alpha^2(a^2+e^2+g^2)- a^2\Big)\cos^2\theta \,,\nn
  Q&= \Big(r^2-2mr+a^2+e^2+g^2 \Big) (1-\alpha^2r^2)  +(a^2+r^2)r^2 \,,
\end{align}
and we note that $P=P(\theta)$ and $Q=Q(r)$ while $\Omegach$ and $\Sigma$ depend on both $r$ and $\theta$.
The gauge field is given by
\begin{align} \label{PDgauge}
A=-e\frac{r}{\Sigma}(dt-a\sin^2\theta d\phi)+g\frac{\cos\theta}{\Sigma}(adt-(r^2+a^2)d\phi)\,.
\end{align}
The solution depends on five free parameters $m,e,g,a$ and $\alpha$, which we can loosely associate with mass, electric charge, magnetic charge, rotation and 
acceleration, respectively. 
We note that we have set a possible NUT (Newman-Unti-Tamburino) parameter in the PD metrics to zero, since we want to avoid closed timelike 
curves. We have fixed the cosmological constant to be $\Lambda=-3$, as in \eqref{act}, and finally we note that we have changed the sign of $g$ compared with \cite{Podolsky:2006px}.
 
We will assume $m>0$. Note that there are various coordinate changes $\theta\to \pi-\theta$, $t\to-t$, $\phi\to-\phi$ as
 well as $A\to -A$, which change the signs of the pairs $(\alpha,g)$, $(a,e)$, $(a,g)$ and $(e,g)$, respectively. If $\alpha=0$ we
 can thus choose $e,g,a\ge 0$ without loss of generality. If $\alpha\ne0$ we can choose 
 $m,\alpha>0$ and $e,g\ge 0$
 with either $a\ge 0$ or $a\le 0$. However, in order to get supersymmetric, extremal black holes one should take $a\ge 0$. Thus, in the sequel we
 focus on 
 \begin{align} \label{signchoice}
 m>0\, , \qquad \alpha,e,g, a\ge 0\,.
  \end{align}

 The range of the $\theta$ coordinate is taken to be $0\le\theta\le\pi$. We will assume that $P(\theta)>0$ in this range.\footnote{As in \cite{Anabalon:2018qfv}, 
 this can be achieved if $m\alpha<\frac{1}{2}\Xi$ for $\Xi\in (0,2]$ and $m\alpha<\frac{1}{2}(\Xi-1)^{1/2}$ for $\Xi >2$,
 where $\Xi\equiv 1+\alpha^2(a^2+e^2+g^2)- a^2$.}
 It will be convenient to define 
\begin{align}
\theta_-=0\, ,  \qquad \mbox{and}  \qquad \theta_+=\pi\ .
\end{align}
 On slices of constant $t,r$, we can examine the behaviour of the metric as we approach $\theta=\theta_\pm$ and we find
\begin{align}\label{metthetphid4}
 ds^2_{\theta,\phi}\approx \left[\frac{\Sigma}{P\Omegach^2}\right]_{\theta=\theta_\pm} \left[d\theta^2+P_\pm^2(\theta-\theta_\pm)^2 d\phi^2\right]\,,
 \end{align}
 where 
 \begin{align}\label{ppmexp}
 P_\pm\equiv P(\theta_\pm)= 1\pm2\alpha m +\alpha^2(a^2+e^2+g^2)- a^2\,,
 \end{align}
 are constants. With $\alpha m \ne0$ it
 is not possible to choose a period for $\phi$ so that we obtain a smooth metric on a round two sphere as there will always be conical deficits
 at one or both of the poles. Thus, as is well known, the mismatch of these conical deficits at the two poles is directly connected with the non-vanishing of the acceleration parameter. 
 
 A simple observation, which will turn out to be important in obtaining regular solutions in $D=11$, is that we can suitably constrain the parameters in the metric, and choose a corresponding period for $\phi$, so that the conical defects give rise to an orbifold 
known as a spindle, or equivalently a weighted projective space.
We can demand that the ratio $P_+/P_-$ is a rational number, which we write as 
\begin{align}\label{PpPm}
\frac{P_+}{P_-}=\frac{n_-}{n_+}\,,
\end{align}
with $n_\pm\in\mathbb{N}$,
and then choose the period of $\phi$ to be
  \begin{align}\label{delphi}
  \Delta\phi= \frac{2\pi}{n_+P_+}= \frac{2\pi}{n_-P_-}\,,
  \end{align}
and hence from \eqref{ppmexp} we can also write $\Delta\phi=\frac{\pi}{2m\alpha}\frac{n_--n_+}{n_- n_+}$.
The two-dimensional space parametrized by $\theta$ and $\phi$ is then the 
weighted projective space  $\mathbb{WCP}^1_{[n_-,n_+]}$ (for some further
discussion on  $\mathbb{WCP}^1_{[n_-,n_+]}$, see appendix \ref{app:WCP}).
We will see that we are always able to desingularize the singularities of this spindle after uplifting to $D=11$.
As we will see, our procedure will require that we impose the condition $m\alpha=g$; we will also find that 
$m\alpha=g$ is implied by supersymmetry.
It will also be useful in the sequel to introduce a coordinate $\varphi$ on the spindle that has period $2\pi$:
\begin{align}\label{canangcoord}
\varphi\equiv \left(\frac{2\pi}{\Delta\phi}\right)\phi\,.
\end{align}

In order to have a spacetime with a black hole horizon, located at $r=r_+$, 
we demand that $Q(r_+)=0$ and the region exterior to the black hole has $r\ge r_+$. We also require $\Omegach> 0$ and hence the radial coordinate is constrained via
\begin{align} 
\alpha r \cos\theta<1\,,
\end{align} 
and so, in particular, $\alpha r_+<1$. 
The conformal boundary is approached when $\alpha r \cos\theta\to 1$ and one finds 
that the conical defects are still present when $\alpha\ne 0$.
There can also be an acceleration horizon. We discuss this in 
section \ref{sec:conboundary}, where we study the global structure of the 
non-rotating black holes in more detail. In particular, although $Q$ has no other roots for $r>r_+$, 
one can continue the radial coordinate $r$ past $r=\infty$, and there can effectively be another root 
beyond this, corresponding to an acceleration horizon.  A detailed analysis 
of the various cases and their Penrose diagrams may be found in \cite{Dias:2002mi}.
Some
further discussion of these black holes may also be found in \cite{Anabalon:2018qfv} which calculates various thermodynamic quantities (when $g=0$).
Here we record that the entropy of the black holes is given by
\begin{align}\label{entgenexp}
S_{BH} & = \frac{1}{4 G_{(4)}}A = \frac{(r_+^2 +a^2)\Delta\phi}{2G_{(4)}(1-\alpha^2 r_+^2) } \,.
\end{align}
In the next section we show how the PD metrics can be desingularized by uplifting on certain $SE_7$ manifolds described in the previous section.
This analysis will fix $\Delta\phi$, and furthermore the entropy can then be expressed in terms of either $N_{SE_7}$ using \eqref{4dnewt}, 
or the free energy of the $d=3$ SCFT using \eqref{4dnewtfree}.
 For the special class of supersymmetric extremal black holes we then obtain the expression~\eqref{blackholent}.

We also record that the magnetic flux through the horizon, $Q_m$, is defined by 
\begin{align}\label{magflux}
G_{(4)}Q_m\equiv \frac{1}{4\pi }\int_{r=r_+} F=g\,\frac{\Delta\phi}{2\pi}\,
\end{align}
and the electric flux, $Q_e$, is defined by
\begin{align}\label{elecflux}
G_{(4)}Q_e\equiv \frac{1}{4\pi }\int_{r=r_+} *F=e\,\frac{\Delta\phi}{2\pi}\,.
\end{align}
Finally, we can introduce the angular momentum of the black hole.
Rather than a simple Komar integral, as often used, we define the angular momentum as a conserved quantity 
that can be equally evaluated either at the conformal boundary or at the black hole horizon 
(see e.g. \cite{Papadimitriou:2005ii}). Associated with the Killing vector $k=\partial_{\varphi}$, where $\varphi\in[0,2\pi)$
was introduced in \eqref{canangcoord}, we introduce the two-form 
\begin{align}
\Hch=dk+4(A\cdot k)F\,,
\end{align}
and then define the angular momentum via 
\begin{align}
J(A)=\frac{1}{16\pi G_{(4)}}\int_{r=r_+}*\Hch\,.
\end{align}
 The angular momentum is then a kind of Page charge that depends on the gauge.\footnote{Note that when $\alpha=0$ one can explicitly check that the expression for the angular momentum, in the gauge we are using, 
agrees with a Komar integral \eqref{komdef} evaluated at the conformal boundary, which is an expression that is often used to define the angular momentum. We have not verified whether or not this is also the case when $\alpha\ne 0$.} We explore this gauge dependence in appendix \ref{sec:angmom}, where we also highlight a subtle difference with a natural definition of the angular momentum of the near horizon 
 $AdS_2\times Y_9$ limit for the supersymmetric and extremal black holes. Using the gauge as in \eqref{PDgauge} we obtain
 \begin{align}\label{angmtmdef}
G_{(4)}J=m\,a\,\left(\frac{\Delta\phi}{2\pi}\right)^2\,,
\end{align}
which agrees with formulas in the literature for the non-accelerating limit.

\section{Desingularising via the uplift to $D=11$}\label{sec:uplift}

We now consider the PD metrics uplifted to $D=11$ on a Sasaki-Einstein manifold as in \eqref{lift}. 
In this section we will assume that we have non-zero acceleration in $D=4$: 
\begin{align} 
\alpha\ne0\,. 
\end{align} 
In section \ref{sec:liftmetric} we will first analyse the regularity of the metric; as this is somewhat involved, we have included a summary section 
at the end. We then analyse flux quantization in section \ref{ssfq}.

\subsection{Metric}\label{sec:liftmetric}

In analysing the regularity of the $D=11$ supergravity solution, it is convenient to first perform\footnote{Notice that, effectively, the same thing is achieved  by shifting the $\psip$ coordinate via $\psip\to \psip+2c\phi$.}  
a local gauge transformation on the gauge field 
of the form $A\rightarrow A+c\, d\phi$, where $c$ is a constant which will be fixed shortly. 
In the associated $D=11$ metric, we can focus on the metric on a constant $r,t$ slice which is given by 
\begin{align} \label{metricY9}
ds^2_{\theta,\phi,KE_6,\psi}&= 
\frac{1}{4}\left[ g_{\theta\theta}d\theta^2+g_{\phi\phi} d\phi^2\right]+ 
(\tfrac{1}{4}d\psip+\sigma+\tfrac{1}{2}\tilde{A}_\phi d\phi)^2+ds^2(KE_6)\,, 
\end{align} 
where 
\begin{align} 
\tilde{A}_\phi&=\frac{1}{\Sigma}\left[era\sin^2\theta-g\cos\theta(r^2+a^2)\right]+c\,,\nn
g_{\phi\phi}&=\frac{\sin^2\theta}{\Omegach^2\Sigma}\left[-Qa^2\sin^2\theta+P(r^2+a^2)^2\right]\,,\nn
g_{\theta\theta}&=\frac{\Sigma}{\Omegach^2 P}\,,
\end{align}
with $\tilde{A}_\phi=A_\phi + c$.
The most general construction will show that the $(\theta,\phi,\psip)$ part of the metric can be taken to parametrize a smooth $S^3$, or more
generally a Lens space $S^3/\Z_\pp$, that is then fibred over the $KE_6$. 

This $S^3$ can be embedded as $S^3\subset \C^2$, and we would like to find the $U(1)$ generators that rotate the two copies of $\C$. That is, 
we introduce $\ell_\pm =\partial_{\alpha_\pm}$ where $\alpha_\pm$ are $(2\pi)$-period polar coordinates for each copy of $\C$. 
By definition, then $\|\ell_+\|^2(\theta_+)=0$ and $\|\ell_-\|^2(\theta_-)=0$, and moreover the surface gravity for 
$\ell_\pm$ satisfies $\kappa_\pm^2=1$ at $\theta=\theta_\pm$. 
Choosing the constant $c$ appearing in the gauge field to be
\begin{align}
c= \frac{g}{4m\alpha}(P_++P_-)~,
\end{align}
implies that $ \ell_\pm $ have the same coefficient of $\partial_{\psip}$,
with 
\begin{align}
\partial_{\alpha_\pm} = \ell_\pm = \frac{g}{m\alpha}\partial_{\psip} -\frac{1}{P_\pm}\partial_\phi~.
\end{align}
Equivalently
\begin{align}\label{psiphi}
\psip= \frac{g}{m\alpha}\left(\alpha_++\alpha_-\right)~, \qquad \phi = -\frac{1}{P_+}\alpha_+ - \frac{1}{P_-}\alpha_-~.
\end{align}
Using the comments in  section \ref{sec:gensetting}, we then conclude that the $D=11$ Killing spinor will have charge 
$g/2m\alpha$ under both $\partial_{\alpha_\pm}$. Furthermore, later we will see that one of the BPS equations which ensures that the $D=4$ or $D=11$ solutions preserve supersymmetry is precisely $g/m\alpha=1$.

To proceed we next define the functions
\begin{align}
A_\pm & = \frac{1}{4P_\pm^2}g_{\phi\phi} + \Big(\frac{g}{4m\alpha}-\frac{1}{2P_\pm}\tilde{A}_\phi\Big)^2~,\nn
B & = \frac{1}{4P_+ P_-}g_{\phi\phi} + \Big(\frac{g}{4m\alpha}-\frac{1}{2P_+}\tilde{A}_\phi\Big)\Big(\frac{g}{4m\alpha}-\frac{1}{2P_-}\tilde{A}_\phi\Big)~,
\end{align}
and the connection one-forms
\bea
D\alpha_\pm = d\alpha_\pm \pm \frac{P_\pm}{g}\sigma~,
\eea
to find that the metric can be written
\begin{align}
ds^2_{\theta,\phi,KE_6,\psi} & = \frac{g_{\theta\theta}}{4}d\theta^2 + A_+(D\alpha_+)^2 + A_-(D\alpha_-)^2 + 2B(D\alpha_+)(D\alpha_-) 
+ ds^2(KE_6)~.
\end{align}
Here $A_+(\theta_+)=0=A_-(\theta_-)$, and $B(\theta_\pm)=0$.

\newcommand{\alc}{\gamma}

It is next convenient to  introduce the coordinates
\begin{align}
\alc = \alpha_+ + \alpha_-~, \qquad \beta = \alpha_+-\alpha_-\, ,
\end{align}
so that $\partial_\alc = \frac{1}{2}(\partial_{\alpha+}+\partial_{\alpha-})$, $\partial_\beta=\frac{1}{2}(\partial_{\alpha+}-\partial_{\alpha-})$. 
We then make the periodic identifications
\begin{align}\label{periodbetagamma}
\Delta\alc=2\pi\, , \qquad \Delta\beta=4\pi/\pp\, .
\end{align}
This leads to a smooth $S^3$ fibre when $\pp=1$, and more generally a Lens space $S^3/\Z_{\pp}$. 
Here we are using the fact that a Lens space $S^3/\Z_{\pp}$ may be viewed as the total 
space of a circle fibration, here with circle fibre coordinate $\beta$, over a two-sphere. Specifically, 
in the construction above, the two-sphere has standard spherical polar coordinates $(\theta,\alc)$.
Notice from our comments above that 
the spinors are charged under $\partial_\alc$, but not under $\partial_\beta$, so we may quotient 
the period of $\beta$ by $\pp$ and preserve supersymmetry. For bosonic solutions more generally 
we could in principle take further/different quotients but we will not investigate this further here. 

Having established that the fibre is a Lens space, we now examine the fibration over the $KE_6$. 
We will do this in two steps, firstly discussing an $S^2$ bundle over the $KE_6$, with the $S^2$ parametrized by $(\theta,\alc)$,
and then discussing the circle bundle over this, with the circle parametrized by $\beta$.
After
writing $D\alpha_\pm\equiv \frac{1}{2}D\alc\pm\frac{1}{2}D\beta$,
the connection forms are given by
\begin{align}\label{connforms}
D\alc = d\alc + \frac{1}{g}(P_+-P_-)\sigma~, \qquad D\beta = d\beta + \frac{1}{g}(P_++P_-)\sigma~.
\end{align}
Focussing on $D\gamma$ we recall from \eqref{contactform} that $d\sigma = \rho/4$, where $\rho$ is the Ricci form for the $KE_6$. 
We can also compute
\begin{align}
\frac{1}{4g}(P_+-P-) = \frac{m\alpha}{g}~.
\end{align}
Thus, since $\alc$ has period $2\pi$, we will obtain an $S^2$ bundle over the $KE_6$ with fibre parametrized by $(\theta,\alc)$ being the Riemann sphere compactification 
of a well-defined line bundle, provided that $m\alpha/g = n/I$, where $I$ is the Fano index of the $KE_6$ and $n$ is an integer. 
The line bundle is then $\mathcal{L}^{n/I}$. However, recall we also commented earlier that $g/2m\alpha$ is  precisely the charge 
of the $D=11$ Killing spinors under $\partial_{\alpha_\pm}$, and hence $\partial_\alc$, and we shall later find 
that one of the BPS equations is precisely $g/m\alpha=1$, so these charges are all 1/2. Thus, we will
impose
\begin{align}\label{BPStop}
\frac{1}{4g}(P_+-P_-) = \frac{m\alpha}{g} = 1\,,
\end{align}
and this implies that the $(\theta,\alc)$ $S^2$ bundle over the $KE_6$ 
is necessarily that associated to the canonical bundle. 
For the bosonic solutions more generally we could instead take powers of the canonical bundle, with different $n$ above,
but here we do not. 

It remains to ensure that we have a well-defined circle bundle, with circle fibre coordinate $\beta$, over the above 
$S^2$ bundle over the $KE_6$. As described in \cite{Gauntlett:2004hh}, a necessary and sufficient 
condition for this is to verify that the corresponding curvature two-form has appropriately quantized periods over a basis of two-cycles. 
One such two-cycle is a copy of the fibre $S^2$, at a fixed point on the $KE_6$, and 
this has already fixed the period $\Delta\beta=4\pi/\pp$ in \eqref{periodbetagamma}, to obtain 
a Lens space fibre $S^3/\Z_\pp$ over the $KE_6$. 
The remaining two-cycles may be taken to be two-cycles in the  copy of the $KE_6$ base at 
either $\theta=\theta_+$ or $\theta=\theta_-$. For example, in the  former case the corresponding circle bundle 
over the $KE_6$ has connection term $D\alpha_-$, and quantizing the periods of the associated curvature two-form leads 
to setting
\begin{align}\label{Pminusexp}
\frac{P_-}{4g} = \frac{\kp}{\pp I }~,
\end{align}
where $\kp\in\mathbb{N}$. Specifically, setting $\theta=\theta_+=\pi$ then gives a circle, parametrized by $\alpha_-$, 
inside the Lens space fibre. Since on this circle 
 $\alpha_-$ has period $2\pi/\pp$, the choice \eqref{Pminusexp} implies that we obtain the circle bundle associated to 
$\mathcal{L}^{\kp/I}$. It then follows that 
\begin{align}\label{Pplusexp}
\frac{P_+}{4g} = 1 +\frac{\kp}{\pp I} = \frac{\pp I+\kp}{\pp I}~,
\end{align}
so that the 
$\alpha_+$ circle bundle at $\theta=\theta_-=0$ is that associated to $\mathcal{L}^{(\pp I+\kp)/I}$. 
We also deduce that
\begin{align}\label{PpPmagain}
\frac{P_+}{P_-} = \frac{\pp I+\kp}{\kp}~.
\end{align}
With these choices we have thus constructed a regular Lens space $S^3/\Z_\pp$ fibration 
over the $KE_6$, where the positive integer parameter $\kp$ determines the twisting. 
The total space $Y_9$ of this fibration is simply-connected if we further 
require $\mathrm{hcf}(\kp,\pp)=1$. We shall henceforth also assume this condition, but note that 
more generally the topology is simply a free $\mathbb{Z}_{\mathrm{hcf}(\kp,\pp)}$ quotient 
of the solution with parameters $(\kp,\pp)/\mathrm{hcf}(\kp,\pp)$, so there is essentially no loss of generality.

Recall that as originally presented in \eqref{metricY9}, the space $Y_9$ is a fibration of a $SE_7$ 
over the two-dimensional space parametrized by $(\theta,\phi)$. It will be important to understand 
this fibration structure also. 
In particular, 
the  Reeb vector $\partial_{\psip}$ that rotates the $SE_7$ $U(1)$ fibre over the $KE_6$ may be 
computed to be
\begin{align}\label{trickypsi}
\partial_{\psip} &= \partial_{\alpha_+} + \frac{\kp}{I} \cdot \frac{1}{\pp}\left(\partial_{\alpha_+} - \partial_{\alpha_-}\right)\, .
\end{align}
Note that moving $2\pi $ along the orbits of both $\partial_{\alpha_+}$ and $\frac{1}{\pp}\left(\partial_{\alpha_+} - \partial_{\alpha_-}\right)$ 
returns to the same point on the Lens space fibre, in the latter case precisely because we took a $\Z_\pp$ quotient 
of $S^3$ along the $U(1)$ generated by $\partial_{\alpha_+} - \partial_{\alpha_-}$. 
To determine the period of $\psip$ it is useful to rewrite \eqref{trickypsi} as
\begin{align}\label{periodpain}
\frac{I}{k}\partial_{\psip} = \frac{I}{k}\partial_+ + \frac{\kp}{k}\cdot \frac{1}{\pp}\left(\partial_+ - \partial_-\right)\, ,
\end{align}
where we have defined
\begin{align}\label{kdef}
k = \mathrm{hcf}(I,\kp)~.
\end{align}
This ensures that moving $2\pi$ along the orbit of the vector field on the right hand side of \eqref{periodpain} 
closes, and moreover this is the minimal period. But this  shows 
 that on the Lens space $\psip$ has period $2\pi I/k$, 
\begin{align}\label{psipperiod}
\Delta\psip = \frac{2\pi I}{k}\, ,
\end{align}
precisely as in \eqref{psipp}, where specifically $k$ is fixed via \eqref{kdef}. 

Recalling that 
\begin{align}
\psip = \alpha_+ + \alpha_-~, \qquad \phi = -\frac{1}{P_+}\alpha_+ - \frac{1}{P_-}\alpha_-\, ,
\end{align}
and that the torus  made up of $\alpha_+$, $\alpha_-$ has volume $(2\pi)^2/\pp$, 
we deduce from the Jacobian of this transformation that $\phi$ has period 
\begin{align}
\Delta\phi = \left(\frac{1}{P_-}-\frac{1}{P_+}\right)\frac{2\pi}{\pp}\frac{k}{I}~.
\end{align}
From the discussion below \eqref{metthetphid4} and specifically \eqref{delphi}, we see that $(\theta,\phi)$ on the base of the $\psip$ fibration 
at fixed point on the $KE_6$ will be a spindle/weighted projective space
$\mathbb{WCP}^1_{[n_-,n_+]}$, where 
 we may identify
\begin{align}\label{nplusminus}
n_- = \frac{\pp I+\kp}{k}~, \qquad  n_+ = \frac{\kp}{k}~.
\end{align}
Notice that these are indeed integers, due to the definition \eqref{kdef} of $k$, and moreover since 
$\mathrm{hcf}(\kp,\pp)=1$ we also note that $\mathrm{hcf}(n_+,n_-)=1$. 

To complete the viewpoint of the $SE_7$ as the fibre, we may also look directly at the 
 $\psip$ circle fibration over the weighted projective space. 
Recalling that $\psip$ has period $2\pi I/k$ and 
using $A_\phi(\theta_\pm)=\pm g$, we calculate the Chern number of this fibration as
\begin{align}\label{chernnumber}
\frac{1}{\pi I/2k} \int_{\mathbb{WCP}^1} \frac{1}{2}d(\tilde{A}_\phi\diff\phi) = \frac{k^2 \pp }{\kp^2+ \kp \pp I} = \frac{k}{I}\frac{n_--n_+}{n_+n_-} = \frac{\pp}{n_+ n_-}~.
\end{align}
Here we have used the fact that 
\begin{align}
\mathtt{q} = \frac{k}{I}(n_--n_+)\, .
\end{align}
Since $\mathtt{q}$ is an integer, notice that the difference of the weights $n_--n_+$ is necessarily divisible by the integer $I/k$.
The orbifold line bundle over $\mathbb{WCP}^1_{[n_-,n_+]}$ with Chern number \eqref{chernnumber}, denoted $O(\mathtt{q})$, is 
 discussed in  appendix \ref{app:WCP}. In that appendix it is shown that the total space of the associated circle bundle
is indeed a Lens space $S^3/\Z_\pp$, completing the circle of arguments. 

\

\noindent {\bf Summary:}  We can summarize the results of this section as follows. The $D=4$ PD metrics of interest depend on five parameters
$m,a,e,g,\alpha$. We uplift to $D=11$ using a $SE_7$ manifold in the regular class, which is a circle bundle  over a $KE_6$ manifold.
We then obtain a regular $D=11$ solution after imposing the following two
constraints on the five parameters:
\begin{align}\label{allconds}
\frac{P_+}{4g}=1+\frac{\kp}{I\pp}\,,
\qquad
\frac{P_-}{4g}=\frac{\kp}{I\pp}\,,
\end{align}
where $\pp\in\mathbb{N}$ and $\kp\in\mathbb{N}$, and we take $\mathrm{hcf}(\kp,\pp)=1$ so that 
the total space is simply-connected. Here $I$ is the Fano index for the $KE_6$ associated with the regular $SE_7$ and
$P_\pm$ are given in \eqref{ppmexp}. Defining
\begin{align}\label{kpick}
k = \mathrm{hcf}(I,\kp)~,
\end{align}
we then choose the  periods of $\psip$ and $\phi$ to be
\begin{align}\label{defperiods}
\Delta\psip = \frac{2\pi I}{k}\,,\qquad
\Delta\phi = \left(\frac{1}{P_-}-\frac{1}{P_+}\right)\frac{2\pi}{\pp}\frac{k}{I}\,.
\end{align}
The  nine-dimensional manifold $Y_9$ at fixed $t,r$ is then a Lens space $S^3/\Z_\pp$ bundle over the $KE_6$, 
while the $SE_7$ at fixed point in $D=4$ spacetime has fundamental group $\Z_k$, and 
is the circle bundle over the $KE_6$ associated to the line bundle $\mathcal{L}^{k/I}$, 
where $\mathcal{L}$ is the canonical bundle over the $KE_6$.
Moreover, the base space of the $SE_7$ fibration, at fixed $t$, $r$, 
is parametrized by $\theta$, $\phi$ of the $D=4$ metric, which is 
topologically a spindle, a weighted projective space
$\mathbb{WCP}^1_{[n_-,n_+]}$. Here 
\begin{align}\label{nplusminussummary}
n_- = \frac{\pp I+\kp}{k}~, \qquad  n_+ = \frac{\kp}{k}\,,
\end{align} 
are relatively prime integers, where $\theta=\theta_-=0$ is 
an $\R^2/\Z_{n_-}$ orbifold singularity, while $\theta=\theta_+=\pi$ is 
an $\R^2/\Z_{n_+}$ orbifold singularity. 
The magnetic flux in \eqref{magflux} through the spindle horizon in the $D=4$ spacetime is given by the rational number
 \begin{align}\label{magnpnm}
 G_{(4)}Q_m=\frac{n_--n_+}{4 n_+n_-}\,.
 \end{align}

Conversely, we can begin with a weighted projective space $\mathbb{WCP}^1_{[n_-,n_+]}$, 
with arbitrary coprime integers $n_->n_+$. We then set
\begin{align}\label{pickqp}
\pp = \frac{n_--n_+}{I/k}\, , \qquad \kp = k n_+\, .
\end{align}
Here we choose the integer 
\begin{align}\label{kalt}
k=\frac{I}{\mathrm{hcf}(I,n_--n_+)}\, .
\end{align}
With this definition of $k$ 
we have that $I/k$ is an integer that  divides $n_--n_+$, which ensures 
that $\kp$ and $\pp$ in \eqref{pickqp} are integer. 
Moreover, note that we also have $k=\mathrm{hcf}(I,\mathtt{p})$, as in \eqref{kpick}.\footnote{To see this, 
write $\mathtt{f}=\mathrm{hcf}(I,n_--n_+)$, so that $k=I/\mathtt{f}$. Then compute 
$\mathrm{hcf}(I,\kp)= \mathrm{hcf}\left((I/\mathtt{f})\mathtt{f},(I/\mathtt{f})n_+\right) = 
(I/\mathtt{f})\mathrm{hcf}(\mathtt{f},n_+) = I/\mathtt{f} = k$, where 
in the penultimate step we have used $\mathrm{hcf}(n_--n_+,n_+)=\mathrm{hcf}(n_-,n_+)=1$.} 
The above 
 construction then leads to a 
Lens space fibration over the $KE_6$.

Since we have imposed two conditions \eqref{allconds}, we are left with a three-parameter family of non-singular, rotating and accelerating black hole solutions with spindle horizon. The three parameters correspond to the three independent 
physical conserved quantities, namely mass, electric charge $Q_e$ and angular momentum $J$.
The above conditions are consistent\footnote{As we noted just below \eqref{BPStop}, if one is just interested in a purely
bosonic solution, then one can relax the conditions a little and still maintain regularity.}
with preservation of supersymmetry as we discuss in the next section. 
The entropy of the black holes is given by
\eqref{entgenexp}, after using \eqref{4dnewt}.
In particular,
from \eqref{ppmexp}, the conditions \eqref{allconds} imply 
\begin{align}
\frac{m\alpha}{g} = 1\,.
\end{align}

Finally, we further illustrate with a concrete example. We take $KE_6=\mathbb{CP}^3$ and there are then three cases. Firstly, we have an $S^7$ fibration over
$\mathbb{WCP}^1_{[n_-,n_+]}$ 
for $\mathtt{p}=1,3,5,\dots$ and relatively prime $\mathtt{q}$, with $n_+=\mathtt{p}$ and $n_-=4\mathtt{q}+\mathtt{p}$. Secondly,
we have an $S^7/\mathbb{Z}_2$ fibration
for $\mathtt{p}=2,6,10,\dots$ and relatively prime $\mathtt{q}$, with $n_+=\mathtt{p}/2$ and $n_-=2\mathtt{q}+\mathtt{p}/2$. Finally,
we have an $S^7/\mathbb{Z}_4$ fibration
for $\mathtt{p}=4,8,12,\dots$ and relatively prime $\mathtt{q}$, with $n_+=\mathtt{p}/4$ and $n_-=\mathtt{q}+\mathtt{p}/4$.

\subsection{Flux quantization}\label{ssfq}
We can also quantize the flux in the $D=11$ solution \eqref{lift}. There is no quantization condition 
on the four-form $G$ as there are no non-trivial four-cycles. We therefore consider the dual seven-form $*_{11}G$ as given in
\eqref{dualflux}.
We have already seen  in \eqref{fluxse7} that the flux through the $SE_7$ fibre over a point in the $D=4$ spacetime 
gives  $N_{SE}$. 
In this section we present a general analysis 
for ensuring the fluxes through an appropriate basis of seven-cycles are quantized, by determining the constant $L$ in \eqref{lift}.

Note first that as for the previous subsection we may restrict to a constant $r$, $t$ slice, 
since these directions span $\R^2$ and hence don't contribute to any non-trivial cycles. 
The resulting nine-manifold $Y_9$ is a Lens space $S^3/\Z_\pp$ fibred over the $KE_6$. 
This is the same topology as the solutions discussed in appendix D.2 of 
\cite{Gauntlett:2019pqg}, and indeed later in the paper we shall see that those solutions
are precisely the near horizon limit of the black holes we are discussing, when the rotation parameter is set to zero. 

Setting $\theta=\theta_\pm$ gives rise to two seven-cycles that we call $D_\pm$.\footnote{These
were called $\tilde{D}_0$ and ${D}_0$ in \cite{Gauntlett:2019pqg}, respectively.} 
There are also seven-cycles $D_a$ that arise as the Lens space fibred over four-cycles
$\Sigma_a \in H_4(KE_6,\Z)$, where by definition these form a basis for the free part of the 
latter homology group. We may then write
\begin{align}
c_1 = c_1(KE_6) = I s_a\Sigma_a~,
\end{align}
where recall that $I$ is the Fano index and the $s_a$ are then coprime integers, and we have 
identified $H^2(KE_6,\Z)\cong H_4(KE_6,\Z)$ using Poincar\'e duality. As discussed in  \cite{Gauntlett:2019pqg}, 
we then have the homology relation
\begin{align}\label{homrel}
D_- = D_+ - I s_a D_a~.
\end{align}

Writing 
\begin{align}
N(D) \equiv \frac{1}{(2\pi \ell_p)^6}\int_{D}*_{11}G\,,
\end{align}
as the flux through the seven-cycle $D$, from \eqref{dualflux} we compute
\begin{align}
N(D_\pm ) = \frac{L^6}{(2\pi\ell_p)^6}6\mathrm{vol}(KE_6)\int_{S^1_{\theta=\theta_\pm}} \eta~, 
\end{align}
where recall that $\eta$ is the contact one-form of the Sasaki-Einstein manifold. 
A short computation shows that
\begin{align}
\eta\mid_{\theta=\theta_\pm} = \frac{1}{4}d\psi + \frac{1}{2}\tilde{A}_\phi(\theta_\pm) d\phi =  \mp\frac{I\pp}{4 k n_\pm}d\alpha_\mp~,
\end{align}
where $\alpha_\pm$ are the coordinates introduced in the previous subsection in \eqref{psiphi}, and recall that $n_\pm$ 
are defined in terms of $\pp$, $\kp$, $I$ and $k$ via \eqref{nplusminussummary}. Since 
$\alpha_\pm$ have period $2\pi/\pp$ through their respective circles $S^1\subset S^3/\Z_\pp$, we deduce that (choosing orientations to give a positive flux)
\begin{align}\label{NDpm}
N(D_\pm) = \frac{L^6}{(2\pi\ell_p)^6}\frac{I M\pi^4}{128k  n_\pm}~,
\end{align}
where we used \eqref{volumes}.
Similarly, we compute
\begin{align}
N(D_a) =  \frac{L^6}{(2\pi\ell_p)^6}\frac{1}{4}g \Delta\phi \Delta\psi\frac{1}{2!}\int_{\Sigma_a} J^2 ~.
\end{align}
Using
\begin{align}
\frac{1}{2!}\int_{\Sigma_a} J^2 = \frac{\pi ^2}{2^5}I^2n_a~,
\end{align}
where $n_a\equiv \int_{\Sigma_a}(c_1/I)^2$ are coprime integers, and inserting the periods 
$\Delta\phi=2\pi/(n_+P_+)$, $\Delta\psi=2\pi I/k$ from the previous section, we find
\begin{align}\label{NDa}
N(D_a) =  \frac{L^6}{(2\pi\ell_p)^6}\frac{I^4\pp\pi^4}{128 k^2 n_+n_-}n_a~.
\end{align}
Using $s_an_a=M/I^3$, one can check that the fluxes (\ref{NDpm}), \eqref{NDa} satisfy
the homology relation \eqref{homrel}. 

Since the cycles we have introduced form a basis of seven-cycles, we can now introduce 
a minimal flux number, that we call $N$, such that all fluxes are integer multiples of $N$. 
Specifically, this fixes $L$ via\footnote{Note that this is consistent with (D.18) of \cite{Gauntlett:2019pqg} after taking into account a
difference of $32/3$ in the $L^2$ between here and there, as a result of this factor in \eqref{bbmet}.}
\begin{align}\label{defL}
\frac{L^6}{(2\pi\ell_p)^6} = \frac{128 k^2 n_+n_-}{I^4h\pi^4}N~,
\end{align}
where we have introduced $h=\mathrm{hcf}(M/I^3,\pp)$, where recall that $\pp=(k/I)(n_--n_+)$ 
is an integer.
We then find 
\begin{align}
N(D_\pm) = \frac{M }{I^3h} k n_\mp N~, \qquad N(D_a) = \frac{\pp}{h}  n_a N~,
\end{align}
where the factors are all integers. Moreover, the expression for $N_{SE}$ given in \eqref{fluxse7}
can then be written as
\begin{align}\label{nseexpr}
N_{SE} = n_- N(D_-) = n_+ N(D_+) = \frac{M}{I^3h}k n_+ n_- N~.
\end{align}
Notice that this is indeed an integer.

We have thus shown that the $D=4$ PD black hole metrics of interest 
uplift to smooth $D=11$ solutions with properly quantized flux, provided that we impose
\eqref{allconds}--\eqref{nplusminussummary} and fix $L$ via \eqref{defL}.

\section{Supersymmetric and extremal limits}\label{sec:BPS}
In this section we analyse the additional conditions for supersymmetry, as well as the conditions required to have an extremal
black hole horizon with vanishing surface gravity. In general, the extremality condition is not implied by supersymmetry. 
We also derive the black hole entropy formulae \eqref{blackholent}.

By definition the supersymmetric (BPS) limit occurs when the solutions
admit solutions to the $D=4$ Killing spinor equation of minimal gauged supergravity given by 
\begin{align}\label{kseq}
\nabla_{\mu}\epsilon=\left(\ii A_{\mu}-\tfrac{1}{2}\gamma_{\mu}-\tfrac{\ii}{4}F_{\alpha\beta}\,\gamma^{\alpha\beta}\,\gamma_{\mu}\right)\,\epsilon\, ,
\end{align}
where $\epsilon$ is a $D=4$ Dirac spinor (see appendix \ref{appeee}).
 The conditions for the PD solutions to admit Killing spinors were determined
in \cite{Klemm:2013eca}. 
Our primary interest in this paper is when $\alpha\ne0$, and we will continue with this, but for reference, in appendix \ref{nonacc}
we briefly discuss the PD black holes when $\alpha=0$ where the supersymmetry analysis is different.  
The non-rotating solutions with $a=0$ are discussed in more detail in section~\ref{appaz}.

By examining the integrability conditions for \eqref{kseq}, as in \cite{AlonsoAlberca:2000cs,Klemm:2013eca} 
we find that supersymmetry implies that the five parameters are constrained 
by the following two conditions
\begin{align}
&0 = 2 a e g \alpha + g^4 \alpha^4 + g^2 (-1 + \alpha^2 + (a^2 + 2 e^2) \alpha^4) + e^2 \alpha^2 (1 + e^2 \alpha^2 + a^2 (-1 + \alpha^2))\, , \nn
&0 =2 a^2 g^2 \alpha +  2 e^4 \alpha^3 + 2 g^4 \alpha^3 -  2 a e g (-1 + a^2 \alpha^2)
- 2 e^2 \alpha (a^2 -  2 g^2 \alpha^2) \nonumber \\
 & - \alpha (m + a^2 m \alpha^2)^2 - (e^2 + g^2) \alpha (1 + a^2 \alpha^2) (-1 + e^2 \alpha^2 + g^2 \alpha^2 + a^2 (-1 + \alpha^2))\,.
 \label{bps2}
\end{align}
With $\alpha\ne 0$, and $m>0$ we must have $g>0$. With the signs of the parameters
as in \eqref{signchoice}, we can solve these conditions to obtain
\begin{align}\label{solBPSgeneral}
m &= \frac{g}{\alpha}\, , \nn
a &=\frac{-e g \alpha + (e^2 + g^2) \
\alpha^2 \sqrt{
  1 + (-1 + e^2) \alpha^2 - (e^2 + g^2) \
\alpha^4}}{g^2 \alpha^4 + e^2 \alpha^2 (-1 + \alpha^2)}\,.
\end{align}
Recall that we imposed the first of these two conditions in our construction of regular uplifted $D=11$ solutions (it is implied
by \eqref{allconds}).

We now consider the additional conditions imposed by extremality when $r=r_+$ becomes a double root of $Q(r)$. 
Assuming that  
both equations in \eqref{bps2} are satisfied we find
\begin{align}\label{extremality_general}
a^3 \alpha ^2 e g^3+a^2 \alpha  g^2 (e-g) (e+g)-a e g^3+\alpha ^3 e^2 \left(e^2+g^2\right)^2=0\, .
\end{align}

In principle, one could solve these three constraints in terms of two independent parameters, and then solve \eqref{PpPm} to find the relation between these and $n_{\pm}$. However, this is a little cumbersome to do in practice, and it is clearer to keep all the parameters in our expressions, where it is then understood that they must solve the constraints given by \eqref{bps2} and \eqref{extremality_general}. For example, a simple expression for the horizon radius of the supersymmetric and extremal black hole is given by
\begin{align}\label{Pietrorp}
r_+=\sqrt{\frac{a\,(m-a\,e)}{e+a\,m\,\alpha^2}}\,,
\end{align}
which would not be so simple if one were to use the explicit solutions of the constraints above. 
However, for the special case when we set the rotation parameter $a=0$, we will find simple and explicit expressions as we discuss in
section \ref{appaz}. For supersymmetric and extremal black holes with $a=0$ we also have
$e=0$, and the expression \eqref{Pietrorp} doesn't apply directly.

Notice that substituting \eqref{Pietrorp} into the general black hole entropy formula \eqref{entgenexp} immediately gives
\begin{align}\label{BHagain}
S_{BH} = \frac{am}{2G_{(4)}e}\Delta \phi\, ,
\end{align}
as long as the parameter $e\neq 0$. Further using the expressions for the black hole electric charge $Q_e$ in \eqref{elecflux}
and angular momentum $J$ in \eqref{angmtmdef} then leads to 
\begin{align}\label{BHintro}
S_{BH} = \frac{J}{Q_e}\frac{\pi}{G_{(4)}}\, ,
\end{align}
which is the first expression in \eqref{blackholent}. This formula for the entropy 
holds for the subfamily of supersymmetric extremal Kerr-Newman black holes, as 
discussed in appendix~\ref{nonacc}  (see equation \eqref{entgenexp2}). We have shown that,
remarkably, exactly the same formula holds also when we 
turn on acceleration. On the other hand, as we discuss in
section \ref{sec:conboundary}, for the supersymmetric extremal 
black holes with $J=Q_e=0$, or equivalently $a=e=0$, neither \eqref{BHagain} nor \eqref{BHintro} 
apply directly, although we shall see later in section \ref{sec:conboundary} that the second expression in \eqref{blackholent} 
is valid in this limit.

\subsection{Near horizon limit: $AdS_2\times \mathbb{WCP}^1_{[n_-,n_+]}$}\label{sec:nearhorizon}
We now elucidate the near horizon limit of these supersymmetric extremal black holes. The result is a new class
of supersymmetric $AdS_2\times \mathbb{WCP}^1_{[n_-,n_+]}$ solutions of $D=4$ gauged supergravity. These uplift to regular $AdS_2\times Y_9$ solutions,
which generalize those of \cite{Gauntlett:2006ns} by an extra rotation parameter.

A convenient way to find the near horizon solution is to implement the following coordinate transformation,
\begin{align}
r\to r_++\lambda\,s\,\rho\, ,\quad
t \to \lambda^{-1}\,s\,\tau\, ,\quad
\phi\to\phi'+\lambda^{-1}\,s\,W\frac{\Delta\phi}{2\pi}\,\tau\, ,
\end{align}
where $s$ is a constant, and then take the $\lambda\to 0$ limit. Here $W$ is given by
\begin{align}
W=\frac{a}{r_+^2+a^2}\frac{2\pi}{\Delta\phi}\, ,
\end{align}
with $\partial_t+W\frac{\Delta\phi}{2\pi} \partial_\phi=\partial_t+W\partial_\varphi$ a null generator of the horizon.

Some details of the limiting procedure are given in appendix \ref{appnh}. After carrying out various coordinate 
and gauge transformations, as well as redefining the parameters, we eventually end up with the following class of
$AdS_2\times\mathbb{WCP}^1_{[n_-,n_+]}$ solutions:
\begin{align} \label{epsilonsols}
\diff s^2&=\tfrac{1}{4}(y^2+\qp^2)\left( -\rho^2\,d\tau^2+\frac{d\rho^2}{\rho^2} \right)+\frac{y^2+\qp^2}{q(y)}d y^2+\frac{q(y)}{4(y^2+\qp^2)}(d\sigmap+\qp \, \rho\,d \tau)^2\, , \nn
A&=\gf(y)\,(d\sigmap+\qp \, \rho\,d \tau)\, ,
\end{align}
where we have defined
\begin{align}
q(y)&=\left({y^2+\qp^2}\right)^2-4\left(1-\qp^2\right) y^2+4\Qp\,\sqrt{1-\qp^2} y- \Qp^2\,,\nn
\gf(y)&=\frac{\sqrt{1-\qp^2}}{2}-\frac{1}{2(y^2+\qp^2)}(\Qp y+2\qp^2\sqrt{1-\qp^2})\,.
\end{align}
The solutions depend on two free parameters $\qp$, $\Qp$, which are  functions of the original $m,\alpha,e,g,a$, given implicitly in appendix
\ref{appnh}. Indeed, it is remarkable how simple the solution is in the above parametrization. 
The parameter $\qp$ is a rotation parameter and $\qp=0$ is the non-rotating limit. In fact if we set
$\qp=0$ then we precisely recover the $AdS_2\times Y_9$
solutions of \cite{Gauntlett:2006ns}, after dimensional reduction on a $SE_7$,
as we explain in appendix \ref{oldsols}.

Continuing\footnote{Note that we can change the sign of $\qp$ by changing the sign of $\tau$, $z$ and also the gauge
field $A$.}
with $\qp\ge 0$, we need ${\qp}\in[0,1]$ in order to get a real solution. We next analyse
the roots of $q(y)$, which are given by
$y = \sqrt{1-{\qp}^2}\pm \sqrt{1-\Qp -2{\qp}^2}$ and $y= -\sqrt{1-{\qp}^2}\pm \sqrt{1+\Qp -2{\qp}^2}$.
Given that $q(y)$ is a quartic in $y$ with positive coefficient of $y^4$, in order for $q(y)\geq 0$ we need to 
choose $y$ to lie in between the middle two roots of the quartic, with all roots real, which fixes $y\in [y_2,y_3]$ with
\begin{align}\label{yroots}
y_2 = -\sqrt{1-{\qp}^2}+ \sqrt{1+\Qp -2{\qp}^2}~, \qquad y_3 =\sqrt{1-{\qp}^2}- \sqrt{1-\Qp -2{\qp}^2}~.
\end{align}
For these to be real we need to take
\begin{align}
\qp\in [0,\tfrac{1}{\sqrt{2}})\,.
\end{align}

By analysing how the metric behaves at the roots, we demand that the $(y,z)$ part of the metric becomes 
$\mathbb{WCP}^1_{[n_-,n_+]}$, which fixes 
\begin{align}
\sqrt{1+\Qp-2{\qp}^2}\cdot \Delta z = \frac{2\pi}{n_+}~, \qquad \sqrt{1-\Qp-2{\qp}^2}\cdot \Delta z = \frac{2\pi}{n_-}~,
\end{align}
where we have identified $y=y_2, y_3$ with $\theta_+,\theta_-$ of previous sections, respectively,
which has solution
\begin{align}\label{nhcps}
\Qp=\frac{(1-2{\qp}^2)(n_-^2-n_+^2)}{n_+^2+n_-^2}\,,
\qquad \Delta z = \frac{\sqrt{2}\sqrt{n_+^2+n_-^2}}{n_+n_-\sqrt{1-2{\qp}^2}}\pi\,.
\end{align}

We may now compute the magnetic flux \eqref{magflux} of the gauge field and find 
\begin{align}\label{QmAdS}
G_{(4)}Q_m=\frac{1}{4\pi }\int_{\Sigma} dA     =\frac{n_--n_+}{4n_+n_-}~,
\end{align}
where $\Sigma=\mathbb{WCP}^1_{[n_-,n_+]}$ is the spindle horizon,
precisely as we had for the general PD black holes. In particular notice this result is independent of the continuous rotation parameter $\qp$.
We can also derive a very useful expression for the electric charge $Q_e$ \eqref{elecflux} by calculating it
directly in the near horizon solution. Doing so we find 
\begin{align}
G_{(4)}Q_e=\frac{1}{4\pi }\int_{\Sigma} *F = \qp \frac{\Delta z}{4\pi } \, .
\end{align}
Given the expression \eqref{nhcps} for $\Delta z$, we may now solve for 
$\qp$ in terms of the physical black hole parameter $Q_e$:
\begin{align}\label{jQe}
\qp = \frac{2\sqrt{2} n_-n_+ (G_{(4)}Q_e)}{\sqrt{16n_-^2n_+^2(G_{(4)}Q_e)^2+n_-^2+n_+^2}}\, .
\end{align}

The area of the horizon is then: $\mbox{Area}  = \frac{1}{2}(y_3-y_2)\Delta z$. Substituting for 
$\qp$ in terms of $Q_e$ using \eqref{jQe}, we find the entropy is
\begin{align}\label{gensbh}
S_{BH} &= \frac{1}{4G_{(4)}}\mbox{Area}=\frac{\sqrt{2} \sqrt{8 n_-^2 n_+^2(G_{(4)}Q_e)^2+n_-^2+n_+^2}-(n_-+n_+)}{n_- n_+}\frac{\pi}{4G_{(4)}}  ~.
\end{align}
This is precisely the second expression in \eqref{blackholent}. Notice that 
setting $\qp=0$ is equivalent to $Q_e=0$, which gives the non-rotating 
limit. The expression \eqref{gensbh} with $Q_e=0$ gives the entropy 
of the non-rotating but accelerating extremal supersymmetric black holes,
 studied in more detail in section \ref{sec:conboundary}. We also 
note that the angular momentum $J$ and electric charge $Q_e$ 
are related by 
\begin{align}\label{JQe}
J = Q_e \frac{\sqrt{2} \sqrt{8 n_-^2 n_+^2(G_{(4)}Q_e)^2+n_-^2+n_+^2}-(n_-+n_+)}{4 n_- n_+}\, ,
\end{align}
for these extremal solutions. Formally setting $n_-=n_+=1$ into the relation 
\eqref{JQe} gives the corresponding relation for the supersymmetric 
extremal Kerr-Newman-AdS black holes, discussed in appendix \ref{nonacc} (see equation 
\eqref{JQeapp}).

We may also rewrite \eqref{gensbh} using  equations \eqref{4dnewt}, \eqref{nplusminus}, and \eqref{nseexpr}, which gives
\begin{align}\label{SBHbarry}
S_{BH} =
\frac{2^{7/2} \pi  M \sqrt{\pJ (I \qJ+\pJ)} \left[
\sqrt{\frac{(1-\qp^2)}{(1-2\qp^2)}}  \sqrt{(I \qJ+2 \pJ)^2+I^2 \qJ^2}-(I \qJ+2 \pJ) \right]}
{3 I^5 h^{3/2}}N^{3/2}\,.
\end{align}
Notice that all quantities appearing, except for ${\qp}$, are integers. We also note that if we set ${\qp}=0$ then we precisely recover the expression for the entropy of the $AdS_2\times Y_9$ solutions as given in (D.21) of \cite{Gauntlett:2019pqg}. 

We note that we can express the black hole entropy in yet another way, namely as 
\begin{align}\label{anotherexp}
S_{BH} = \left(\frac{J_{AdS_2}}{Q_e}-\frac{1}{4}\chi(\Sigma)\right)\frac{\pi}{G_{(4)}}\, .
\end{align}
Here $\chi(\Sigma)$ is the Euler number of the spindle 
horizon $\Sigma=\mathbb{WCP}^1_{[n_-,n_+]}$
 given by
\begin{align}\label{eulernumber}
\chi(\Sigma) = \frac{1}{4\pi}\int_{\Sigma} R_{2}\, \vol_2 = \frac{n_-+n_+}{n_+n_-}\, ,
\end{align}
where $R_2$ denotes the Ricci scalar of the spindle, and $J_{AdS_2}$ 
is the angular momentum that is defined naturally for the near horizon 
$AdS_2$ solutions described in this subsection. Specifically, $J_{AdS_2}$ is invariant 
under the $AdS_2$ symmetries. We refer the reader to appendix 
\ref{sec:angmom} for further details, as well as a derivation of the formula we gave in
\eqref{angmomsurprise}:
\begin{align}\label{angmomsurprise2}
J_{AdS_2}-J = \frac{Q_e}{4}\chi(\Sigma)\, .
\end{align}
This formula as well as \eqref{anotherexp} are 
also valid for non-accelerating Kerr-Newman-AdS black holes 
upon setting $n_-=n_+=1$.

We can also express the entropy in another form\footnote{It is interesting to wonder if this formula can also be used for supersymmetric black holes with no acceleration, electric charge or rotation. 
The answer is no. However, we note that
for the so called universal twist black holes, with horizon consisting of Riemman surface with genus $g>1$, one obtains the correct entropy formula
(e.g as in \cite{Azzurli:2017kxo}), after setting $Q_e=0$ and formally taking $G_{(4)}Q_m\to \ii\chi/4$ with $\chi=2(1-g)$.}, 
by replacing the orbifold parameters $n_\pm$ with the magnetic flux 
 $Q_m$, given in \eqref{QmAdS}, and the Euler number \eqref{eulernumber}:
 \begin{align}\label{blackholentnicerbulk}
S_{BH} =  \frac{\pi}{4G_{(4)}}\left(\sqrt{\chi^2+16 \left[(G_{(4)}Q_e)^2+(G_{(4)}Q_m)^2\right]}-\chi\right)\, .
\end{align}
Correspondingly, this implies that the near horizon angular momentum can be written in the form
\begin{align}\label{blackholentnicerbulkjads2}
J_{AdS_2} = \frac{Q_e}{4} \sqrt{\chi^2+16 \left[(G_{(4)}Q_e)^2+(G_{(4)}Q_m)^2\right]} \, . 
\end{align}

Finally, we point out that the local metric appearing in \eqref{epsilonsols} was 
also used
in \cite{Anabalon:2020loe} in a completely different context of constructing supersymmetric wormholes in $AdS_4$.
To do this, the authors used ranges of the parameters and the coordinates so that, in particular, $q(y)>0$, in contrast to what we have done here.

\subsection{Killing spinors for $AdS_2\times \mathbb{WCP}^1_{[n_-,n_+]}$}\label{sec:spinors}
We now construct the $D=4$ Killing spinors associated with the rotating, magnetically charged $AdS_2\times \mathbb{WCP}^1_{[n_-,n_+]}$ solutions
given in \eqref{epsilonsols} with \eqref{nhcps} and $\qp\in [0,\tfrac{1}{\sqrt{2}})$. 
The fact that these solutions describe M2-branes wrapped on a surface 
$\Sigma=\mathbb{WCP}^1_{[n_-,n_+]}$, with a magnetic flux \eqref{QmAdS} through the surface, 
looks similar to a topological twist. However, in the latter case one instead needs the 
flux to be proportional to the Euler number $\chi(\Sigma)$ of the spindle $\Sigma=\mathbb{WCP}^1_{[n_-,n_+]}$ given in \eqref{eulernumber}.
The flux \eqref{QmAdS} instead leads to spinors 
that are sections of non-trivial line bundles over $\mathbb{WCP}^1_{[n_-,n_+]}$, which we shall describe 
explicitly, rather than the constant spinor solutions one obtains for the topological twist. 

We first introduce the following orthonormal frame for the near horizon metric \eqref{epsilonsols}:
\begin{align}\label{4dframe}
e^0 & = \tfrac{1}{2}\sqrt{y^2+\qp^2}\, \rho\,  d\tau\, , \qquad e^1 = \tfrac{1}{2}\sqrt{y^2+\qp^2}\, \frac{d\rho}{\rho}\, , \nn
e^2 & = \sqrt{\frac{y^2+\qp^2}{q(y)}}\, d y\, , \qquad \quad e^3 = \tfrac{1}{2}\sqrt{\frac{q(y)}{y^2+\qp^2}}\, (dz + \qp \, \rho\,  d\tau)\, .
\end{align}
For this frame, we then take the four-dimensional gamma matrices\footnote{Explicitly,
$\gamma_0=\left(
\begin{array}{cc}
 0 &  1\\
 -1 & 0
\end{array}
\right)$, $\gamma_1=\left(
\begin{array}{cc}
 0 &  1\\
 1 & 0
\end{array}
\right)$, 
$\gamma_2=\left(
\begin{array}{cc}
 \sigma^1 & 0\\
 0 & -\sigma^1
\end{array}
\right)$, 
$\gamma_3=\left(
\begin{array}{cc}
 \sigma^2 &  0\\
 0 & -\sigma^2
\end{array}
\right)$.} to be
\begin{align}\label{4dgammas}
\gamma_a&=\beta_a\otimes 1_2,  \qquad \qquad \qquad \quad a=0,1\,,\nn
\gamma_2&=\beta_3\otimes \sigma^1\, , \quad 
\gamma_3=\beta_3\otimes \sigma^2\, ,
\end{align}
where the two-dimensional gamma-matrices $\beta_a$ are defined by
\begin{align}
\beta_0=\ii\,\sigma^2\, , \quad
\beta_1=\sigma^1\, , \quad
\beta_3\equiv\beta_0\,\beta_1=\sigma^3\,,
\end{align}
and the $\sigma^i$ are Pauli matrices.

We next recall that the Killing spinor equation for $AdS_2$ is
\begin{align}
\nabla_a \theta=\frac{\ii}{2}n\,\beta_a\, \beta_3\,\theta\, ,
\end{align}
with $n=\pm 1$. This is solved  by Majorana spinors that can be decomposed as $\theta_{1,2}=\theta^{(+)}_{1,2}+\theta^{(-)}_{1,2}$, with the
Majorana-Weyl spinors $\theta^{(\pm)}_{1,2}$ of chirality $\beta_3\theta^{(\pm)}_{1,2}=\pm\theta^{(\pm)}_{1,2}$,
given by 
\begin{align}
\theta^{(+)}_1& =\begin{pmatrix}
\sqrt{\rho}\\
0
\end{pmatrix}, \qquad \qquad 
\theta^{(-)}_1=  \begin{pmatrix}
0\\
\ii\, n \sqrt{\rho}
\end{pmatrix}, \nn
\theta^{(+)}_2 & =\begin{pmatrix}
\sqrt{\rho } \tau -\frac{1}{\sqrt{\rho }}\\
0
\end{pmatrix}, \quad 
\theta^{(-)}_2=\begin{pmatrix}
0\\
\ii\, n \left(\sqrt{\rho } \tau +\frac{1}{\sqrt{\rho }}\right)
\end{pmatrix}\, .
\end{align}
After a lengthy calculation we find that the $D=4$ Killing spinors for the near horizon limit of the supersymmetric, extremal
PD black hole, satisfying \eqref{kseq}, can be expressed in the remarkably simple form
\begin{align}
\label{AdSspinors}
\epsilon_1&=\theta^{(+)}_1\otimes \chi_1+\theta^{(-)}_1\otimes \chi_2\, ,\nn
\epsilon_2&=\theta^{(+)}_2\otimes \chi_1+\theta^{(-)}_2\otimes \chi_2\, ,
\end{align}
where $\chi_{1,2}$ are two
two-dimensional spinors, given by
\begin{align}\label{M2spinors}
\chi_1=
\begin{pmatrix}
\frac{\mathcal{Y}^{1/2}_1(y)}{\sqrt{y-\ii\,\qp}}\\
-\frac{\mathcal{Y}^{1/2}_2(y)}{\sqrt{y+\ii\,\qp}}
\end{pmatrix}, 
\quad 
\chi_2=-n\,\ex^{\ii\,\xi}\,
\begin{pmatrix}
\frac{\mathcal{Y}^{1/2}_1(y)}{\sqrt{y+\ii\,\qp}}\nn
\frac{\mathcal{Y}^{1/2}_2(y)}{\sqrt{y-\ii\,\qp}}
\end{pmatrix}\, .
\end{align}
Here
\begin{align}
\mathcal{Y}_1(y)&=y^2-2 \sqrt{1-\qp^2} y+\qp^2+\Qp\, ,\nn
\mathcal{Y}_2(y)&=y^2+2 \sqrt{1-\qp^2} y+\qp^2-\Qp\, ,
\end{align}
which satisfy $\mathcal{Y}_1(y)\,\mathcal{Y}_2(y)=q(y)$, and the phase $\xi$ appearing in the Killing spinor is given by
\begin{align}
\xi=\arccos \qp\, .
\end{align}

Let us look more carefully at the global structure of the Killing spinors $\epsilon_i$ in (\ref{AdSspinors}). 
The spinors $\theta_i^{(\pm)}$ are simply the standard Killing spinors on $AdS_2$, where $i=1,2$, 
so our focus will be on the 
 two-dimensional spinors $\chi_i$ \eqref{M2spinors} on $\mathbb{WCP}^1_{[n_-,n_+]}$. 
Note first that the two components of $\chi_i$ 
have chiralities $\pm 1$ under $\sigma_3$, which is the two-dimensional chirality operator on $\mathbb{WCP}^1_{[n_-,n_+]}$.
Thus we can write $\chi_i=\chi_i^{(+)} + \chi_i^{(-)}$, $i=1,2$, where 
\begin{align}
\chi_1^{(+)}& =
\begin{pmatrix}
\frac{\mathcal{Y}^{1/2}_1(y)}{\sqrt{y-\ii\,\qp}}\\
0
\end{pmatrix}\,, 
\qquad \qquad \ \  \ 
\chi_1^{(-)}=
\begin{pmatrix}
0\\
-\frac{\mathcal{Y}^{1/2}_2(y)}{\sqrt{y+\ii\,\qp}}
\end{pmatrix}\,, \nn
\chi_2^{(+)}& =-n\,\ex^{\ii\,\xi}\,
\begin{pmatrix}
\frac{\mathcal{Y}^{1/2}_1(y)}{\sqrt{y+\ii\,\qp}}\\
0
\end{pmatrix}\,,
\qquad 
\chi_2^{(-)}=-n\,\ex^{\ii\,\xi}\,
\begin{pmatrix}
0\\
\frac{\mathcal{Y}^{1/2}_2(y)}{\sqrt{y-\ii\,\qp}}
\end{pmatrix}\, .
\end{align}
We also note that $\mathcal{Y}_1(y_3)=0=\mathcal{Y}_2(y_2)$, where $y_2$, $y_3$ are the roots 
\eqref{yroots}, so that the positive chirality components $\chi_i^{(+)}$ are zero 
at $y=y_3$, while the negative chirality  components $\chi_i^{(-)}$ are zero 
at $y=y_2$. 

Both the frame \eqref{4dframe} and the R-symmetry $U(1)$ gauge field $A$ in \eqref{epsilonsols} 
are singular at the roots $y=y_2$, $y_3$. Let us first look at the gauge field. The 
four-dimensional Killing spinors $\epsilon_i$ have charge $+1$ under $A$, as we 
see from the Killing spinor equation \eqref{kseq}. A gauge transformation 
$A\rightarrow A + d\gamma$ then leads to a   $U(1)$ rotation $\epsilon_i\rightarrow \ex^{\ii\gamma}\epsilon_i$. 
The magnetic flux of this gauge field through $\mathbb{WCP}^1_{[n_-,n_+]}$ is given by \eqref{QmAdS}. 
As explained further in appendix \ref{app:WCP}, this identifies $2A$ as a connection on the 
complex line bundle $O(n_--n_+)$. When $n_--n_+$ is not divisible by 2, this is a spin$^c$ gauge field on 
the weighted projective space $\mathbb{WCP}^1_{[n_-,n_+]}$. We shall 
comment on this further below when we describe the spin structure more explicitly. 

Note that we may write the gauge field in \eqref{epsilonsols}, restricted to $\mathbb{WCP}^1_{[n_-,n_+]}$, as 
\begin{align}\label{AM2}
A\mid_{\mathbb{WCP}^1_{[n_-,n_+]}} = h(y)\frac{\sqrt{n_+^2+n_-^2}}{\sqrt{2}n_+n_-\sqrt{1-2\qp^2}} d\jphi\, ,
\end{align}
where we have defined
\begin{align}
\jphi= \frac{\sqrt{2}n_+n_-\sqrt{1-2\qp^2}}{\sqrt{n_+^2+n_-^2}}z\, .
\end{align}
Here $\jphi$ is the same as the coordinate introduced in \eqref{canangcoord}, and
 has the canonical period of $2\pi$. However, $d\jphi$ is not 
defined at the roots $y=y_3$, $y=y_2$, so that the gauge field \eqref{AM2} 
is singular at the roots. 
We may then introduce open sets 
$U_-$ and $U_+$ on $\mathbb{WCP}^1_{[n_-,n_+]}$ that cover hemispheres containing the roots 
$y=y_3$ and $y=y_2$, respectively. Here $y=y_3$ is a $\C/\Z_{n_-}$ orbifold singularity,
while $y=y_2$ is a  $\C/\Z_{n_+}$ orbifold singularity.
One can then check that to obtain a well-defined connection in each patch, we
need to make the local gauge transformations
\begin{align}\label{Rrot}
U_-:&  \qquad A \rightarrow A  + \frac{1}{2n_-}d\jphi \equiv A_- \, ,\nn
U_+:&  \qquad A \rightarrow A + \frac{1}{2n_+}d\jphi \equiv A_+\, .
\end{align}
The gauge fields $A_\pm$ are now smooth one-forms in their respective patches $U_\pm$, 
and on the overlap $U_-\cap U_+$ they are related by
\begin{align}\label{ApAm}
A_+ = A_-  + \frac{n_--n_+}{2n_+n_-}d\jphi\, .
\end{align}
This again identifies the complex line bundle on which $2A$ is a connection as $O(n_--n_+)$, where 
by definition the transition function defining this line bundle over $\mathbb{WCP}^1_{[n_-,n_+]}$  is given by the gauge transformation 
in \eqref{ApAm}, cf. \eqref{QmAdS}. 
We discuss this further towards the end of this subsection.

Next let us look at the two-dimensional frame $\{e^2, e^3\}$ for $\mathbb{WCP}^1_{[n_-,n_+]}$ in \eqref{4dframe}, which is again
singular at the roots. Specifically,
\begin{align}
\mbox{$U_-$, near $y=y_3$}: \qquad & e^2\sim - d\varrho_-\,  \qquad e^3 \sim \varrho_-\,  \frac{d\jphi}{n_-}\, ,\nn
\mbox{$U_+$, near $y=y_2$}: \qquad & e^2\sim + d\varrho_+\, \qquad e^3 \sim \varrho_+\,  \frac{d\jphi}{n_+}\, ,
\end{align}
where $\varrho_\pm$ is geodesic distance measured from each root, to leading order. 
Note here that $y$ is increasing as one approaches $y=y_3>y_2$, while $\varrho_-$ is decreasing, 
hence the minus sign. We may then introduce a complex coordinate 
$z_-\equiv -\varrho_-\ex^{-\ii \jphi/n_-}$ in the patch $U_-$, which defines 
a smooth one-form $dz_-$ on the orbifold; that is, $dz_-$ is a smooth 
one-form on the covering space $\C$ in which $\jphi$ has period $2\pi n_-$. 
We may then write $z_- \equiv  x_-+\ii y_- = -\varrho_- \cos \frac{\jphi}{n_-} + \ii \varrho_- \sin\frac{\jphi}{n_-}$, and 
rotate the frame
\begin{align}
\begin{pmatrix} e^2 \\ e^3\end{pmatrix}\rightarrow \left(\begin{array}{cc} \cos\frac{\jphi}{n_-} & \sin \frac{\jphi}{n_-}\\ -\sin \frac{\jphi}{n_-} & \cos\frac{\jphi}{n_-}\end{array}\right)
\begin{pmatrix} e^2 \\ e^3\end{pmatrix} \sim \begin{pmatrix}dx_- \\ dy_-\end{pmatrix}\, . 
\end{align}
This is an $SO(2)\cong U(1)$ rotation of the frame on the patch $U_-$, which leads to a corresponding 
spinor $U(1)$ rotation $\exp(\pm \ii \jphi/2n_-)$, where the sign is correlated with two-dimensional chirality of the spinor. 
This follows from exponentiating the spinor representation 
of the infinitesimal version of the above $SO(2)$ rotation, namely
\begin{align}\label{spinrotmat}
\exp\left(-\frac{1}{2}\sigma_2\sigma_1\jphi/n_-\right) =\left( \begin{array}{cc}
\ex^{\ii\jphi/2n_-} & 0 \\ 0 & \ex^{-\ii\jphi/2n_-}\end{array}\right)\, .
\end{align}
We may then rotate the spinors $\chi_i$, $i=1,2$, in the patch $U_-$, noting there is both a spinor 
rotation and an R-symmetry rotation \eqref{Rrot}:
\begin{align}\label{smoothchim}
\mbox{$U_-$, near $y_3$}: \qquad \chi_i & = \chi_i^{(+)} +  \chi_i^{(-)}
\rightarrow \left(\ex^{\ii \jphi/2n_-}\chi_i^{(+)} +  \ex^{-\ii \jphi/2n_-}\chi_i^{(-)} \right)\ex^{\ii \jphi/2n_-} \nn
& =
\ex^{\ii \jphi/n_-}\chi_i^{(+)} + \chi_i^{(-)}\, .
\end{align}
The coordinate $\jphi$ is not defined at the root $y=y_3$, but on the other hand $\chi_i^{(+)}(y_3)=0$. Thus the above spinor 
is smooth and well-defined near to $y=y_3$, in the patch $U_-$. Note in particular that the spinor rotation and R-symmetry rotation cancel
each other for the non-vanishing negative chirality component $\chi_i^{(-)}$. 

A similar calculation goes through at the other root $y=y_2$, in the patch $U_+$. We introduce 
a coordinate $z_+\equiv \varrho_+\ex^{\ii\jphi/n_+} = x_+ + \ii y_+ = \varrho_+ \cos\frac{\jphi}{n_+} + \ii \varrho_+\sin\frac{\jphi}{n_+}$. 
The rotation is now 
in the opposite direction
\begin{align}
\begin{pmatrix} e^2 \\ e^3\end{pmatrix}\rightarrow \left(\begin{array}{cc} \cos\frac{\jphi}{n_+} & -\sin \frac{\jphi}{n_+}\\ \sin \frac{\jphi}{n_+} & \cos\frac{\jphi}{n_+}\end{array}\right)
\begin{pmatrix} e^2 \\ e^3\end{pmatrix} \sim \begin{pmatrix}dx_+ \\ dy_+\end{pmatrix}\, . 
\end{align}
The corresponding spinor rotation and R-symmetry rotation is thus
\begin{align}\label{smoothchip}
\mbox{$U_+$, near $y_2$}: \qquad \chi_i & = \chi_i^{(+)} +  \chi_i^{(-)}
\rightarrow \left(\ex^{-\ii \jphi/2n_+}\chi_i^{(+)} +  \ex^{\ii \jphi/2n_+}\chi_i^{(-)} \right)\ex^{\ii \jphi/2n_+} \nn
& =
\chi_i^{(+)} + \ex^{\ii \jphi/n_+}\chi_i^{(-)}\, .
\end{align}
Now $\chi_i^{(-)}(y_2)=0$, and we see the spinor is smooth and well-defined near the root.

The above analysis shows that the two-dimensional spinors $\chi_i$ on $\mathbb{WCP}^1_{[n_-,n_+]}$ are smooth and well-defined, 
in the appropriate orbifold sense. Notice that the above computations show that the spinor transition function in going from the patch $U_-$ to the patch $U_+$ is, 
similarly to  \eqref{ApAm}, given by
\begin{align}
\left( \begin{array}{cc}
\ex^{-\ii\jphi/2n_+} & 0 \\ 0 & \ex^{\ii\jphi/2n_+}\end{array}\right)\cdot 
\left(\begin{array}{cc}\ex^{-\ii\jphi/2n_-} & 0 \\ 0 & \ex^{\ii\jphi/2n_-}\end{array}\right) = 
\left(\begin{array}{cc} \ex^{-\ii\jphi \frac{n_++n_-}{2n_+n_-}} & 0 \\ 0 & \ex^{\ii\jphi \frac{n_++n_-}{2n_+n_-}}\end{array}\right)\, .
\end{align}
Here the original spinor rotation \eqref{spinrotmat} is inverted, since we begin with the smooth spinor 
in the patch $U_-$. This identifies the positive and negative chirality spin bundles 
$\mathcal{S}^{(\pm)}$ on $\mathbb{WCP}^1_{[n_-,n_+]}$ as $O(\mp (n_++n_-)/2)$. 
Notice these are well-defined as line bundles when $n_++n_-$ is divisible by 2, 
which is the case if and only if $n_+-n_-$ is divisible by 2, when the gauge field 
$A$ is a connection on a well-defined line bundle. We may understand this 
more abstractly as follows. On any oriented two-manifold (or orbifold)
the spinor bundles are 
\begin{align}
\mathcal{S}^{(+)} = \mathcal{K}^{1/2}~, \qquad \mathcal{S}^{(-)} = \mathcal{K}^{-1/2} = \Lambda^{(0,1)}\otimes \mathcal{K}^{1/2}~,
\end{align}
where $\mathcal{K}\cong \Lambda^{(1,0)}$ is the canonical bundle, namely the cotangent bundle, and $\Lambda^{(0,1)}$ denote 
$(0,1)$-forms with respect to the canonical complex structure. 
Again, the above computations explicitly show that the cotangent bundle is $\mathcal{K}=O(-(n_++n_-))$.
Instead, our spinors $\chi_i$ are spin$^c$ spinors that are also charged under the gauge field $A$. 
Denoting the line bundle on which $2A$ is a connection as $\mathcal{L}$, the spinors 
$\chi_i^{(\pm)}$ are hence sections~of
\begin{align}
\chi_i^{(+)} & : \quad \mathcal{S}^{(+)}\otimes \mathcal{L}^{1/2} = 
O(-\tfrac{1}{2}(n_++n_-))\otimes O(\tfrac{1}{2}(n_--n_+)) = O(-n_+)~, \nn
 \chi_i^{(-)} & : \quad \mathcal{S}^{(-)}\otimes \mathcal{L}^{1/2} = 
O(\tfrac{1}{2}(n_++n_-))\otimes O(\tfrac{1}{2}(n_--n_+)) = O(n_-)~.
\end{align}
Again, these also follow directly from composing the transition functions we worked out explicitly above. 
Notice that these chiral spin$^c$ bundles $O(-n_+)$ and $O(n_-)$ are always well-defined as line bundles, irrespective 
of whether $n_+\pm n_-$ is divisible by two.

\subsection{R-symmetry Killing vector}\label{sec:Rsymmetry}

The supersymmetric $AdS_2\times Y_9$ solutions we have constructed 
should have a holographic dual description in terms of a $d=1$ superconformal quantum mechanics (SCQM). 
This has an Abelian R-symmetry, that is realized in the supergravity solution as a 
canonically defined Killing vector field under which the Killing spinors are charged. 
As usual, this R-symmetry Killing vector $R$ may be constructed as a bilinear 
in the Killing spinors as we explain below. For the solutions in this paper we find 
\begin{align}\label{R2d}
R = R_{3d} + 2 \sqrt{1-\qp^2}\partial_z = R_{3d}+ 2\sqrt{2}\frac{n_+n_-\sqrt{8n_-^2n_+^2(G_{(4)}Q_e)^2+n_-^2+n_+^2}}{\sqrt{16n_-^2n_+^2(G_{(4)}Q_e)^2+n_-^2+n_+^2}}\partial_\varphi\, ,
\end{align}
where in the second expression 
we have used \eqref{jQe}, and we have also defined $R_{3d}=2\partial_\psi$.
 The latter is precisely the R-symmetry 
Killing vector for the corresponding $AdS_4\times SE_7$ solutions, normalized so that  the 
Killing spinor on the $SE_7$ has unit charge under $R_{3d}$ (see appendix \ref{appeee}). 
This is the geometric counterpart to the superconformal R-symmetry of the dual 
 $d=3$, $\mathcal{N}=2$ SCFT. 
We note that \eqref{R2d} reduces to \eqref{R2dnorot} on setting $\qp=0$, 
and that $\partial_\varphi$ generates the $U(1)$ isometry of the spindle 
$\mathbb{WCP}^1_{[n_-,n_+]}$, normalized so that $\varphi$ has period $2\pi$ (see \eqref{canangcoord}). 
We discuss the physical interpretation of \eqref{R2d} in the discussion section \ref{sec:discussion}. 

One way to identify the R-symmetry Killing vector is to construct bilinears of the Killing spinors on $Y_9$, as was essentially done in
\cite{Gauntlett:2006ns} for the case of $\qp=0$. Instead here we will construct bilinears of the $D=11$ Killing spinors for the $AdS_2\times Y_9$ solution.
With the conventions of appendix \ref{appeee}, we can obtain the $D=11$ Killing spinors as a tensor product of
the $D=4$ Killing spinors \eqref{AdSspinors} with the Killing spinor $\chi$ on $SE_7$. The solution preserves four $D=11$ Majorana spinors which
we can package into two complex $D=11$ spinors via
\begin{align}
\varepsilon_1=\epsilon_1 \otimes \chi\,,\quad
\varepsilon_2=\epsilon_2 \otimes \chi\,.
\end{align}
We then obtain 
the following bilinears
\begin{align}
\ii\,\bar{\varepsilon}_1\,\Gamma_M\,\varepsilon_1&=P_M\,,\nn
\ii\,\bar{\varepsilon}_2\,\Gamma_M\,\varepsilon_2&=-K_M\,,\nn
\ii\,\bar{\varepsilon}_1\,\Gamma_M\,\varepsilon_2&=D_M+\tfrac{\ii}{2}\,R_M\,,
\end{align}
where 
\begin{align}
P&=\partial_{\tau}\,,\nn
D&=\tau\,\partial_{\tau}-\rho\,\partial_{\rho}\,,\nn
K&=-(\tau^2+\rho^{-2})\,\partial_{\tau}+2\,\tau\,\rho\,\partial_{\rho}+2\,\qp\,\rho^{-1}\,\partial_z\,,\nn
R&=2\,\partial_{\psi}+2\,\sqrt{1-\qp^2}\,\partial_z\,.
\end{align}
The Killing vectors $P$, $D$ and $K$ generate the $\mathfrak{sl}(2)$ symmetry algebra of $AdS_2$:
\begin{align}
[D,\,P]=-P\,, \quad
[D,\,K]=K\,, \quad
[P,\,K]=-2\,D\,,
\end{align}
and hence we can identify the Killing vector $R$ to be the R-symmetry Killing vector, as claimed above.

\section{The conformal boundary for non-rotating solutions: $a=0$}\label{sec:conboundary}

An ultimate goal for holography would be to reproduce the black hole 
entropy \eqref{blackholent} for the extremal supersymmetric solutions 
via a dual field theory computation. The dual theory lives on the conformal 
boundary three-manifold of the full black hole solution. In this 
section we therefore turn to looking at this conformal boundary and,
for simplicity, we will now set the rotation parameter
$a=0$. We will see that for the supersymmetric extremal black holes we must have $e=0$, hence
$J=Q_e=0$, and the near horizon solutions can be obtained by setting 
$\qp=0$ in the near horizon metric of section \ref{sec:nearhorizon}.

We shall see that the global black hole geometry of the solutions with $a=0$ have some interesting features, 
including an acceleration horizon 
beyond $r=\infty$. For the supersymmetric and extremal
solution this acceleration horizon intersects the conformal boundary, 
effectively dividing the latter in half.  Moreover, we shall 
find that the Killing spinor on this conformal boundary 
is given by a topological 
twist, so that the spinor is constant, but 
it is a \emph{different} constant spinor 
on each half of the space. 

These exotic features will make a dual field theory 
calculation more challenging, but in the remaining subsections we show 
that the features arise as a limit of more well-behaved solutions, still with $a=0$. In particular in 
section \ref{sec:nonextreme} we relax the extremality 
condition, while preserving supersymmetry. The boundary 
three-manifold is now a smooth product of the time direction with a spindle, with a single component, and has a smooth boundary Killing spinor.  
It is therefore natural to perform any field theory localization calculation  
in this setting. One could then take the extremal limit. However, in the bulk of this solution
there is a naked curvature singularity. 
In section~\ref{sec:deform} we discuss a simple way of regulating this feature by also relaxing the requirement of supersymmetry, to obtain
completely regular accelerating black holes with a smooth conformal boundary. 

\subsection{Supersymmetric and extremal solutions}\label{appaz}
In this section
we focus on the solutions in \eqref{PDmetric} with $a=0$, which depend on 
four free parameters $m,e,g$ and $\alpha$.
Explicitly, we then have
\begin{equation}\label{nonrotBH}
d s^2=\frac{1}{\Omegach^2}\bigg[
 -\frac{Q}{r^2}d t^2 +\frac{r^2}{Q}\,d r^2  
+\frac{r^2}{P}d\theta^2 + P r^2 \sin^2\theta d\phi^2
 \bigg]\, ,
\end{equation}
with 
\begin{align}
 \Omegach&=1-\alpha\, r\cos\theta\,,\nn
 P&= 1-2\alpha m\cos\theta +\alpha^2(e^2+g^2)\cos^2\theta \,,\nn
 Q&= (r^2-2mr+ e^2+g^2) (1-\alpha^2r^2)  + r^4\,.
\end{align}
The gauge field is given by
\begin{align}
A=-\frac{e}{r}dt-g\cos\theta d\phi\,.
\end{align}
  
By directly examining the integrability conditions for the Killing spinor equations when $a=0$, we find that supersymmetry requires
\begin{align}\label{BPSa0}
0 =&\, m^2- (e^2+g^2)(1+(e^2+g^2)\alpha^2)  \,,\nn
0 =&\,  m^4 - m^2 (e^2 + g^2) - g^2 (e^2 + g^2)^2 \, .
\end{align}
In fact we get the same system of equations from \eqref{bps2} after setting $a=0$ and assuming that $\alpha\ne0$ as well as
one of $e,g$ to be non-vanishing.
These can be solved to give
\begin{align}
m^2 & = \frac{1-\alpha ^2-2 \alpha ^4 e^2+ \sqrt{(1-\alpha ^2)^2-4 \alpha ^4 e^2}}{2 \alpha ^6}\, , \nn
g^2 & =  \frac{1-\alpha ^2-2 \alpha ^4 e^2+\sqrt{(1-\alpha ^2)^2  -4 \alpha ^4 e^2}}{2 \alpha ^4}\, ,
\label{strangeisntit}
\end{align}
where we have taken the positive square roots in order to continuously connect with the extremal solution 
below. Note that \eqref{strangeisntit} implies 
 that $m=g/\alpha$.

The extremal limit is  given by setting $e=0$. Everything may then be expressed in terms of one parameter, for example $\alpha$, via
\begin{align}\label{BPSextr_nonrotating}
e=0\, , \quad
g=\frac{\sqrt{1-\alpha ^2}}{\alpha ^2}\, , \quad 
m=\frac{\sqrt{1-\alpha ^2}}{\alpha ^3}\, . \quad
\end{align}
The black hole horizon radius is given by the largest
double root $r_+>0$ of $Q(r)$, where
\begin{align}\label{BPSextr_nonrotatingrp}
r_\pm =\frac{-1\pm \sqrt{5-4 \alpha ^2}}{2 \alpha  \sqrt{1-\alpha ^2}}\, .
\end{align}
We require the function $P\ge0$ and so we should restrict the range of $\alpha$ to be
\begin{align}\label{alpharange}
\frac{\sqrt 3}{2}<\alpha<1\,,
\end{align}
and we also observe that in this range $r_+$ is positive and $r_-$ is negative. 
Recall that regularity of the uplifted solution requires the conditions \eqref{allconds} are also imposed, and this fixes the parameter $\alpha$ in terms of the integers $n_-,n_+$, so that there are no remaining free parameters.
In particular, we note that in terms of the integers $n_->n_+$ specifying the orbifold singularities 
at the poles $\theta=0$, $\theta=\pi$, we have
\begin{align}
\alpha = \frac{\sqrt{(3n_-+n_+)(3n_++n_-)}}{2(n_-+n_+)}\, .
\end{align}
Thus the lower limit $\sqrt{3}/2$ for $\alpha$ in \eqref{alpharange} is the limit $n_-\rightarrow \infty$, holding 
$n_+$ fixed, while the upper limit of 1 corresponds\footnote{Note that in this limit we have
$\alpha=1$, $g=m=0$ and hence, in particular, vanishing gauge field, and the metric is locally that of $AdS_4$.} to $n_--n_+\to 0$.

Next we look in more detail at the global structure of this extremal solution. We take 
$r\geq r_+$, where recall that the conformal boundary is at $\alpha r\cos\theta=1$. 
Notice immediately that for $\theta>\pi/2$ this requires $r$ \emph{negative}. 
Globally $r$ is not a good coordinate, and we instead put $y=1/r$. 
The black hole metric \eqref{nonrotBH} now reads 
\begin{align}
ds^2 = \frac{1}{(y-\alpha\cos\theta)^2}\Big[-Y dt^2 + \frac{dy^2}{Y} + \frac{1}{P}d\theta^2 + P\sin^2\theta d\phi^2\Big
]\, ,
\end{align}
where for the extremal supersymmetric solution we have introduced
\begin{align}
Y = Y(y) = (1-\alpha^2)(1-r_-y)^2(1-r_+y)^2\, .
\end{align}
For the original $r$ coordinate we have $r>r_+>0$, which implies
\begin{align}
y\leq \frac{1}{r_+} \equiv y_+\, ,
\end{align}
and we may then continue $y$ past zero to negative values (effectively extending beyond $r=\infty$). 
The $y$ coordinate decreases 
as one moves away from the horizon at $y=y_+$, eventually hitting the conformal boundary 
at $y=\alpha\cos\theta$, with $y-\alpha\cos\theta>0$ in the interior of the spacetime. 
However, although $1-r_+y>0$ for $y< y_+$, for negative $y$ one can 
reach the double root of the metric function $Y$ at $y=1/r_-$, where we recall that $r_-<0$ given \eqref{alpharange}. One can show that
this is an \emph{acceleration horizon}. To emphasize this we write $r_-\equiv r_A$, and then the
acceleration horizon is located at 
\begin{align}
y = \frac{1}{r_A} \equiv y_A \,, \qquad r_A\equiv r_-\,.
\end{align}
One can ask when the acceleration horizon at $y=y_A$ intersects the conformal boundary. This is determined by the equation
\begin{align}
y_A = \frac{1}{r_A} = \alpha \cos\theta\, ,
\end{align}
which can be solved to give
\begin{align}\label{theta0}
\theta = \theta_0 \equiv \arccos\left(\frac{1-\sqrt{5-4\alpha^2}}{2\sqrt{1-\alpha^2}}\right)
= \arccos\frac{n_-+n_+-\sqrt{2n_-^2+2n_+^2}}{n_--n_+}\, .
\end{align}
On the other hand, for given $\alpha$ and fixed $\theta>\theta_0$, note 
$y_A>\alpha\cos\theta$, 
which means that as one approaches from the black hole
one hits the acceleration horizon before the conformal boundary. On the conformal boundary itself the lower half of the 
spindle, with $\theta\in (\theta_0,\pi]$, then effectively lies behind the acceleration horizon. 
Interestingly, the acceleration horizon is also extremal,  and there is an asymptotic 
$AdS_2$ region as one approaches $y\rightarrow y_A$ from above.

\begin{figure}[hbt!]
    \centering
    \begin{subfigure}[b]{0.15\textwidth}
        \includegraphics[width=\textwidth]{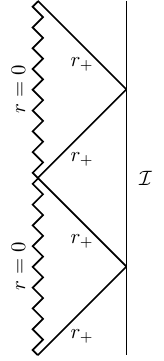}
        \caption{$\theta<\theta_0$}
        \label{fig:PenroseBPSext1}
    \end{subfigure}\qquad\qquad
    \begin{subfigure}[b]{0.20\textwidth}
        \includegraphics[width=\textwidth]{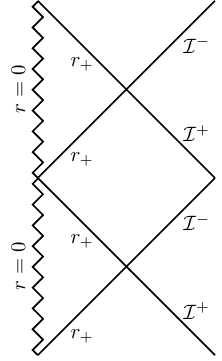}
        \caption{$\theta=\theta_0$}
        \label{fig:PenroseBPSext2}
    \end{subfigure}\qquad\qquad
    \begin{subfigure}[b]{0.23\textwidth}
        \includegraphics[width=\textwidth]{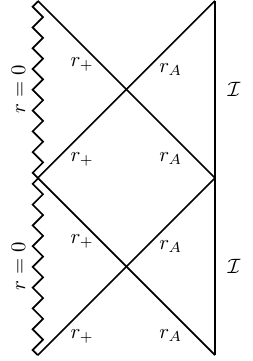}
        \caption{$\theta>\theta_0$}
        \label{fig:PenroseBPSext3}
    \end{subfigure}
\caption{Penrose diagram for the non-rotating supersymmetric extremal black hole, for different slices of constant $\theta$. The black hole horizon is denoted by $r_+$. The left panel is associated with a conformal boundary, consisting of the product of the time direction with half a spindle with $\theta<\theta_0$. In the right panel one reaches the other half of the spindle with $\theta>\theta_0$ at the conformal boundary after passing through the extremal acceleration horizon, denoted by $r_A$. For $\theta=\theta_0$, in the middle panel, a null infinity appears. }\label{fig:extremePenrose}
\end{figure}

An extensive analysis
of the causal structure of the $AdS$ $C$-metrics is presented
in \cite{Dias:2002mi}, for general values of the parameters. 
The Penrose diagram for the extremal supersymmetric black hole solution is shown in Figure~\ref{fig:extremePenrose}.
In particular we note that the lower half of the spindle boundary, with 
$\theta>\theta_0$, lies behind the acceleration horizon $r=r_A$.

Next let us look more closely at the conformal boundary itself. Starting from the 
general  non-rotating black hole metric \eqref{nonrotBH}, the conformal boundary is located at $\Omegach=0$.
Choosing the conformal factor 
so that the timelike Killing vector $\partial_t$ has unit norm on the boundary, we find 
that the general form of the conformal boundary metric is
\begin{align}\label{bmetric}
ds^2_{3d}=-dt^2+ds^2_\Sigma = -dt^2+\frac{d\theta^2}{P\,(1-\alpha^2\,\sin^2\theta\,P)^2}+\frac{P\,\sin^2\theta}{1-\alpha^2\,\sin^2\theta\,P}\,d\phi^2\, .
\end{align}
That is, the conformal boundary is a static product metric, where the induced metric on a constant time slice is $ds^2_\Sigma$. 

We have seen that for the extremal supersymmetric black hole 
 the 
acceleration horizon intersects the conformal boundary at $\theta=\theta_0$, and in fact the boundary 
actually splits in half along this slice. Indeed, 
although $P=P(\theta)>0$ for all $\theta\in[0,\pi]$, we find that for the extremal solution $1-\alpha^2 \sin^2\theta P(\theta)\geq 0$, with 
equality if and only if $\theta=\theta_0$.
The
metric on $\Sigma$ in \eqref{bmetric} is then singular for $\theta=\theta_0$. Introducing the new coordinate
\begin{align}
\rho = \frac{\alpha  \sqrt{1+\sqrt{5-4 \alpha ^2}}}{\sqrt{2} \left(5-4 \alpha ^2\right)}\frac{1}{\theta-\theta_0}\, ,
\end{align}
we find that near to $\theta=\theta_0$ (which is $\rho=\infty$), the metric on $\Sigma$ takes the form
\begin{align}
\diff s^2_\Sigma \simeq d\rho^2 + \frac{5-4\alpha^2}{\alpha^4}\rho^2 d\phi^2\, .
\end{align}
Thus each side of $\theta=\theta_0$ opens out into a non-compact asymptotically locally Euclidean end, 
with 
each of the poles $\theta=0$ and $\theta=\pi$ being an infinite distance from $\theta=\theta_0$.
The conformal boundary thus effectively has two halves, with the lower half $\theta\in (\theta_0,\pi]$ lying 
behind the acceleration horizon. 

We can now discuss the behaviour of the Killing spinor on the conformal boundary.
The bulk Killing spinor equation for minimal $d=4$, $\mathcal{N}=2$ gauged supergravity induces the following conformal 
Killing spinor equation (CKSE) on the conformal boundary~\cite{Cassani:2012ri}:
\begin{align}\label{CKSE}
\nabla_a \zetaJ = \frac{1}{3}\gammathree_a \sla \nabla \zetaJ\, ,
\end{align}
where we have introduced the covariant derivative $\nabla_a = \partial_a + \frac{1}{4}\omega_a^{bc}\gammathree_{bc} -\ii A_a$. 
Here $a=0,1,2$ is a tangent space index, and we may take the gamma matrices to be $\gammathree_0=\ii \sigma^1$, 
$\gammathree_1=\sigma^2$, $\gammathree_2=\sigma^3$, in terms of Pauli matrices. To solve this equation
we begin by introducing the obvious orthonormal frame
\begin{align}\label{bframe}
e^0 = dt\, , \qquad e^1  = \frac{d\theta}{\sqrt{P}(1-\alpha^2\sin^2\theta P)}\, , \qquad e^2 = \sqrt{\frac{P}{1-\alpha^2\sin^2\theta P}}\sin\theta d\phi\, .
\end{align}
For the extremal solution with $e=0$ 
the gauge field is 
\begin{align}
A = -g \cos\theta d\phi = -\frac{\sqrt{1-\alpha^2}}{\alpha^2}\cos\theta d\phi\, .
\end{align}
It is convenient to define the gauge-equivalent gauge field
\begin{align}\label{A0gauge}
A_0 \equiv A + \frac{1}{2\alpha^2}d\phi\, .
\end{align}
Then remarkably we find that this gauge field is equal to plus or minus $\tfrac{1}{2}\omega^{12}$, where 
$\omega^{12}$ is the non-zero spin connection component for $\Sigma$, with the sign depending 
on which half of $\Sigma$ the expressions are compared in:
\begin{align}\label{twistme}
\tfrac{1}{2}\omega^{12} = \begin{cases} \ -A_0 & \theta\in [0,\theta_0)\, , \\ \ +A_0 & \theta\in(\theta_0,\pi]\, . \end{cases} 
\end{align}
We find that 
the solution to \eqref{CKSE} is in fact covariantly constant, so 
$\nabla_a\zetaJ=0$, and moreover due to \eqref{twistme} in the gauge \eqref{A0gauge}
the solution for $\zetaJ$ is in fact constant:
\begin{align}\label{constantzeta}
\zetaJ = \begin{cases}\ \begin{pmatrix} -\ii \\ \ii  \end{pmatrix} &  \theta\in [0,\theta_0)\, , \\ \ \begin{pmatrix} 1 \\ 1 \end{pmatrix} &  \theta\in (\theta_0,\pi]\,  .\end{cases}
\end{align}
There is thus a topological twist on each half of the conformal boundary, 
with the gauge field effectively cancelling the spin connection and leading to a constant spinor,
but with a discontinuity in the spinor as one moves across the slice $\theta=\theta_0$ that intersects the acceleration horizon of the bulk black hole. 
Of course we may multiply each spinor in \eqref{constantzeta} by any constant, and this will still be a solution.
The reason for normalizing the spinors in the way that we have, in particular taking a purely imaginary
spinor in $\theta\in[0,\theta_0)$, will be come apparent in the next subsection.

Finally in this subsection we note that it is possible to derive an explicit expression for the 
entropy of the non-rotating extremal supersymmetric solution directly  
from \eqref{entgenexp}. In terms of $n_\pm$ we find
\begin{align}\label{gandphi}
g & = \frac{2(n_-^2-n_+^2)}{(3n_-+n_+)(n_-+3n_+)}~, \qquad 
\Delta\phi = \frac{(3n_-+n_+)(n_-+3n_+)\pi}{4n_-n_+(n_-+n_+)}~.
\end{align}
The horizon radii $r=r_\pm$ are given by
\begin{align}
r_{\pm}=\frac{2 \left(n_-+n_+\right) \left(\pm\sqrt{2} \sqrt{n_-^2+n_+^2}-\left(n_-+n_+\right)\right)}{\left(n_--n_+\right) \sqrt{\left(3 n_-+n_+\right) \left(n_-+3 n_+\right)}}\, .
\end{align}
Using 
\eqref{entgenexp} we then find the entropy is given by
\begin{align}\label{gensbhaz}
S_{BH} &=
\frac{\sqrt{2} \sqrt{n_-^2+n_+^2}-(n_-+n_+)}{n_- n_+}\frac{\pi}{4G_{(4)}}  \nn
&=
\frac{2^{7/2} \pi  M \sqrt{\pJ (I \qJ+\pJ)} \left[
 \sqrt{(I \qJ+2 \pJ)^2+I^2 \qJ^2}-(I \qJ+2 \pJ) \right]}
{3 I^5 h^{3/2}}N^{3/2}\,.
\end{align}
The first expression agrees with \eqref{gensbh}  after setting $Q_e=0$, where  recall that \eqref{gensbh}   was instead computed using the near horizon metric for the general supersymmetric extremal solution. The second expression agrees with \eqref{SBHbarry} after correspondingly setting $\qp=0$, and is the same as
the expression for the entropy of the $AdS_2\times Y_9$ solutions as given in equation (D.21) of \cite{Gauntlett:2019pqg}.

\subsection{Supersymmetric and non-extremal solutions}\label{sec:nonextreme}

The non-rotating, extremal supersymmetric black hole has some slightly exotic features, 
especially the behaviour of the conformal boundary and its Killing spinor. 
In this section we relax the extremality condition $e=0$, instead imposing 
the BPS relations \eqref{strangeisntit} on the conformal boundary geometry with $e\neq 0$. 
We shall find that the conformal boundary has the same form as \eqref{bmetric}, 
but now with a completely regular metric on the spindle $\Sigma$, apart from the
usual orbifold singularities at $\theta=0$ and $\theta=\pi$, so that $\Sigma\cong \mathbb{WCP}^1_{[n_-,n_+]}$
has the same topology as the black hole horizon in the extremal limit.
The circumference of $\Sigma$ near to $\theta=\theta_0$ grows as 
 $e\rightarrow 0$, 
 as does the distance between the poles and $\theta=\theta_0$,
with the spindle effectively completely splitting in 
half in the extremal limit $e=0$. 
There is correspondingly a smooth solution $\zeta$ to the $d=3$ Killing spinor equation for $e\neq 0$, that approaches 
the piecewise constant solution \eqref{constantzeta} in the extremal limit. 
As we discuss, the non-rotating BPS and non-extremal solutions no longer have a smooth black horizon but a naked singularity.

We first note that with $a=0$, the BPS conditions \eqref{strangeisntit} and the regularity conditions \eqref{PpPm} imply that
\begin{align}
m&=\frac{g}{\alpha},\qquad\qquad
\alpha = \frac{\sqrt{(3n_-+n_+)(3n_++n_-)}}{2(n_-+n_+)\sqrt{1+\frac{e^2(3n_-+n_+)^2(3n_++n_-)^2}{4(n_-^2-n_+^2)^2}    }}\,,\nn
g & = \frac{2(n_-^2-n_+^2)}{(3n_-+n_+)(n_-+3n_+)}~, \qquad 
\Delta\phi = \frac{(3n_-+n_+)(n_-+3n_+)}{4n_-n_+(n_-+n_+)}\pi~.
\end{align}
These expressions can be obtained by solving the regularity condition \eqref{PpPm} for $e$ and then substiuting this into the first line
of \eqref{bps2} to derive the expression for $g$. Then substituting this expression for $g$ into \eqref{PpPm} we get the expression for $\alpha$.
We also note that if we set $e=0$ then we recover the same conditions for the BPS and extremal solutions that we considered in the previous subsection.

To analyse the conformal boundary and the Killing spinors both on the boundary and in the bulk, it is convenient to change 
to PD-type coordinates via 
\begin{align}\label{toPDcoordinates}
t = \alpha \tau\, , \qquad \cos \theta = p\, , \qquad \phi = \alpha^2\sigma\, , \qquad r=-1/(\alpha q)\,,
\end{align}
where $p\in [-1,1]$. We also change parameters by introducing
\begin{align}
g = \frac{\mathsf{P}}{\alpha^2}\, , \qquad  e = \frac{\mathsf{Q}}{\alpha^2}\, , \qquad \mathsf{C} = \mathsf{P}^2 + \mathsf{Q}^2\, .
\end{align}
The BPS conditions \eqref{strangeisntit} imply $m=g/\alpha$, as usual, and $\alpha=(\mathsf{P}^2/\mathsf{C}-\mathsf{C})^{1/2}$.
Using these relations, we can then write the full non-rotating solution \eqref{nonrotBH} in these coordinates:
\begin{align}\label{metric1303app}
\begin{split}
ds^2&=\frac{1}{(p+q)^2}\,\left(-\mathcal{Q}(q)\,d\tau^2+\frac{dq^2}{\mathcal{Q}(q)}+\frac{dp^2}{\mathcal{P}(p)}+\mathcal{P}(p)\,d\sigma^2\right)\,,\\
A&=\mathsf{Q}\,q\,d\tau-\mathsf{P}\,p\,d\sigma\,,
\end{split}
\end{align}
where the metric functions are 
\begin{align}
\mathcal{P}(p)=\mathsf{C}^{-1}\,\mathcal{P}_1(p)\,\mathcal{P}_2(p)\,, \quad
\mathcal{Q}(q)=\mathsf{C}^{-1}\,\mathcal{Q}_1(q)\,\mathcal{Q}_2(q)\,,
\end{align}
and we have introduced
\begin{align}\nonumber
\mathcal{P}_1(p)&=-(1-p) (\mathsf{C} p+\mathsf{C}-\mathsf{P})\,,\quad
&& \mathcal{Q}_1(q)=\mathsf{C} q^2-\mathsf{C}+\mathsf{P} q+\ii \mathsf{Q}\,,\\
\mathcal{P}_2(p)&=-(1+p) (\mathsf{C} p-\mathsf{C}-\mathsf{P})
\,,\quad 
&&\mathcal{Q}_2(q)=\mathsf{C} q^2-\mathsf{C}+\mathsf{P} q-\ii \mathsf{Q}\,.
\end{align}
The regularity conditions on the metric imply that 
\begin{align}\label{peeque}
\mathsf{P} = \frac{n_--n_++\sqrt{(n_--n_+)^2-16(n_-+n_+)^2\mathsf{Q}^2}}{4(n_-+n_+)}\, ,
\end{align}
with $\mathsf{P}<\frac{1}{2}$,
while the period of $\sigma$ can be expressed as
\begin{align}
\Delta\sigma = 2\pi\frac{n_-^2-n_+^2}{n_-n_+(n_--n_++\sqrt{(n_--n_+)^2-16(n_-+n_+)^2\mathsf{Q}^2})}\, .
\end{align}
Notice that we can parametrize this class of solutions in terms of $n_\pm$ and $\mathsf{Q}$, with the extremal limit
obtained when $\mathsf{Q}\to 0$.
The reality of $\mathsf{P} $ in \eqref{peeque} requires that we impose
\begin{align}\label{strongq}
\mathsf{Q}\le\frac{n_--n_+}{4(n_-+n_+)}\, ,
\end{align}

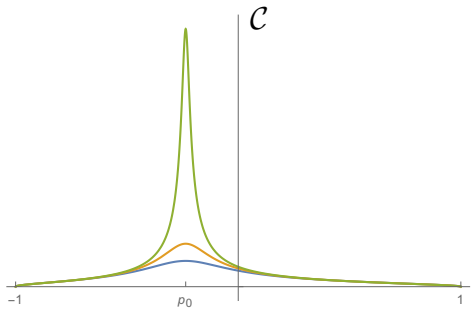
\begin{figure}[hbt!]
    \centering
    \begin{picture}(0.1,0.25)(0,0)
\put(100,110){\makebox(0,0){$\mathcal{C}$}}
\end{picture}
        \includegraphics[width=0.4 \textwidth]{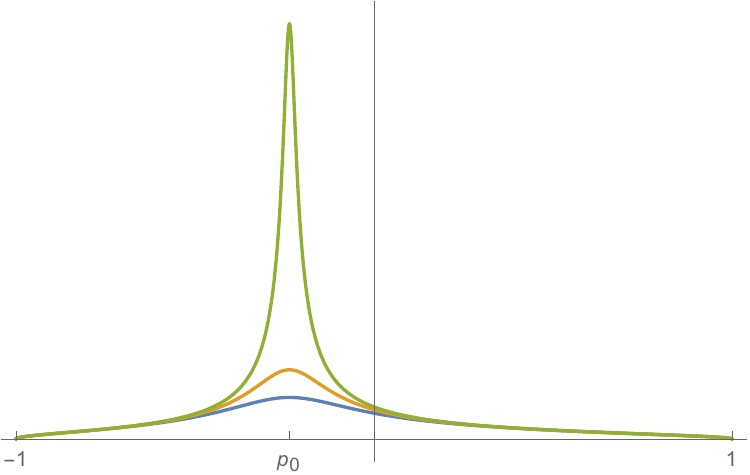}
        \caption{Circumference $\mathcal{C}$ of the metric on the spindle $\Sigma=\mathbb{WCP}^1_{[3,1]}$ on the conformal boundary
      as a function of $p\in [-1,1]$, for the supersymmetric non-extremal black holes. The blue, orange, and green curves have progressively smaller values of $\mathsf{Q}$, namely $\mathsf{Q}=0.04$, $\mathsf{Q}=0.01$ and $\mathsf{Q}=0.0002$, respectively, tending to the extremal 
solution with $\mathsf{Q}=0$. The same values of $\mathsf{Q}$ are plotted also in Figures \ref{fig:args} and 
\ref{fig:spingauge}.}
        \label{fig:circum}
    \end{figure}

The conformal boundary metric \eqref{bmetric}, obtained at $p=-q$, is then (after a rescaling by the constant $\alpha^2$) given by 
\begin{align}\label{pdbmetric}
ds^2_{3d} = -d\tau^2 + ds^2_{\Sigma} =  -d\tau^2 + \frac{dp^2}{\PP(p)(1-\PP(p))^2} + \frac{\PP(p)}{1-\PP(p)}d\sigma^2\, .
\end{align}
Notice that $\PP(p)>0$ for $p\in[-1,1]$ is implied by \eqref{strongq}. 
The circumference $\mathcal{C}$ of the spindle, at fixed $p\in[-1,1]$, is given by the function
\begin{align}
\mathcal{C}=\mathcal{C}(p)=\sqrt{\frac{\PP(p)}{1-\PP(p)}}\Delta\sigma\, .
\end{align}
We have plotted this in Figure \ref{fig:circum} for the spindle  $\Sigma=\mathbb{WCP}^1_{[3,1]}$, 
with progressively smaller values of $\mathsf{Q}$, tending to the extremal 
solution with $\mathsf{Q}=0$. The circumference at $p=p_0$ is infinite 
for $\mathsf{Q}=0$, where $p_0=\cos \theta_0$, with $\theta_0$ given by \eqref{theta0}:
\begin{align}
p_0 = \frac{n_-+n_+-\sqrt{2n_-^2+2n_+^2}}{n_--n_+}\,.
\end{align}
 In the $\mathsf{Q}=0$ limit the spindle has then effectively split in half. 

\begin{figure}[hbt!]
    \centering
        \includegraphics[width=0.27 \textwidth]{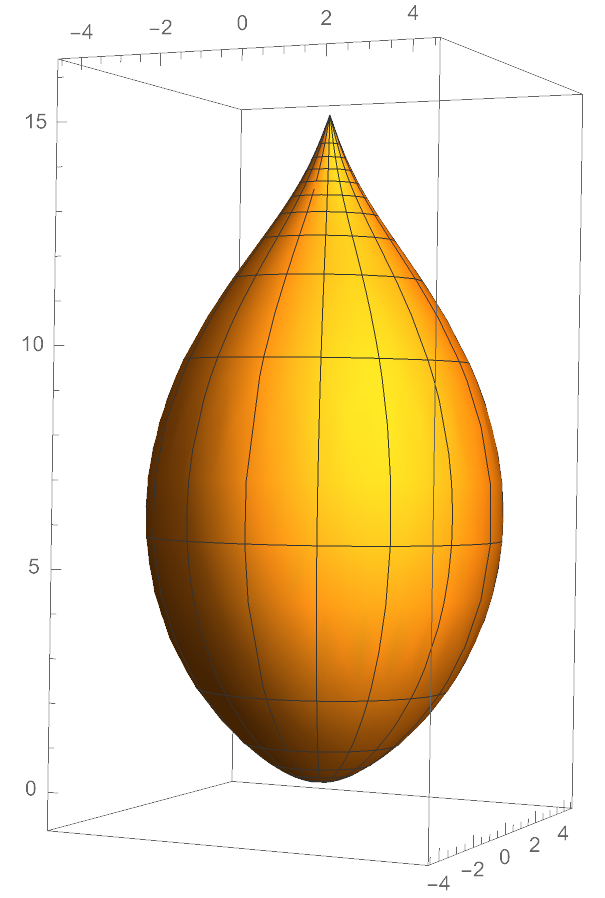}\qquad\qquad\qquad\qquad
                \includegraphics[width=0.14 \textwidth]{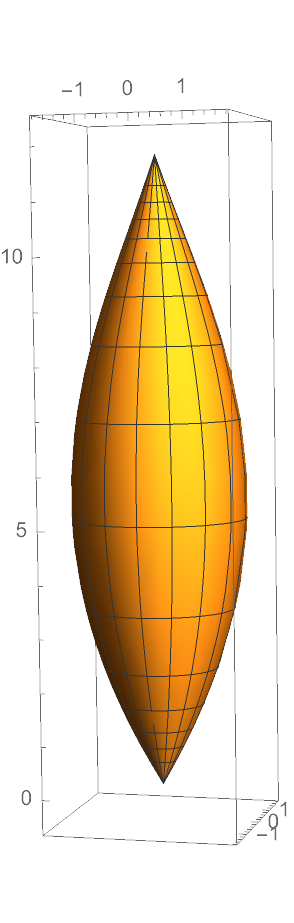}
        \caption{Embedding of two spindles  on the conformal boundary for the supersymmetric non-extremal black holes in three dimensional Euclidean space.
The left plot is for $\Sigma=\mathbb{WCP}^1_{[3,1]}$ and $\mathsf{Q}=0.04$
and the right plot is for $\Sigma=\mathbb{WCP}^1_{[3,2]}$ and $\mathsf{Q}=0.025$.}
        \label{fig:embed}
    \end{figure}

At this point one might picture the geometry as breaking up into two ``pancakes" as $\mathsf{Q}\to 0$. However, this is not correct
and is clarified by calculating the proper distance from $p=\pm 1 $ to $p=p_0$, both of which diverge as $\mathsf{Q}\to 0$.
It is illuminating to present the geometry of the spindle as an embedding in three-dimensional Euclidean space, as in Figure \ref{fig:embed}. 
As $\mathsf{Q}\to 0$, while the circumference at $p_0$ is diverging, so too is the height of the figures.

Introducing the orthonormal frame\footnote{Notice the overall minus sign in $e^1$ is 
so as to match the orientation in the corresponding frame \eqref{bframe}, where $dp = -\sin\theta d\theta$.} 
for the conformal boundary metric \eqref{pdbmetric}
\begin{align} \label{boundaryframePD}
e^0 = d\tau\, , \qquad e^1 = -\frac{dp}{\sqrt{\PP(p)}(1-\PP(p))}\, , \qquad e^2 = \sqrt{\frac{\PP(p)}{1-\PP(p)}}d\sigma\, , 
\end{align}
we may solve the conformal Killing spinor equation \eqref{CKSE} using the same basis of 
gamma matrices as the previous subsection (see below \eqref{CKSE}). We find that
\begin{align}\label{nonextremespinor}
\zetaJ = \ex^{-\ii (\kappa_1 \tau + \kappa_2\sigma)} \begin{pmatrix}
\zetaJ_1(p)\\ \zetaJ_2(p)\end{pmatrix}\, ,
\end{align}
where we have introduced the constants
\begin{align}\label{kappa12}
\kappa_1 = \frac{\mathsf{P}\mathsf{Q}}{2(\mathsf{P}^2+\mathsf{Q}^2)}\, , \qquad \kappa_2 = \frac{\mathsf{P}^2}{2(\mathsf{P}^2+\mathsf{Q}^2)}\, .
\end{align}
Notice that in the extremal limit $\mathsf{Q}=0$ the  phase in \eqref{nonextremespinor} 
is $\ex^{-\ii \sigma/2} = \ex^{-\ii \phi/2\alpha^2}$, which was compensated for 
in the previous subsection by making the gauge transformation \eqref{A0gauge}.
The components
$\zetaJ_1(p)$, $\zetaJ_2(p)$ satisfy the equations
\begin{align}
\begin{split}
\zetaJ_1' & = \frac{\ii (p\mathsf{Q} - \kappa_1)}{\sqrt{\PP(p)}(1-\PP(p))}\zetaJ_1\, ,\\ 
\zetaJ_2 & = \frac{4\ii (p\mathsf{Q}-\kappa_1)\PP(p) - \sqrt{\PP(p)}\PP'(p)}{4(p\mathsf{P}-\kappa_2)\sqrt{\PP(p)(1-\PP(p))}}\zetaJ_1\, .
\end{split}
\end{align}
After some effort, one finds the solution 
\begin{align}\label{boundaryspinorcomponentsPD}
\begin{split}
\zetaJ_1(p) & = \sqrt{\frac{\PP_1(p)-\PP_2(p) - 2\ii \mathsf{Q}\sqrt{\PP(p)}}{2\mathsf{P}\sqrt{1-\PP(p)}}}\, , \\
\zetaJ_2(p) & =\sqrt{\frac{\PP_1(p)-\PP_2(p) + 2\ii \mathsf{Q}\sqrt{\PP(p)}}{2\mathsf{P}\sqrt{1-\PP(p)}}}\, .
\end{split}
\end{align}
Notice that $\zetaJ_1^*=\zetaJ_2$ and $|\zetaJ_1|=|\zetaJ_2|$. In fact we have 
chosen the overall normalization constant of the spinor so that the components 
lie on the unit circle in the complex plane, namely 
$|\zetaJ_1|=|\zetaJ_2|=1$. 

\begin{figure}[hbt!]
    \centering
    \begin{subfigure}[b]{0.4\textwidth}
        \includegraphics[width=\textwidth]{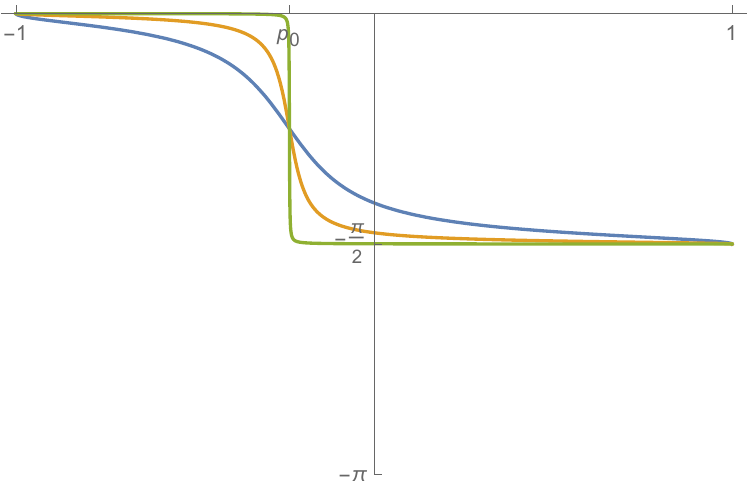}
        \caption{$\mathrm{arg}\, \zetaJ_1(p)$}
        \label{fig:zeta1}
    \end{subfigure}
    \begin{subfigure}[b]{0.4\textwidth}
        \includegraphics[width=\textwidth]{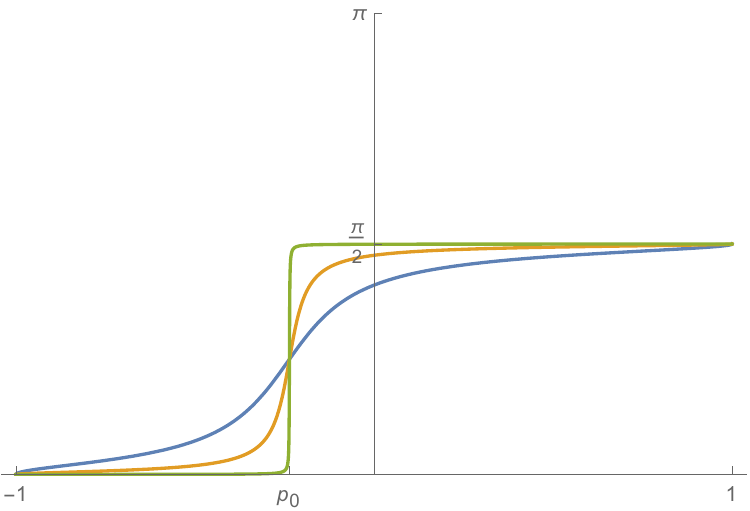}
        \caption{$\mathrm{arg}\, \zetaJ_2(p)$}
        \label{fig:zeta2}
    \end{subfigure}
    \caption{Arguments of $\zetaJ_1(p)$, $\zetaJ_2(p)$, appearing in the conformal Killing spinors 
     on the conformal boundary for the supersymmetric non-extremal black holes, as a function of $p\in[-1,1]$ and plotted for the spindle $\Sigma=\mathbb{WCP}^1_{[3,1]}$ 
The blue, orange, and green curves have progressively smaller values of $\mathsf{Q}$, as given in Figure \ref{fig:circum}, tending to the extremal 
solution with $\mathsf{Q}=0$. For $\mathsf{Q}\ne0$ we have a smooth conformal Killing spinor on the spindle
that approaches two different constant values on each half of the spindle in the extremal limit  $\mathsf{Q}\to 0$.
}\label{fig:args}
\end{figure}

In Figure \ref{fig:args} we have plotted the arguments $\mathrm{arg}\, \zetaJ_1(p)$, 
$\mathrm{arg}\, \zetaJ_2(p)$, for the spindle $\Sigma=\mathbb{WCP}^1_{[3,1]}$ with progressively 
smaller values of $\mathsf{Q}$, tending to the extremal 
solution with $\mathsf{Q}=0$. In the latter case 
notice that in the $\mathsf{Q}\rightarrow 0$ limit  we have
\begin{align}
\mbox{for $\mathsf{Q}=0$:} \qquad \begin{cases}
\ \mathrm{arg}\, \zetaJ_2(p) = - \mathrm{arg}\, \zetaJ_1(p) = \frac{\pi}{2}\, , & p\in (p_0,1]\, ,\\ 
\ \mathrm{arg}\, \zetaJ_2(p) = \mathrm{arg}\, \zetaJ_1(p) = 0\, , & p\in [-1,p_0)\, . \end{cases}
\end{align}
This precisely corresponds to the extremal solution \eqref{constantzeta}.

\begin{figure}[hbt!]
    \centering
        \includegraphics[width=0.4 \textwidth]{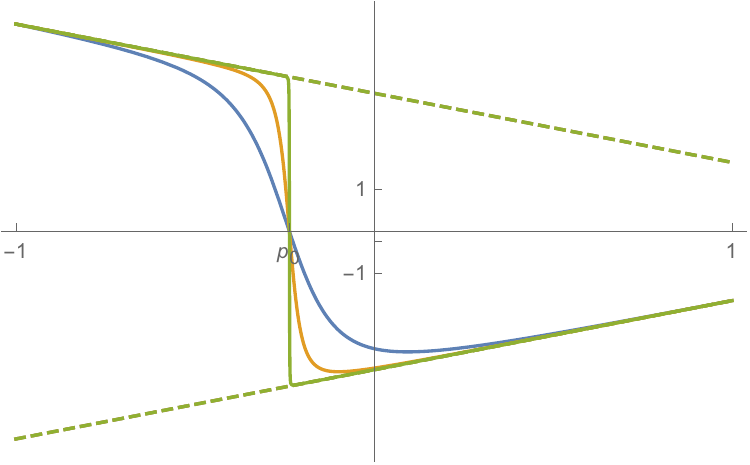}
        \caption{Spin connection and gauge field holonomies around a circle in the spindle $\Sigma$ for the supersymmetric non-extremal black holes, as a function of $p\in[-1,1]$ and plotted for the spindle $\Sigma=\mathbb{WCP}^1_{[3,1]}$.
        The dashed lines are $\pm$ the
holonomy of the gauge field, while the solid blue, orange, and green lines are the corresponding holonomy of the spin connection
for progressively smaller values of $\mathsf{Q}$, as given in Figure \ref{fig:circum}, tending to the extremal 
solution with $\mathsf{Q}=0$. For
$\mathsf{Q}\ne0$ we do not have a topological twist, but in the extremal limit $\mathsf{Q}\to 0$ we get a different topological twist
on each half of the spindle.}
        \label{fig:spingauge}
    \end{figure}

We can also display the way in which the two different topological twists for the extremal case arise in the limit that $\mathsf{Q}\rightarrow 0$.
In Figure \ref{fig:spingauge} we have plotted the spin connection and gauge field 
for the conformal boundary geometries corresponding to Figure \ref{fig:args}. More
precisely, the  non-trivial spin connection component is $\omega^{12}$, and we define
its holonomy around a circle in the spindle $\Sigma$ at constant $p$, parametrized by $\sigma$,~via
\begin{align}
\omega(p) = \frac{1}{2\pi}\int_{S^1} \omega^{12}\, .
\end{align}
The solid curves in Figure \ref{fig:spingauge}
are then $\tfrac{1}{2}\omega(p)$, while the dashed lines are $\pm$ the corresponding 
holonomy of the gauge field $A+\kappa_2 d\sigma$, {\it cf.} equation 
\eqref{twistme}. 

\begin{figure}[hbt!]
\begin{center}
        \includegraphics[width=0.15 \textwidth]{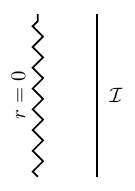}
\caption{Penrose diagram for the non-extremal supersymmetric black holes, with $e\ne 0$.
The black hole horizon has disappeared leaving a naked singularity. There is now a smooth conformal
boundary which consists of the product of the time direction with a spindle.
 }\label{fig:nakedPenrose}
\end{center}
\end{figure}

Although the conformal boundary of this non-extremal supersymmetric solution 
is perfectly regular, in the bulk the black hole horizon has disappeared
and there is a naked curvature singularity. To see this, we return to the 
black hole metric given in \eqref{nonrotBH} in PD-type coordinates and observe
that the function 
$\QQ(q)$ has no real roots for $\mathsf{Q}\neq 0$. There is then a 
naked curvature singularity at $q=\infty$, with 
Penrose diagram given by Figure \ref{fig:nakedPenrose}. 

\subsection{A one-parameter family of non-supersymmetric and non-extremal solutions}\label{sec:deform}

In the previous two subsections we have discussed special cases that lie inside the more general class of non-rotating PD black holes. While interesting because they preserve supersymmetry, as discussed they both have some pathologies: either an acceleration horizon that cuts the conformal boundary, or a naked singularity. These pathologies arise because of the specific restrictions we have imposed on the parameters in those cases. According to the number and value of the roots of the functions $Q$ and $P$ there are many other possibilities. A detailed analysis of the causal structure in various cases can be found in \cite{Dias:2002mi}. In this subsection we will consider another special case that, while allowing some degree of analytic control over the roots of the metric functions, gives a black hole with completely regular conformal boundary, and two ordinary horizons. This configuration is also smoothly connected with the extremal and BPS black hole, thus providing a kind of ``regulator'' of the latter solution, while staying 
within\footnote{We can also regulate the solutions in sections \ref{appaz} and \ref{sec:nonextreme} by
turning on the rotation parameter $a$ as considered in sections \ref{sec:uplift}, \ref{sec:BPS}. In particular, the supersymmetric and extremal rotating black holes have no acceleration horizons.}
the non-rotating family of solutions.

We start again with the general metric \eqref{PDmetric} with $a=0$, and we further restrict the parameters to satisfy 
\begin{align}\label{mgalee}
m=\frac{g}{\alpha}\,, \quad e=0\,,
\end{align}
and hence, in particular, we have $J=Q_e=0$.
The first of the conditions in \eqref{mgalee} is satisfied when {\it both} BPS conditions \eqref{BPSa0} are met, so it amounts to imposing only one of the two conditions. 
In addition, we recall that we imposed this condition in constructing regular uplifted solutions, as we discussed in section \ref{sec:uplift}. The second is the extremality condition in the BPS case. It follows that the black hole that we obtain with these restrictions is neither BPS nor extremal, but it is continuously connected with the case discussed in section \ref{appaz} by taking the limit
\begin{align}
g \rightarrow g_{BPS}=\frac{\sqrt{1-\alpha^2}}{\alpha^2}\,.
\end{align}
Since we are taking $a=e=0$ in \eqref{PDgauge}, we note that the gauge field is simply
\begin{align}
A=-g\,\cos\theta\,d\phi\,.
\end{align}

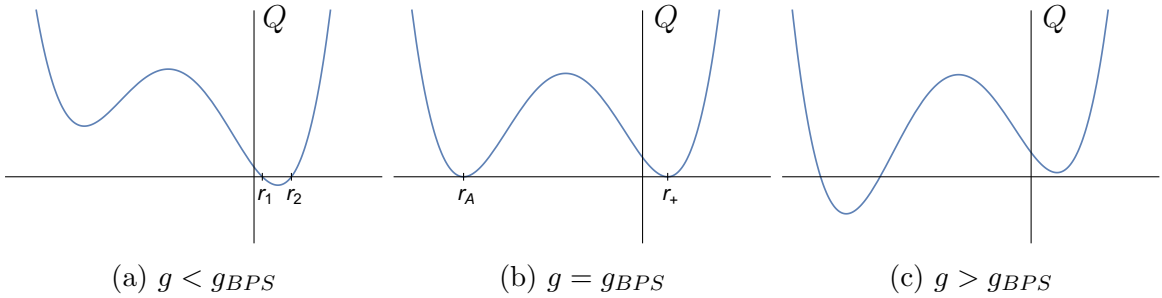
\begin{figure}[hbt!]
    \centering
        \begin{picture}(0.1,0.25)(0,0)
\put(106,105){\makebox(0,0){$Q$}}
\put(254,105){\makebox(0,0){$Q$}}
\put(401,105){\makebox(0,0){$Q$}}
\end{picture}
    \begin{subfigure}[b]{0.325\textwidth}
        \includegraphics[width=\textwidth]{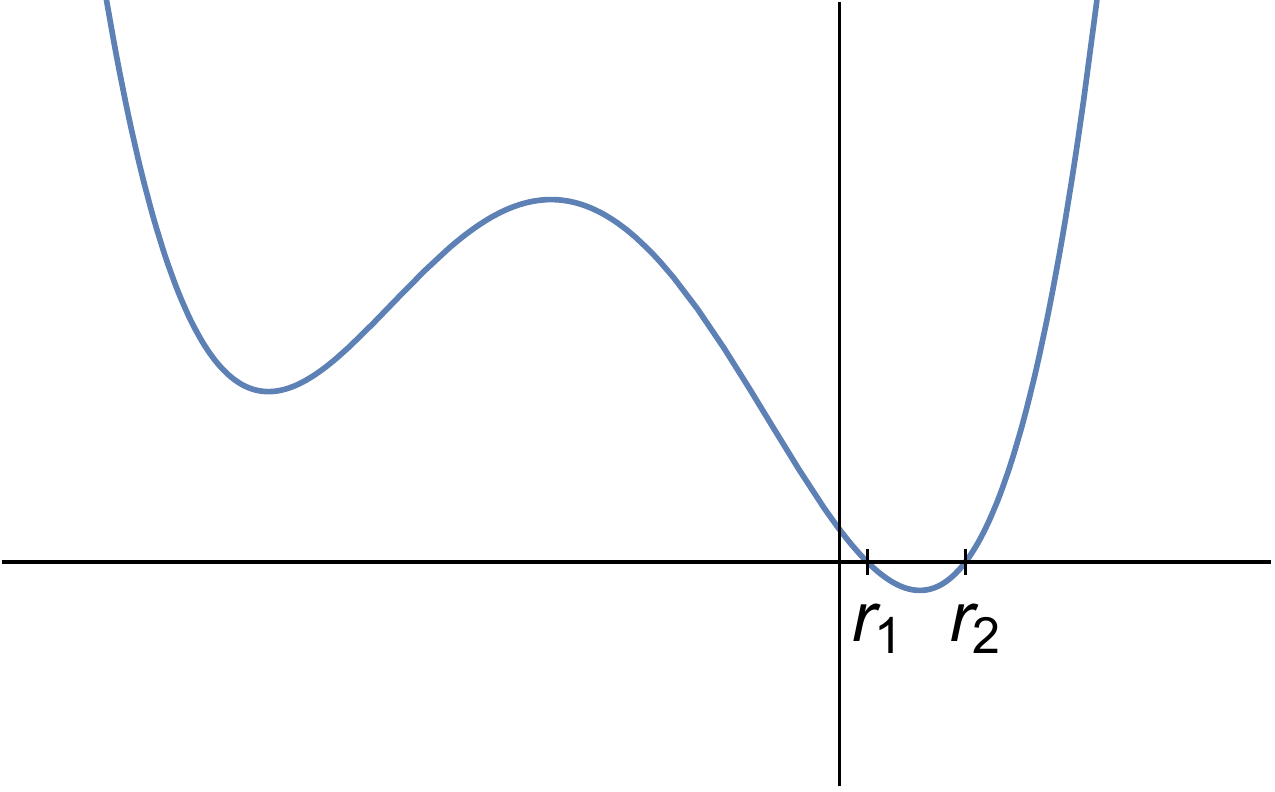}
        \caption{$g<g_{BPS}$}
        \label{fig:gsupBPS}
    \end{subfigure}
  \begin{subfigure}[b]{0.325\textwidth}
        \includegraphics[width=\textwidth]{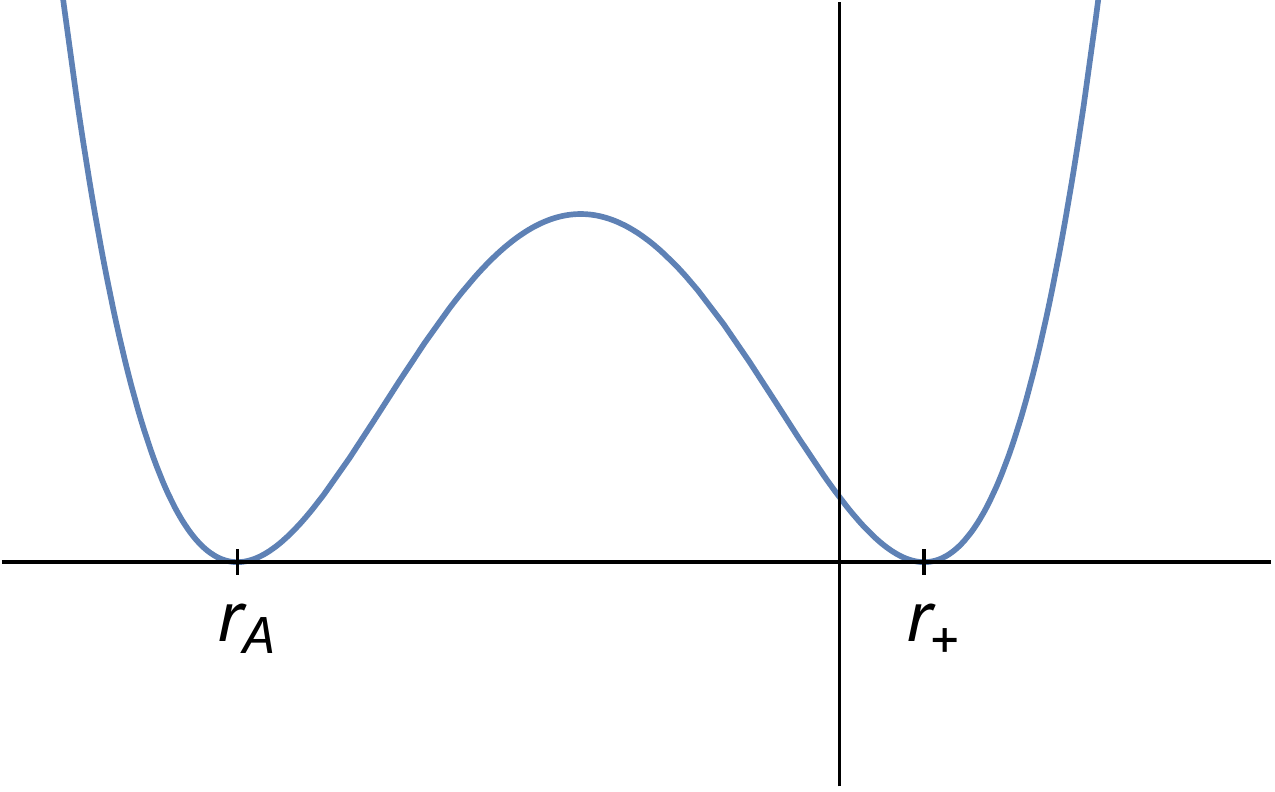}
        \caption{$g=g_{BPS}$}
        \label{fig:gBPS}
    \end{subfigure}
        \begin{subfigure}[b]{0.325\textwidth}
        \includegraphics[width=\textwidth]{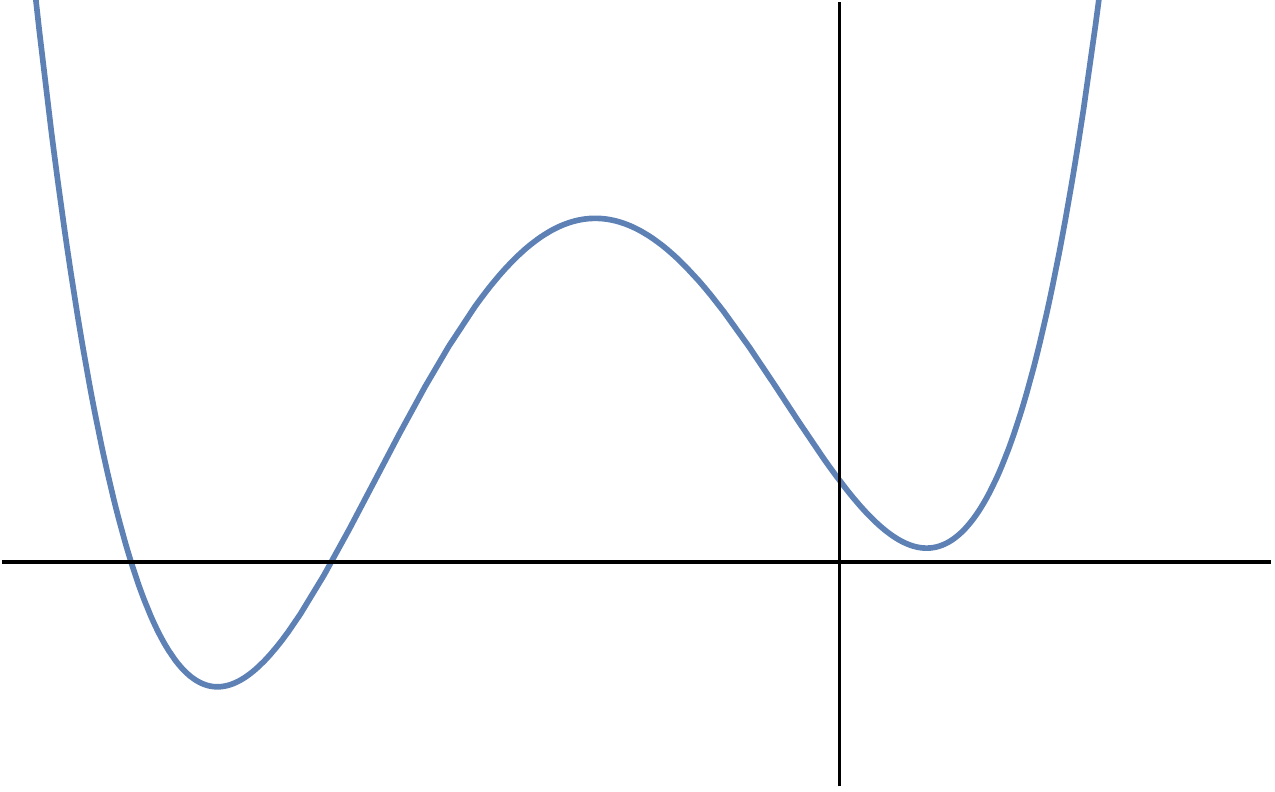}
        \caption{$g>g_{BPS}$}
        \label{fig:gsubBPS}
    \end{subfigure}
    \caption{The function $Q(r)$ for three different cases in the class of 
non-supersymmetric and non-extremal black holes with $m=g/\alpha$ and $a=e=0$. 
We focus on the cases $g<g_{BPS}$ and $g=g_{BPS}$. When $g<g_{BPS}$ the two positive roots, $r_1, r_2$, correspond to an inner and outer black hole horizon.
When $g=g_{BPS}$ these two horizons coalesce to give an extremal horizon, $r_+$, and, in addition, an acceleration horizon, $r_A$, also appears. For 
$g>g_{BPS}$ there are no positive roots of $Q$, and hence no black hole horizon, but instead two negative roots which correspond to two acceleration horizons. 
}\label{fig:gconditions}
\end{figure}

We continue to take $\sqrt{3}/2<\alpha<1$ as in section \ref{appaz}, or equivalently $0<\mathsf{P}<1/2$ as  in section \ref{sec:nonextreme}. For fixed $\alpha$, the roots of $Q(r)$ depend on $g$ in a very simple way, as illustrated in Figure \ref{fig:gconditions}. In particular, we find:
\begin{itemize}
\item When $g>g_{BPS}$, $Q(r)$ has two real roots for {\it negative} $r$ which correspond to acceleration horizons, and no black hole horizon. As in section \ref{appaz}, the former intersect the conformal boundary, which causes pathologies: this can be seen from the fact that in this case the combination $1-\alpha^2\sin^2\theta\,P(\theta)$ appearing in the boundary metric \eqref{bmetric}
has two real roots, and is negative between the two. 
\item When $g<g_{BPS}$, $Q(r)$ has two real roots for {\it positive} $r$, which give two black hole horizons (an inner and an outer horizon) with no acceleration horizons. For this case $1-\alpha^2\sin^2\theta\,P(\theta)$ has no roots for $0\leq \theta\leq \pi$, which means the conformal boundary is a smooth spindle as in section \ref{sec:nonextreme}.
\item Finally, when $g=g_{BPS}$ then $Q(r)$ has two pairs of coincident real roots, which is the case discussed in section \ref{appaz}.
\end{itemize}

\begin{figure}[hbt!]
\begin{center}
       \includegraphics[width=0.2 \textwidth]{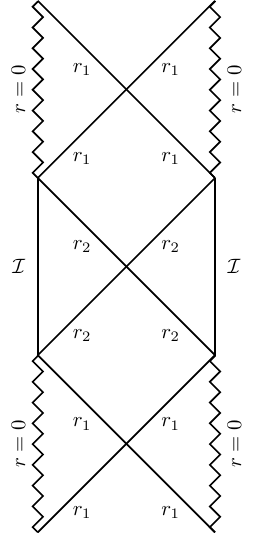}
\caption{\label{fig:Penrose_GeneralDeformation}Penrose diagram for the non-supersymmetric and non-extremal black holes
with $m=g/\alpha$ and $a=e=0$ and $g<g_{BPS}$. In addition to the black hole event horizon, denoted by $r_2$, there is
also an inner horizon, denoted by $r_1$. There is a smooth conformal
boundary which consists of the product of the time direction with a spindle.}
\end{center}
\end{figure}

Since the main point of this section is showing that we can have an ordinary black hole with no acceleration horizons, we shall focus on the case
\begin{align}
g<g_{BPS}\, .
\end{align}
Furthermore, note that while so far we have focused on the roots of $Q(r)$, the restrictions we have put on $\alpha$ and $g$ are also such as to guarantee $P(\theta)>0$ for $0\leq \theta\leq \pi$. Hence this case indeed corresponds to a completely regular black hole, whose Penrose diagram is given in Figure \ref{fig:Penrose_GeneralDeformation}, where we have denoted by $r_{1,2}$ the two positive roots of $Q(r)$, with $r_1<r_2$.
In the supersymmetric and extremal limit $g=g_{BPS}$ we have $r_1=r_2=r_+$.

We can then require that the topology in the $\theta,\phi$ directions is that of a spindle $\Sigma=\mathbb{WCP}^1_{[n_-,n_+]}$, by appropriately quantizing the conical deficits at $\theta=0,\pi$. This gives 
\begin{align}
\alpha=\sqrt{\frac{4\,n_-}{g(n_--n_+)}-\frac{2 g+1}{g^2}}\,,
\end{align}
while the periodicity of $\phi$ is given by
\begin{align}
\Delta\phi=\frac{\pi}{2\,g}\frac{n_--n_+}{n_-\,n_+}\,.
\end{align}
As discussed above, in this case the black hole is completely smooth, and the spindle topology at fixed $t$ and $r$ persists  to the conformal boundary, where the metric is again given by \eqref{bmetric} and has the topology $\mathbb{R}\times \mathbb{WCP}^1_{[n_-,n_+]}$. This is regular for any $0<g<g_{BPS}$, but degenerates as described in section \ref{appaz} when $g$ approaches the BPS value. The behaviour of the circumference of the spindle at the boundary and of the spin connection of the boundary metric are very similar to those given in Figures \ref{fig:circum} and~\ref{fig:spingauge}, respectively. Namely, when $g\to g_{BPS}$, the spindle splits in half at
\begin{align}
\theta = \theta_0 \equiv \arccos\left(\frac{1-\sqrt{5-4\alpha^2}}{2\sqrt{1-\alpha^2}}\right)\, ,
\end{align} 
while the spin connection approaches $\pm$ the gauge field, up to the pure gauge term discussed in section \ref{appaz}.

\section{Discussion}\label{sec:discussion}

In this paper we have studied a very general class of 
four-dimensional dyonically charged, rotating, and accelerating black holes in four-dimensional anti-de Sitter space. 
The acceleration leads to conical deficit 
singularities at the horizon which can be taken to stretch out to the conformal boundary.
When these conical deficits
are appropriately ``quantized'', so that 
the deficit angles are $2\pi(1- 1/n_\mp)$ with 
positive coprime integers $n_\pm$, the resulting 
space is known in the mathematics literature 
as a spindle, or equivalently a weighted 
projective space $\Sigma=\mathbb{WCP}^1_{[n_-,n_+]}$. 
Remarkably, when uplifted to $D=11$ on 
a regular Sasaki-Einstein seven-manifold $SE_7$,
the solutions become completely regular, free from any 
conical deficit singularities whatsoever.  
We have also quantized the flux of these $D=11$ solutions, 
thus showing that they give good M-theory backgrounds. 

We have shown that there is a sub-family 
of both supersymmetric and extremal black hole solutions, 
which interpolate between $AdS_4$ in the 
UV and $AdS_2\times \Sigma$ in the near horizon IR limit. 
These are characterized by the integers $n_\pm$, 
that determine the spindle horizon geometry $\Sigma=\mathbb{WCP}^1_{[n_-,n_+]}$, 
and a continuous parameter, which parametrizes both the electric charge $Q_e$ and the angular momentum $J$.
The entropy of these black holes, which also carry magnetic charge (given in \eqref{magnpnm}), can be expressed simply in terms of $n_\pm$ and $Q_e$.
We have shown that the entropy can be expressed in a number of equivalent ways, generalising previous  expressions applicable for non-accelerating supersymmetric $AdS_4$ black holes. In particular, \eqref{blackholentnicerbulk}
 reduces to the entropy of the extremal Kerr-Newman-AdS black hole upon setting $Q_m=0$ and $\chi=2$ (equivalently setting $n_-=n_+=1$ in \eqref{blackholent}).
The formula \eqref{anotherexp}, which applies also to the Kerr-Newman-AdS 
black hole, highlights the dependence of the entropy on the angular momentum computed at the horizon.
When uplifted, the near horizon limit gives a new class of rotating 
$AdS_2\times Y_9$ solutions, where we have shown that $Y_9$ 
may be viewed as either a regular $SE_7$ fibration over $\Sigma=\mathbb{WCP}^1_{[n_-,n_+]}$, 
or equivalently as a Lens space $S^3/\Z_{\pp}$ fibred over the 
$KE_6$ base of the $SE_7$. Remarkably, setting $Q_e=0$, which also sets $J=0$,
these reduce to a known class of 
supersymmetric $AdS_2\times Y_9$ solutions first 
constructed in \cite{Gauntlett:2006ns}. 
We have thus provided a new physical interpretation 
of those solutions: they are the near horizon limits 
of the accelerating (but non-rotating) black holes described in section \ref{sec:conboundary}, 
and we have generalized those solutions by adding angular momentum, 
preserving supersymmetry and extremality. 
It would be interesting to understand in more generality what kind of singularities
in lower-dimensional supergravity theories can be uplifted to obtain regular solutions in higher dimensions. For example, it would be  interesting to explore this for the $D=4$ black holes of \cite{Gnecchi:2013mja,Klemm:2014rda} which have non-compact horizons, but with finite entropy.

In this paper we have restricted our attention to solutions of minimal $D=4$ gauged supergravity. However, it is very likely
that our constructions can be generalized to more general gauged supergravity theories with various matter content. More specifically, we
expect to be able to construct supersymmetric spinning spindles which would generalize the constructions of \cite{Hristov:2019mqp}, for example,
where it was assumed that the horizon has spherical topology. 
We note the similarity of our formula for the black hole entropy 
\eqref{blackholentnicerbulk} with eq. (54) of \cite{Hristov:2019mqp}.

We now return to the holographic interpretation of the supersymmetric extremal black holes. 
The $D=11$ black hole solutions interpolate between $AdS_4\times SE_7$ in the UV and 
$AdS_2\times Y_9$ in the IR, where $Y_9$ is a $SE_7$ fibration 
over the spindle $\Sigma=\mathbb{WCP}^1_{[n_-,n_+]}$. 
The $AdS_4\times SE_7$ vacuum solution describes $N_{SE}$ 
M2-branes at the Calabi-Yau four-fold singularity with 
conical metric $dr^2+r^2ds^2(SE_7)$, and these
typically have dual field theory descriptions as 
 Chern-Simons quiver gauge theories, with the integer $N_{SE}$ determining 
the ranks of the gauge groups. Physically 
we are then wrapping the world-volume of the M2-branes 
over $\Sigma$. 
We have studied this conformal boundary geometry in some detail in 
section \ref{sec:conboundary} for the non-rotating solutions with $Q_e=J=0$. 
 An important subtlety in this case 
 is that
in the UV the conformal boundary is such that the spindle is split into two components.
Moreover, in this limit supersymmetry is preserved via a different topological twist on each component.
However, we have also shown that this split can be regulated in a family of non-rotating supersymmetric but non-extremal
black holes (or by further relaxing the supersymmetry condition). Moreover, we do not expect  
the generic supersymmetric extremal rotating black holes, with $Q_e\neq 0$, to have this pathology.
Indeed, in section \ref{sec:BPS} we have shown that the formula for the entropy 
of these black holes \eqref{BHintro}, is identical to that for the supersymmetric extremal Kerr-Newman 
family, obtained formally by setting $n_-=n_+=1$. On the other hand, 
the $Q_e=0$ solutions studied in section \ref{sec:conboundary} are a somewhat degenerate 
limit. 

With the above holographic interpretation, it should be possible to reproduce 
the black hole entropy formulae \eqref{blackholent} 
by studying the dual M2-brane field theories. 
Indeed, there has been considerable progress on this topic for 
various classes of 
supersymmetric $AdS_4$ black holes.  
In particular, the first class of black holes for which a dual field theory interpretation has been found
have just magnetic flux through a Riemann surface horizon $\Sigma$. The field theory calculation 
utilizes $I$-extremization, where the index can be identified with the localized partition function of the dual field theory 
on $S^1\times \Sigma$ \cite{Benini:2015eyy,Benini:2016rke,Cabo-Bizet:2017jsl,Azzurli:2017kxo,Hosseini:2019use,Gauntlett:2019roi,Hosseini:2019ddy,Kim:2019umc}.  
More recently, following the approach put forward in \cite{Cabo-Bizet:2018ehj}, progress has also been made in understanding
the class of electrically charged and rotating $AdS_4$ black holes 
from the dual field theory point of view
\cite{Choi:2018fdc,Cassani:2019mms,Bobev:2019zmz,Nian:2019pxj}.  
For the accelerating black hole solutions that we discussed in this paper, it should be straightforward
to now 
compute the suitably regularized on-shell action of the corresponding Euclidean solutions
and reproduce the entropy by extremizing the corresponding entropy function. 
From the field theory side we should then focus on the Euclidean version of the conformal boundary geometry of the 
charged, rotating and accelerating black holes, and compute a certain twisted topological index 
associated with the $d=3$, $\mathcal{N}=2$ SCFTs on the M2-branes, wrapped on the 
spinning spindle $\Sigma$. While some care may be required in taking the BPS and extremal limit, it seems possible that
we can get agreement between these computations using localization techniques in the large $N$ limit. 

The wrapped M2-brane theories flow to a
 $d=1$ 
superconformal quantum mechanics in the IR, that is dual to the 
$AdS_2\times Y_9$ solutions that arise as the near horizon limit of the black holes. 
Equation \eqref{R2d} says that the $d=3$ superconformal R-symmetry mixes 
with the $U(1)$ isometry of the internal space $\Sigma$
in flowing to the $d=1$ superconformal R-symmetry in the IR. It should similarly be possible 
to reproduce this formula via a dual field theory/SCQM calculation by computing and then extremizing a suitable
index. Indeed, in a companion 
paper \cite{Ferrero:2020laf} we study an analogous class 
of supersymmetric $D=5$ $AdS_3\times \Sigma$ solutions, 
with $\Sigma=\mathbb{WCP}^1_{[n_-,n_+]}$ again a spindle, that 
uplift on a regular Sasaki-Einstein five-manifold to 
smooth solutions of type IIB supergravity. These are the holographic 
duals to D3-branes wrapping the spindle, and in this case we 
are able to reproduce both the central charge and 
the mixing of $d=4$ and $d=2$ superconformal R-symmetries in the supergravity solution, 
where in the field theory dual we  
make use of anomaly polynomials and $c$-extremization \cite{Benini:2012cz}.   

\subsection*{Acknowledgments}
This work was supported in part by STFC grants ST/P000762/1, ST/T000791/1 and 
ST/T000864/1. 
JPG is supported as a Korea Institute for Advanced Study (KIAS) Scholar and as a Visiting Fellow 
at the Perimeter Institute for Theoretical Physics.

\appendix

\section{Circle fibrations over spindles}\label{app:WCP}

In section \ref{sec:liftmetric} we showed that the $D=4$ PD black hole metrics uplift to regular $D=11$ solutions 
that on a fixed $t$, $r$ slice
are topologically Lens space $S^3/\Z_{\pp}$ bundles over the $KE_6$. On the other hand, 
we explained that this same space
may also be viewed as an $SE_7$ fibration over a spindle/weighted projective space $\mathbb{WCP}^1_{[n_-,n_+]}$. 
Fixing a point on the $KE_6$ implies that the Lens space fibre $S^3/\Z_\pp$ is a circle bundle over the weighted 
projective space, where recall $\pp=(k/I)(n_--n_+)$. In this appendix we spell this out in a little more detail, discussing circle bundles 
over spindles more generally.

We begin with $S^3$, embedded inside $\C^2$ as the unit sphere $S^3=\{|z_1|^2 + |z_2|^2 = 1 \}\subset \C^2$, with $z_1,z_2$ standard 
complex coordinates. We may then consider the weighted circle action 
\begin{align}\label{weightedaction}
(z_1,z_2) \rightarrow (\lambda^{n_+} z_1, \lambda^{n_-} z_2)\, ,
\end{align}
where $\lambda = \ex^{\ii\theta}\in U(1)$, $n_\pm \in\mathbb{N}$, and note that for the action to be 
effective we need $\mathrm{hcf}(n_+,n_-)=1$. The quotient $S^3/U(1)=\mathbb{WCP}^1_{[n_-,n_+]}$ 
is by definition a weighted projective space. This is a complex orbifold which is topologically a two-sphere 
with conical angles $2\pi/n_\pm$ at the poles. This is also known as a spindle.
In terms of the action \eqref{weightedaction}, the 
poles arise from $z_2=0$ and $z_1=0$, respectively, where in the first case all powers of the primitive $n_+-$th root of unity $\lambda=\ex^{2\pi \ii /n_+}$
act trivially, while in the second case this is true for $\lambda=\ex^{2\pi \ii/n_-}$. 

It is a straightforward exercise to compute the Chern number of this fibration. That is, introduce a $(2\pi)$-period coordinate $\anotherangle$
along the weighted $U(1)$ action, and corresponding connection one-form $\mathscr{A}$. This can be done starting from the round 
metric on $S^3$, with corresponding term $(\diff \anotherangle + \mathscr{A})^2$ for the metric on the circle fibre. The result\footnote{Write the metric on $S^3$ as $ds^2=d\theta^2+\cos^2\vartheta d\phi_1^2+\sin^2\vartheta d\phi_2^2$ with $\vartheta\in[0,\pi/2]$ and $\Delta\phi_i=2\pi$. The weighted $U(1)$ action is $V=n_+\partial_{\phi_1}+n_-\partial_{\phi_2}$. Introduce new coordinates $\phi_1=n_+\anotherangle$ and $\phi_2=n_-\anotherangle+(n_+)^{-1}\mu$ with $\Delta \anotherangle=\Delta \mu=2\pi$. In the new coordinates 
$V=\partial_\anotherangle$, 
and the metric can be written as a $U(1)$ fibration over $\mathbb{WCP}^1_{[n_-,n_+]}$ as follows:
$ds^2=\Lambda(d\anotherangle+\mathscr{A}d\mu)^2+d\vartheta^2+\frac{\cos^2\vartheta \sin^2\vartheta}{\Lambda}d\mu^2$, where
$\mathscr{A}=\frac{n_-}{n_+}\frac{\sin^2\vartheta}{\Lambda} $ and $\Lambda=n_+^2\cos^2\vartheta + n_-^2\sin^2\vartheta$.       
}  is 
\begin{align}
\int_{\mathbb{WCP}^1_{[n_-,n_+]}} \frac{\diff\mathscr{A}}{2\pi} = \frac{1}{n_+n_-}\, ,
\end{align} 
where notice the overall sign is a matter of convention. 
By construction, the total space of this circle fibration over $\mathbb{WCP}^1_{[n_-,n_+]}$ is $S^3$. 

We may now consider a more general fibration with 
\begin{align}
\int_{\mathbb{WCP}^1_{[n_-,n_+]}} \frac{\diff\mathscr{A}}{2\pi} = \frac{\kapp}{n_+n_-}\, ,
\end{align} 
and $\kapp\in\Z$, and we will take $\kapp>0$ in what follows. We shall denote the corresponding complex line bundle, on which $\mathscr{A}$ is a connection, by
$O(\kapp)$. In terms of the original construction of $\mathbb{WCP}^1_{[n_-,n_+]}$ as a quotient, notice 
that $\anotherangle$ has period $2\pi/\kapp$. The total space of this circle fibration is then 
 $S^3/\Z_{\kapp}$, where the $\Z_{\kapp}$ action is generated by
\begin{align}\label{actor}
(z_1,z_2) \rightarrow (\omega_{\kapp}^{n_+} z_1,\omega_{\kapp}^{n_-} z_2)\, ,
\end{align} 
where $\omega_{\kapp}\equiv \ex^{2\pi \ii /\kapp}$ is a primitive $\kapp$th root of unity. In general the action \eqref{actor} is not free. 
Specifically, the circle $S^1=\{z_2=0\}$ is fixed by $\Z_{\mathrm{hcf}(\kapp,n_+)}$, while the circle 
$S^1=\{z_1=0\}$ is fixed by $\Z_{\mathrm{hcf}(\kapp,n_-)}$. Thus the total space is a smooth manifold (rather than an orbifold) if and only if $\kapp$ has no common factor with either $n_+$ or $n_-$.  

The main case of interest in the main text is when 
\begin{align}
\kapp=\pp= \frac{k}{I}( n_- - n_+)\, .
\end{align}
Recall here that $I/k$ is an integer that divides $n_--n_+$.
The gauged supergravity connection $2A$ is 
a connection on $O(\pp)$, as one sees for example in equation \eqref{QmAdS}. 
Since $\mathrm{hcf}(n_+,n_-)=1$, it immediately follows that $\mathrm{hcf}(\pp, n_\pm) = 1 $, and 
the total space of the circle fibration is smooth.  In fact the construction in section~\ref{sec:liftmetric}
indirectly implies that this is diffeomorphic to the Lens space $S^3/\Z_{\pp}=L(\pp,1)$. 
We may see this directly as follows. Since $\mathrm{hcf}(\pp,n_+)=1$, we can find integers $a,b\in \Z$ such that 
$an_+ + b \pp = 1$. It follows that $a\frac{n_+}{\pp} = - b + \frac{1}{\pp}$, and hence
\begin{align}
(\omega_{\pp}^{n_+})^a = \ex^{2\pi \ii a n_+/ \pp} = \omega_{\pp}\, .
\end{align}
Thus the $\Z_{\pp}$ action \eqref{actor} is equivalent to 
\begin{align}
(z_1,z_2) \rightarrow (\omega_{\pp} z_1, \omega_{\pp}^{a n_-} z_2)\, ,
\end{align}
which is the definition of the Lens space $L(\pp,an_-)$. On the other hand, we have $an_-= a\left(n_++ \frac{I}{k}\pp\right) = 1  +  \pp (\frac{I}{k}a- b)\cong  1$ mod $\pp$. 
Thus $L(\pp ,an_-)\cong L(\pp,1)=S^3/\Z_{\pp}$, as we wanted to show.

\section{Non-accelerating case: $\alpha=0$}\label{nonacc}

The principal focus of this paper is to study accelerating black holes with $\alpha\ne 0$. However, for completeness
we discuss here the case when $\alpha=0$, also known as the Kerr-Newman-AdS spacetime \cite{Carter:1968ks}.

We set $\alpha=0$ in \eqref{PDmetric} and then rescale 
\begin{equation}
\phi = \varphi\, \Xi^{-1}\, ,
\end{equation}
with $\Xi=1-a^2$, to ensure that the metric is well-defined at $\theta=0$ and $\theta=\pi$, with $\varphi \in [0, 2\pi)$. The metric 
now reads\footnote{We can compare with metric as given in \cite{Bobev:2019zmz}, which has $g=0$. We should make the identification
$m^{\mathrm{there}}(1+2\sinh^2\delta)=m$, $m^{\mathrm{there}}\sinh 2\delta=e$ and $\tilde r=r$ as well as $A^{\mathrm{there}}=-2 A$.}
 \begin{align} \label{PDmetricalphz}
d s^2=
 -\frac{Q}{\Sigma}\big(d t- \frac{a\sin^2\theta}{\Xi}\,d\varphi \big)^2 +\frac{\Sigma}{Q}\,d r^2  
+\frac{\Sigma}{P}d\theta^2 +\frac{P}{\Sigma}\sin^2\theta \big( ad t -\frac{r^2+a^2}{\Xi}d\varphi \big)^2  , 
\end{align}
with 
\begin{align}
\Sigma&=r^2+a^2\cos^2\theta \,, \nn
 P&= 1 - a^2 \cos^2\theta \,,\nn
 Q&= (r^2+a^2)(1+r^2)-2mr +e^2+g^2 \,,
\end{align}
and the gauge field given by
\begin{align}
A=-e\frac{r}{\Sigma}\big(dt-\frac{a\sin^2\theta}{\Xi} d\varphi\big)+g\frac{\cos\theta}{\Sigma}\big(adt-\frac{r^2+a^2}{\Xi}d\varphi\big)\,.
\end{align}

When $\alpha=0$, the conditions for preservation of supersymmetry need to be carried out again. We find that
the integrability conditions now give the BPS constraints
\begin{align}
0 =&\, m \, g \,, \\
0 =&\, e^4 - 2 a^2 e^4 + a^4 e^4 + 2 e^2 g^2 - 4 a^2 e^2 g^2 + 2 a^4 e^2 g^2 - 4 e^4 g^2 + g^4 - 2 a^2 g^4 + a^4 g^4 \nn
&- 8 e^2 g^4 - 4 g^6 - 2 e^2 m^2 - 2 a^2 e^2 m^2 - 2 g^2 m^2 - 2 a^2 g^2 m^2 + m^4 \, .
\end{align}
With $m>0$ we therefore\footnote{As in \cite{Klemm:2013eca} setting  $m=0$ is a solution to the BPS equations, but it gives no real positive roots to $Q$ and hence no event horizon.} take $g=0$. The second equation can then be solved and with $a,e\ge 0$ we conclude
that for BPS black holes we should take
\begin{align}
g=0\,,\qquad
m=(1 + a) e \,.
\end{align}

By studying the roots of $Q$ we can determine that we have an extremal BPS black hole provided that
\begin{equation}
e = \sqrt{a} (1+a) \,,
\end{equation}
and hence
\begin{align}\label{massextbpsaz}
m=\sqrt{a} (1+a)^2\,,
\end{align}
with the horizon at $r_+=\sqrt{a}$. From \eqref{elecflux} we can then write the total electric charge as 
\begin{align}
G_{(4)}Q_e=\frac{\sqrt{a}}{1-a}\,,
\end{align}
and from \eqref{exprangmom2} the total angular momentum is
\begin{align}\label{JQeapp}
G_{(4)}J & = \frac{a\sqrt{a}}{(1-a)^2}  = \frac{G_{(4)}Q_e}{2}\left(\sqrt{1+4(G_{(4)}Q_e)^2}-1 \right) \, .
\end{align}
Note that when $\alpha=0$, this expression for the angular momentum, in the gauge we are using, can also
be obtained from a Komar integral \eqref{komdef}, as is often used in the literature. 
Finally, from \eqref{entgenexp} the entropy of these black holes is given by  
\begin{align}\label{entgenexp2}
S_{BH} & = \frac{\pi}{G_{(4)}}\frac{a}{1-a} =\frac{\pi}{2G_{(4)}}(\sqrt{1+4(G_{(4)}Q_e)^2}-1) =   \frac{\pi} {G_{(4) }}\frac{J}{Q_e} \,,
\end{align}
in agreement with the literature.

\section{Near horizon limit}\label{appnh}

In this appendix we shall consider the near horizon limit of the PD solution (\ref{PDmetric})--(\ref{PDgauge}), in the BPS and extremal case. To this end, we first write the function $Q(r)$ as
\begin{align}
Q(r)=(r-r_+)^2\,\left(x_0+x_1\,r+x_2\,r^2 \right),
\end{align}
where the constants $x_i$ are defined by
\begin{align}\label{defxs}
x_0&= -\alpha ^2 \left(a^2+e^2+g^2\right)+a^2+4 \alpha\,  g\, r_+-3 \left(\alpha ^2-1\right) r_+^2+1\, ,\nn
x_1&= 2 \alpha\,  g-2 \left(\alpha ^2-1\right) r_+\, ,\nn
x_2&= 1-\alpha ^2\, .
\end{align}
The horizon radius, $r_+$, is subject to the following condition:
\begin{align}
&-4 \alpha  \left(a^2+e^2\right)+4 \alpha ^3 \left(a^2+e^2+g^2\right)-5 \alpha  g^2\nn
&+g  \left(\left(a^2-5\right) \alpha ^2+\alpha ^4 \left(-\left(a^2+e^2+g^2\right)\right)+6\right)\,r_+\nn
&+\alpha   \left(-2 a^2 \left(\alpha ^2-1\right)^2-2 \alpha ^4 \left(e^2+g^2\right)+\alpha ^2 \left(2 e^2+5 g^2+2\right)-2\right)\,r_+^2=0\, ,
\end{align}
with the parameters obeying the BPS constraints.\footnote{In the non-accelerating case, with $\alpha=g=0$, this condition degenerates and one instead has $r_+=\sqrt{a}$.}

A convenient way to find the near horizon solution is to implement the following coordinate transformation,
\begin{align} \label{toNH}
r\to r_++\lambda\,s\,\rho\, ,\quad
t \to \lambda^{-1}\,s\,\tau\, ,\quad
\phi\to\phi'+\lambda^{-1}\,s\,W\frac{\Delta\phi}{2\pi}\,\tau\, ,
\end{align}
where $s$ is a constant, and then take the $\lambda\to 0$ limit. Here $W$ is given by
\begin{align}
W=\frac{a}{r_+^2+a^2}\frac{2\pi}{\Delta\phi}\, ,
\end{align}
with $\partial_t+W\frac{\Delta\phi}{2\pi} \partial_\phi=\partial_t+W\partial_\varphi$ a null generator of the horizon.
It is convenient to choose 
\begin{align}
s=\sqrt{\frac{r_+^2+a^2}{x_0+x_1\,r_++x_2\,r_+^2}}\,,
\end{align}
and we then find that the near horizon metric reads, after dropping the primes from the new coordinates,
\begin{align}\label{NHmetric}
ds^2_4={\lambda}(\theta)\,\left( -r^2\, dt^2+\frac{dr^2}{r^2}\right)+\frac{d\theta^2}{\tilde{P}(\theta)}+\frac{(r_+^2+a^2)^2\,\sin^2\theta\,\tilde{P}(\theta)}{(1-\alpha\,r_+\,\cos\theta)^4}\,\left( d\phi+vr\,dt \right)^2,
\end{align}
where
\begin{align}
\tilde{\lambda}(\theta)&=\frac{a^2 \cos ^2 \theta +r_+^2}{\left(1-\alpha  r_+ \cos \theta \right){}^2 \left(x_0+x_1\,r_++x_2\,r_+^2\right)}\, ,\nn
\tilde{P}(\theta)&=\frac{ \left(1-\alpha  r_+ \cos \theta \right){}^2}{a^2 \cos ^2\theta +r_+^2}\,P(\theta )\, ,\nn
v&=\frac{2 a r_+}{\left(a^2+r_+^2\right) \left(x_0+x_1\,r_++x_2\,r_+^2\right)}\, .
\end{align}
For the gauge field, we find that there is a piece that is singular in the $\lambda\to 0$ limit, but it can be removed by a gauge transformation.
We thus implement the gauge transformation
\begin{align} \label{NHgaugetransf}
A\to A -\frac{s \,r_+\,e}{\lambda  \left(a^2+r_+^2\right)}\,dt- \frac{e (a^2-r_+^2) }{2 a r_+}d\phi\,,
\end{align}
where the second term is included for convenience. Taking the $\lambda\to 0$ limit we then 
obtain the near horizon gauge field
\begin{align} \label{NHgauge}
A=-\frac{\left(a^2+r_+^2\right) \left(a^2 e \cos ^2\theta +2 a g r_+ \cos \theta -e r_+^2\right)}{2 a r_+ \left(a^2 \cos ^2\theta +r_+^2\right)}\,\left( d\phi+v\,r\,dt \right)\,.
\end{align}

We can now show that the near horizon metric \eqref{NHmetric} and gauge field \eqref{NHgauge} are equivalent to the solution \eqref{epsilonsols}. The easiest way to see this is to start from \eqref{NHmetric}, and perform the following coordinate transformations: \begin{align} 
\tau \to -t\, , \quad 
\rho \to r\, , \quad
y \to \frac{c_1+c_2\,\cos\theta}{1-\alpha\,r_+\,\cos\theta}\, , \quad
z \to -\kappa\, \phi\, ,
\end{align}
where 
\begin{align}
c_1&=\frac{2\,\alpha\,r_+^3}{\sqrt{\left( x_0+x_1\,r_++x_2\,r_+^2\right)\left( a^2+\alpha^2\,r_+^4 \right)}}\, ,\nn
c_2&=\frac{2\,a^2}{\sqrt{\left( x_0+x_1\,r_++x_2\,r_+^2\right)\left( a^2+\alpha^2\,r_+^4 \right)}}\, ,\nn
\kappa&=\sqrt{\frac{ x_0+x_1\,r_++x_2\,r_+^2 }{ a^2+\alpha^2\,r_+^4 }}\,\left( r_+^2+a^2 \right)\, .
\end{align}
The parameters are then identified as follows:
\begin{align}
\label{pararelations}
\qp&=\frac{2\,a\,r_+}{\sqrt{\left( x_0+x_1\,r_++x_2\,r_+^2\right)\left( a^2+\alpha^2\,r_+^4 \right)}}\, ,\nn
\Qp&=4\frac{-a^2\,g+2\,a\,e\,\alpha\,r_+^2+g\,\alpha^2\,r_+^4}{\left( x_0+x_1\,r_++x_2\,r_+^2 \right)\left( a^2+\alpha^2\,r_+^4 \right)}\, .
\end{align}

Let us now turn to the non-rotating case, $a=0$. As discussed in 
section \ref{appaz}, 
in the BPS and extremal limit we can express all parameters and horizon radius in terms of $\alpha$ via
\begin{align}\label{BPSextr_nonrotatingapp}
e=0\, , \quad
g=\frac{\sqrt{1-\alpha ^2}}{\alpha ^2}\, , \quad 
m=\frac{\sqrt{1-\alpha ^2}}{\alpha ^3}\, , \quad
r_+=\frac{\sqrt{5-4 \alpha ^2}-1}{2 \alpha  \sqrt{1-\alpha ^2}}\, .
\end{align}
With $Q(r)=(r-r_+)^2\,(x_0+x_1\,r_++x_2\,r_+^2)$, we can now obtain simple expressions for $x_0,x_1,x_2$:
\begin{align}
x_0&= \frac{-2 \alpha ^2+\sqrt{5-4 \alpha ^2}+3}{2 \alpha ^2}\, , \quad 
x_1=\frac{\sqrt{1-\alpha ^2} \left(\sqrt{5-4 \alpha ^2}+1\right)}{\alpha }\, ,\nn
x_2& =1-\alpha ^2\, .
\end{align}
We then take the near horizon limit with the coordinate transformation \eqref{toNH}, where we note that now since $a=0$ we have $W=0$, as expected since we have switched off rotation. Taking the limit $\lambda\to 0$, and again
dropping the primes from the new coordinates, we get the near horizon metric
\begin{align}
ds^2& =\frac{r_+^2}{\left( 1-\alpha\,r_+\,\cos\theta \right)^2}\bigg[\frac{1}{x_0+x_1\,r_++x_2\,r_+^2}\,\left( -r^2\,dt^2+\frac{dr^2}{r^2} \right)\nn
& \qquad \qquad \qquad \qquad \qquad +\frac{d\theta^2}{P(\theta)}+P(\theta)\,\sin^2\theta\,d\phi^2\bigg]\, ,
\end{align}
and gauge field
\begin{align}
A=-g\,\cos\theta\,d\phi\, ,
\end{align}
where we recall that all parameters are constrained by the BPS and extremality conditions \eqref{BPSextr_nonrotatingapp}. It can be shown that this solution is equivalent to the non-rotating case ($\qp=0$) of the metric \eqref{epsilonsols}. To see this, starting from \eqref{epsilonsols} with $\qp=0$, one needs to change coordinates and identify the parameters as follows: 
\begin{align}
\tau&\to -t\, , \quad
\rho\to r\, , \quad
y\to \frac{2 \left(\sqrt{5-4 \alpha ^2}-1\right)}{\sqrt{5-4 \alpha ^2} \left(\left(1-\sqrt{5-4 \alpha ^2}\right) \cos \theta +2 \sqrt{1-\alpha ^2}\right)}\, , \nn
z &\to-\frac{\sqrt{5-4 \alpha ^2}}{\alpha ^2}\,\phi\, , \quad
\Qp\to\frac{4 \sqrt{1-\alpha ^2}}{5-4 \alpha ^2}\, .
\end{align}
Recall that $\sqrt{3}/2<\alpha<1$, which corresponds to the range $0<\mathtt{a}<1$.

\section{The $AdS_2\times Y_9$ solutions of \cite{Gauntlett:2006ns}}\label{oldsols}
We briefly outline how the $D=11$ $AdS_2\times Y^9$ solutions found in \cite{Gauntlett:2006ns} and further discussed in appendix D of \cite{Gauntlett:2019pqg} can also be viewed as solutions of $D=4$ minimal gauged supergravity. The analysis of regularity and flux quantization
which was carried out in \cite{Gauntlett:2019pqg} is different to what we have done in this paper, and we explain how they are related.

Consider the metric $ds^2(Y_9)$ given in (3.36) of \cite{Gauntlett:2006ns}:
\begin{align}\label{Y9oldfriendorig}
ds^2(Y_9)=\hch(y)Dz^2+\frac{q(y)}{y^6 \hch(y)} D\psi^2+\frac{4}{q(y)} dy^2+\frac{16}{y^2}ds^2_{KE_6}\, ,
\end{align}
where $Dz=dz-g(y) D\psi$, $D\psi= d\psi+4 B$, with $dB=2J$ and
\begin{align}
q(y)&=y^4-4y^2+4 {\mathtt{a}} y-{\mathtt{a}}^2\,,\nn
g(y)&=\frac{{\mathtt{a}}-y}{y^3-3y+2{\mathtt{a}}}\,,\nn
\hch(y)&=\frac{y^3-3y+2{\mathtt{a}}}{y^3}\,.
\end{align}

If we complete the square using the $\psi$ coordinate 
we
immediately find
\begin{align}\label{Y9oldfriend}
ds^2(Y_9)=\frac{1}{y^2}\Big[D\psi+(1-\frac{\mathtt{a}}{y} )dz\Big]^2+\frac{4}{q} dy^2+\frac{q}{y^4} dz^2+\frac{16}{y^2}ds^2_{KE_6}\,.
\end{align}
Assembling the $D=11$ metric, as described in \cite{Gauntlett:2006ns}, we obtain
\begin{align}\label{bbmet}
ds^2_{11}&=\frac{2}{3}\left[ y^2ds^2(AdS_2)+ \frac{4y^2}{q} dy^2+\frac{q}{y^2} dz^2+\Big[D\psi+(1-\frac{\mathtt{a}}{y} )dz\Big]^2
+16ds^2_{KE_6}\right]\,,\nn
&=\frac{32}{3}\Bigg[ \frac{1}{4}\Big\{\frac{y^2}{4}ds^2(AdS_2)+ \frac{y^2}{q} dy^2+\frac{q}{4y^2} dz^2\Big\}\nn
& \qquad \qquad \qquad +
\Big\{\Big[\frac{1}{4}d\psi+B+\frac{1}{4}(1-\frac{\mathtt{a}}{y} )dz\Big]^2
+ds^2_{KE_6}\Big\}\Bigg]\,.
\end{align}
 Comparing this with \eqref{lift} we see this is precisely of the form to give a solution of minimal $D=4$ gauged supergravity with
\begin{align} \label{AdS2fromJerome06}
ds^2_4&=\frac{y^2}{4}ds^2(AdS_2)+ \frac{y^2}{q} dy^2+\frac{q}{4y^2} dz^2,\nn
A&=\frac{1}{2}(1-\frac{\mathtt{a}}{y})dz\,,
\end{align}
after choosing $L=(32/3)^{1/2}$.
Finally, as a check, the four-form flux in \cite{Gauntlett:2006ns} can be written as
\begin{align}
G=\left(\frac{32}{3}\right)^{3/2}\left[\tfrac{3}{8}\vol_4 - \tfrac{1}{2}*_4 F \wedge J\right]\,,
\end{align}
in agreement with \eqref{lift}.

We now return to the metric as written in \eqref{Y9oldfriendorig} and recall the analysis demonstrating regularity, as discussed in \cite{Gauntlett:2006ns}. 
The analysis of \cite{Gauntlett:2006ns} begins by showing that after taking $\Delta\psi=2\pi$, then $\psi,y$ parametrize a smooth two-sphere. Then one shows that
this two-sphere can be fibred over the $KE_6$ space to give an eight-dimensional manifold.
Next, by choosing the
period of $z$ to be $2\pi l$, for suitably defined $l$,  we obtain a good circle fibration over the eight-dimensional manifold.
To implement the latter one shows that the periods over a basis of two-cycles on the eight-dimensional space are suitably quantized. A basis can be taken to be the $S^2$ fibre at a fixed point on the $KE_6$ together with a basis of two-cycles on
$KE_6$ sitting at the one of two poles on the $S^2$, say $y=y_2$.
The conditions are satisfied if and only if
\begin{align}
g(y_3) -g(y_2)=l\pp\, ,\qquad
g(y_2)=l\kp/I\, ,\qquad
\end{align}
with integers $\kp, \pp$. If $\kp, \pp$ have no common factor  then the nine-dimensional space $Y_9$ is simply-connected.
As shown in \cite{Gauntlett:2006ns} these conditions imply 
\begin{align}
\mathtt{a} &=\frac{ I\pp(2\kp + I \pp)}{ 2\kp^2 + 2I\kp\pp + I^2\pp^2}\,,\nn
l^2&=\frac{I^2(2\kp^2+2I \kp \pp+ I^2 \pp^2}{2\kp^2(\kp+I\pp)^2}\,.
\end{align}

This construction can also be viewed as a fibration of a Lens space $S^3/\mathbb{Z}_\mathtt{q}$ over the $KE_6$. To see this more explicitly we 
can use canonical coordinates on $S^3/\mathbb{Z}_\mathtt{q}$, with the $\mathbb{Z}_\pp$ identification acting on a Hopf fibre coordinate. 
We consider\footnote{To compare with \cite{Gauntlett:2006ns} we should identify $\bar\alpha=\alpha^{there}$.}
 $\psi=\gamma$ and $z=\bar\alpha+\lambda \gamma$ and we fix the constant $\lambda$  momentarily. Observe that this implies $\partial_\psi=\partial_\gamma-\lambda\partial_{\bar\alpha}$ and $\partial_z=\partial_{\bar \alpha}$. In the new coordinates the metric reads 
\begin{align}\label{Y9oldfriendorigagain}
ds^2_{Y_9}=\hch(y)(d\bar \alpha-\tilde g D\gamma- 4\lambda B)^2+\frac{q}{y^6 \hch(y)} D\gamma^2+\frac{4}{q} dy^2+\frac{16}{y^2}ds^2_{KE_6}\,,
\end{align}
with $D\gamma= d\gamma+4 B$ and $\tilde g(y)\equiv g(y)-\lambda$. If we take $\Delta\gamma=2\pi$ then $\gamma,y$ again 
parametrize an $S^2$, which is
fibred over the $KE_6$ manifold. We now consider the $\bar\alpha$ circle fibred over this eight-dimensional base. To see the Lens space structure, we fix 
a point on the $KE_6$ base and fix the constant $\lambda$ by demanding $\tilde g(y_3)+\tilde g(y_2)=0$, which ensures that $\bar\alpha$ is a Hopf fibre coordinate on the resulting $S^3/\mathbb{Z}_\mathtt{q}$. More specifically, we choose $\lambda=\frac{l}{2}(\mathtt{q}+2\mathtt{p}/I)$ and then we have
\begin{align}
\tilde g(y_3)=-\tilde g(y_2)=\frac{l\mathtt{q}}{2}\,.
\end{align}
Taking $\Delta\gamma=2\pi$ and $\Delta\bar\alpha=2\pi l$ then implies that we have an $S^3/\mathbb{Z}_\mathtt{q}$ fibre. That the $S^3/\mathbb{Z}_\mathtt{q}$ is suitably fibred over the $KE_6$ base is easily demonstrated: taking a basis of two-cycles, $\Sigma_a(y_2),$ to be again located
at $y=y_2$, for example we calculate
\begin{align}
\frac{1}{2\pi l}\int_{\Sigma_a(y_2)}d[d\bar\alpha-\tilde g D\gamma-4\lambda B]&=-\frac{1}{2\pi l}(\tilde g(y_2)+\lambda)\int_{\Sigma_a(y_2)}\rho\nn
&=-\mathtt{p}s_a\,,
\end{align}
for integers $s_a$.

We can also introduce coordinates on $S^3/\mathbb{Z}_\mathtt{q}$ via $\bar\alpha_\pm=\gamma\mp\frac{1}{l\mathtt{q}}\bar\alpha$. In these coordinates
we can write the vectors in the original $\psi, z$ coordinates as $\partial_z=-\frac{1}{lq}(\partial_+-\partial_-)$ and
$\partial_\psi=\partial_++\frac{\mathtt{p}}{I\mathtt{q}}(\partial_+-\partial_-)$, with the latter directly analogous to \eqref{trickypsi}.
This underscores that while in the regularity construction using  the $z,\psi$ coordinates, $\Delta\psi=2\pi$ was required
to ensure that $y,\psi$ formed an $S^2$, it is not the case, in general, that $\psi$ is a periodic coordinate on the three-dimensional
Lens space, i.e. under the motion of $\partial_\psi$ by $2\pi$ one does not return to the same point.\footnote{This simple, but potentially confusing,
point can be made even more explicit. If we forget about the $KE_6$ we note that for {\it any} value of $\lambda$
we still obtain a Lens space parametrized by $\bar\alpha,\gamma$ with $\Delta\gamma=2\pi$ and $\Delta \bar\alpha=2\pi l$. However, since
$\partial_\psi=\partial_\gamma-\lambda\partial_{\bar\alpha}$ it is clear that the orbits of $\partial_\psi$ will not close in general.}

A final point is that the R-symmetry Killing vector can be obtained from \cite{Gauntlett:2006ns}, and we find that it is given
by
\begin{align}\label{R2dnorot}
R=2\partial_\psi+2\partial_z\, .
\end{align}
This agrees with our general formula \eqref{R2d} in the $\qp=0$ limit.

\section{Angular momentum}\label{sec:angmom}

In this section we discuss the angular momentum of the PD black holes \eqref{PDmetric}, \eqref{PDgauge}. We will also calculate the angular momentum of the near horizon $AdS_2$ solutions that we constructed in appendix \ref{appnh}, and explain how these two quantities differ in general. 
Finally, we shall derive the formula \eqref{angmomsurprise}. 

A common way of defining the angular momentum is via a Komar integral associated to the spacelike Killing vector $k=\partial_{\varphi}$, where
$\varphi$ is the angular coordinate on the spindle with period $2\pi$ (see \eqref{canangcoord}).
For the associated one-form $k$ one takes
\begin{align}\label{komdef}
J_K(r)&\equiv \frac{1}{16\pi G_{(4)}}\int_{\Sigma_r}\ast dk\,,
\end{align}
where the integral is over the surface $\Sigma_r$, parametrized by $\theta,\phi$ at fixed $r$ and $t$.
While this integral is gauge invariant, it clearly depends on the radial coordinate $r$ since, in general, $d*dk\ne 0$. 

Instead we will adopt a different definition which does not depend on the radial position where the integral is done, and hence
it can equally be evaluated at the black hole horizon. While this an attractive feature, and is the definition that
is expected to appear in the First Law \cite{Papadimitriou:2005ii}, 
the price we pay is that the integral changes under gauge transformations. We first define the two-form  
\begin{align}\label{expforG}
\Hch=dk+4(A\cdot k)F\,,
\end{align}
for an arbitrary Killing vector $k$ which satisfies, on-shell,
\begin{align}
\nabla_\mu \Hch^{\mu\nu}=-k^\nu\mathcal{L}\,,
\end{align}
where $\mathcal{L}$ is the Lagrangian. For $k=\partial_{\varphi}$ we can then define the angular momentum via the horizon integral
\begin{align}\label{capjaydef}
J(A)=\frac{1}{16\pi G_{(4)}}\int_{\Sigma_{r_+}}*\Hch\,.
\end{align}
With the gauge field as in \eqref{PDgauge}, we obtain 
\begin{align}\label{exprangmom}
J\equiv J(A)=\frac{1}{G_{(4)}}{m\,a }\,\left(\frac{\Delta\phi}{2\pi}\right)^2\,.
\end{align}
For the simpler case of the Kerr-Newman-AdS black holes with $\alpha=0$ as in \eqref{PDmetricalphz}, we then have
\begin{align}\label{exprangmom2} 
G_{(4)}J=\frac{m a}{(1-a^2)^2}\,,
\end{align}
in agreement with the results given in \cite{Papadimitriou:2005ii, Bobev:2019zmz}, for example. Note also that for the case of $\alpha=0$, the expressions \eqref{exprangmom}, \eqref{exprangmom2} can also be obtained from the Komar integral \eqref{komdef}; we have not checked whether this also happens to be true when $\alpha\ne 0$ (and one should be aware that the conformal boundary is not obtained simply by taking $r\to \infty$).

We would also like to calculate the angular momentum using the near horizon $AdS_2$ solution \eqref{epsilonsols}.
However, \eqref{capjaydef} is not gauge invariant and since in the derivation of \eqref{epsilonsols} we performed some gauge transformations (see in particular eq. \eqref{NHgaugetransf}), some care is required.
We first note that if we consider gauge transformations of the form
\begin{align}\label{Atilde}
 \tilde{A}=A+\alpha_1\,dt+\alpha_2\,d\phi\,,
 \end{align}
 for constant $\alpha_i$, then 
\begin{align}
J( \tilde{A})=J(A)+\alpha_2\,Q_e\frac{\Delta \phi}{2\pi}\,.
\end{align}
Notice, in particular, that this does not depend on $\alpha_1$. Demanding that the 
gauge field of the full black hole solution is regular on the horizon effectively fixes $\alpha_1$, as we explain in
section \ref{secf1}. However, the freedom in choosing $\alpha_2$ is associated with how one defines the electric potential, as
we discuss in section \ref{secf2}. In section \ref{secf3} we then calculate the angular momentum for the near horizon
$AdS_2$ solution \eqref{epsilonsols} and explain how it is related to that of the black hole \eqref{exprangmom}.
Finally, in section  \ref{secf4} we briefly make some comments concerning the dual conformal field theory (CFT).
\stoptocwriting
\subsection{Regularity of the gauge field on the horizon}\label{secf1}

We now investigate the gauge transformation that is required in order to ensure that the full black hole solution
\eqref{PDmetric}, \eqref{PDgauge} is well-defined on the horizon of the black hole. To do so, we Wick rotate $t\to \ii\,t_E$, as well as $a\to \ii \,a_E, \,\, e\to \ii\,e_E$, and define the Euclidean angular velocity
\begin{align}\label{angvel}
\Omega_E&\equiv \ii\,W\equiv -\frac{a_E}{r_+^2-a_E^2}\frac{2\pi}{\Delta\phi}\, .
\end{align}
We next change coordinates $t_E\to \tilde{t}_E$ and $\phi \to \tilde{\phi}+\frac{\Delta\phi}{2\pi}\Omega_E\,t_E$, and consider the metric near the horizon $r\sim r_+$, where $r_+$ is the outer horizon, defined by the largest positive root of ${Q}(r)$. We find
\begin{align}\label{appheappears}
ds^2\simeq(r-r_+)f(\theta)d\tilde{t}_E^2+\frac{f(\theta)}{4\kappa^2\,(r-r_+)}dr^2+h(\theta)d\tilde{\phi}^2+(r-r_+)\,g(\theta)\,d\tilde{\phi}\,d\tilde{t}_E\, ,
\end{align}
with $\kappa$ the surface gravity. Changing variables via
\begin{align}
x=\sqrt{r-r_+}\,\cos(\kappa \,\tilde{t}_E)\,, \qquad
y=\sqrt{r-r_+}\,\sin(\kappa \,\tilde{t}_E)\, ,
\end{align}
with $(x,y)$ parametrising $\mathbb{R}^2$, we get
\begin{align}
ds^2\simeq\frac{f(\theta)}{\kappa^2}(dx^2+dy^2)+\frac{g(\theta)}{\kappa}(x\,dy-y\,dx)d\tilde{\phi}+h(\theta)d\tilde{\phi}^2\, ,
\end{align}
which is clearly regular near the origin $x=y=0$. 

For the gauge field given in \eqref{PDgauge}, we find
\begin{align}
A&=f_1(r,\theta)d\tilde{t}_E+f_2(r,\theta)d\tilde{\phi}\,,\nn
&=\frac{f_1(r,\theta)}{\kappa}\frac{x\,dy-y\,dx}{x^2+y^2}+f_2(r,\theta)d\tilde{\phi}\, .
\end{align}
Since $f_2(r_+,\theta)$ is finite, the last term is well behaved at the horizon. However, in the gauge \eqref{PDgauge}
we find that $f_1(r,\theta)$ does not vanish fast enough as $r\to r_+$ to ensure regularity. Let us consider
the gauge transformation given in \eqref{Atilde} and choose
\begin{align}\label{alpha1}
\alpha_1=\frac{r_+\,e}{r_+^2+a^2}-\frac{\Delta \phi}{2\pi}W\alpha_2\, ,
\end{align}
with $\alpha_2$ arbitrary.
We then find
\begin{align}
\tilde{A}=\frac{\tilde f_1(r,\theta)}{\kappa}\frac{x\,dy-y\,dx}{x^2+y^2}+\left(f_2(r,\theta)+\alpha_2\right)\,d\tilde{\phi}\, ,
\end{align}
with 
$\tilde f_1(r,\theta)\equiv f_1(r,\theta)-\frac{r_+\,e_E}{r_+^2-a_E^2}$ and, as $r\to r_+$,
\begin{align}
\tilde f_1(r,\theta)\sim (r-r_+)F(\theta)=(x^2+y^2)\,F(\theta)\, .
\end{align}
Thus, this gauge field $\tilde A$ is regular on the horizon, for any value of $\alpha_2$. 
As we discussed above, the choice of $\alpha_1$ does not affect the value of the angular momentum of the black hole, $J(A)$ as
defined in \eqref{capjaydef}, but the choice of $\alpha_2$ will.

Notice that the value of $\alpha_1$ in \eqref{alpha1} with $\alpha_2$ given by
\begin{align}\label{alpha2}
\alpha_2=\frac{e(a^2-r_+^2)}{2a r_+}\,,
\end{align}
exactly matches the gauge transformation that we performed in \eqref{NHgaugetransf}. That the value of $\alpha_1$ agrees is 
exactly as expected, since this is the gauge transformation which was required in that analysis in order to have a regular gauge field in the near horizon limit. The last term in \eqref{NHgaugetransf} corresponds to the specific choice of $\alpha_2$ given in \eqref{alpha2}
and, as noted below  \eqref{NHgaugetransf}, was added for convenience, a point we return
to in section \ref{secf3}.

\vskip 3mm

\subsection{The electric potential}
\label{secf2}

To further illuminate the role of gauge transformations parametrized by $\alpha_2$, we discuss the electric potential $\Phi$. 
We define $\Phi=\Phi_{\infty}-\Phi_H$, where \cite{Anabalon:2018qfv}
\begin{align}
\Phi_{\infty}=\frac{1}{4\pi\, G_{(4)}Q_e\,\beta}\int_{\partial M}\sqrt{-h}\,n_a\,F^{ab}\,A_b\,,
\end{align}
is the potential as $r\to \infty$. Here $h_{ab}$ is the induced metric and $n_a$ the normal vector to the hypersurface $r\to \infty$, while $\beta=1/T$ is the inverse temperature. On the other hand we define
\begin{align}
\Phi_H&=\ell\cdot A|_{r\to r_+}\,,
\end{align}
to be the potential on the horizon $r=r_+$, where $\ell=\partial_t+\frac{\Delta\phi}{2\pi}W\,\partial_{\phi}=
\partial_t+W\,\partial_{\varphi}$ is the null generator of the horizon and $W$ was defined 
in \eqref{angvel}. In general, $\Phi$ depends on the choice of gauge for the gauge field.

For the black hole solutions in the gauge \eqref{PDgauge} we find $\Phi_{\infty}(A)=0$, $\Phi_{H}(A)=-\frac{r_+\,e}{r_+^2+a^2}$ and hence
 \begin{align}
\Phi(A)&=\frac{r_+\,e}{r_+^2+a^2}\, .
\end{align}

On the other hand after a general gauge transformation of the form \eqref{Atilde}, with general $\alpha_1$ and $\alpha_2$, we find
$\Phi_{\infty}(\tilde{A})=\alpha_1$, $\Phi_H(\tilde{A})=\alpha_1+\frac{\Delta\phi}{2\pi}W\alpha_2 -\frac{er_+}{a^2+r_+^2}$ with
\begin{align}
\Phi(\tilde{A})=\frac{r_+\,e}{r_+^2+a^2}-\frac{a}{r_+^2+a^2}\,\alpha_2\, .
\end{align}
Thus $\Phi(\tilde{A})$ depends on $\alpha_2$ but not $\alpha_1$. We also observe that the choice of $\alpha_1$ that
makes the gauge field regular on the horizon given in \eqref{alpha1} has the feature that $\Phi_H(\tilde{A})=0$.

It is interesting to observe that while $\Phi(A)$ and $J(A)$ both
depend on the choice of gauge transformations parametrized by $\alpha_2$, the combination $WJ+\Phi Q_{e}$ does not. This 
will similarly be true for the first law, which necessarily should be gauge invariant.

\vskip 3mm

\subsection{The angular momentum from the near horizon solution}\label{secf3}

Let us now consider computing the angular momentum for the near horizon solution \eqref{epsilonsols}. 
Suppose that in taking the near horizon limit in appendix \ref{appnh}, we had instead started with the regular gauge field given by
\eqref{Atilde}, with $\alpha_1$ given by \eqref{alpha1} and $\alpha_2=0$. Then there would be no need to perform the singular gauge transformation in \eqref{NHgaugetransf}, and we would arrive at a gauge field
for the near horizon solution which we will call $A_1$, which we give below. On the other hand in the expression 
for the gauge field given in \eqref{NHgaugetransf}, which we will call $A_2$, we did an additional gauge transformation associated with 
$\alpha_2$ as given in \eqref{alpha2}.
Specifically, we have
\begin{align}\label{aoneatwo}
A_2&=-\frac{\left(a^2+r_+^2\right) \left(a^2 e \cos ^2\theta +2 a g r_+ \cos \theta -e r_+^2\right)}{2 a r_+ \left(a^2 \cos ^2\theta +r_+^2\right)}\,\left( d\phi+v\,r\,dt \right)\,,\nn
&=h(y)\,(dz+\qp\,\rho\,d\tau)\,,\nn
A_1&=A_2+\frac{e (a^2-r_+^2)}{2 a r_+}d\phi\,.
\end{align}
Notice that $A_2$ (as in \eqref{NHgauge}) has the appealing feature that it is clearly $AdS_2$ 
invariant\footnote{Let $P=\partial_t$, $D=t\,\partial_t-r\,\partial_{r}$ and $K=-\tfrac{1}{2}(t^2+r^{-2})\partial_t+t\,r\,\partial_{r}$ be the standard generators for the isometries of $AdS_2$. We can lift this to an action on $AdS_2\times S^1$ using $P$, $D$ and $\tilde K\equiv K  +\frac{1}{v\,r}\partial_{\phi}$, which satisfy the same algebra. We then notice that the one form $D\phi=d\phi+v\,r\,dt$, which appears in \eqref{aoneatwo}, is invariant, satisfying
$\mathcal{L}_{P,D,\tilde K}D\phi=0$.}; indeed this was the motivation for carrying out the additional gauge transformation
in \eqref{NHgaugetransf}.

Now from the discussion in this appendix we know that $J(A_1)$ will be the angular momentum that agrees with that of the black hole solution in \eqref{exprangmom}. Indeed we find
\begin{align}
J \equiv J(A_1)=\frac{1}{G_{(4)}}{m\,a }\,\left(\frac{\Delta\phi}{2\pi}\right)^2\,.
\end{align}
However, this {\it differs} from the angular momentum calculated in the $AdS_2$ invariant gauge $A_2$. Specifically, 
if we denote this by $J_{AdS_2}$ we have 
\begin{align}\label{angmtm_NH}
J_{AdS_2}\equiv J(A_2)&=\frac{1}{G_{(4)}}\frac{1}{4}\qp\,\sqrt{1-\qp^2}\left(\frac{\Delta z}{2\pi}\right)^2\, ,\nn
& = \frac{1}{G_{(4)}}\frac{\qp \sqrt{1-\qp^2} \left(n_-^2+n_+^2\right)}{8 \left(1-2 \qp^2\right) n_-^2 n_+^2}\,,
\end{align}
where $\Delta z$ is given in \eqref{nhcps}. 
Using this, together with \eqref{jQe}, we find that 
\begin{align}
J_{AdS_2} = Q_e \frac{\sqrt{8 n_-^2 n_+^2 (G_{(4)}Q_e)^2+n_-^2+n_+^2}}{2 \sqrt{2}n_-n_+}\, .
\end{align}
Comparing to the similar expression \eqref{JQe} for $J$, we deduce that
\begin{align}
J_{AdS_2} - J = \frac{Q_e}{4}\frac{n_-+n_+}{n_-n_+}\, ,
\end{align}
which is precisely \eqref{angmomsurprise}.

\vskip 3mm

\subsection{Dual field theory point of view}\label{secf4}

We conclude this appendix by briefly noting that in the context of AdS/CFT, the gauge ambiguity in defining the angular momentum of the black holes has an
analogue in the dual CFT. We consider the dual CFT, which has a global $U(1)$ symmetry dual to the bulk gauge field,
to be defined on a background with metric $g_{ab}$ and background gauge field strength $F_{ab}=2\partial_{[a}A_{b]}$.
The CFT will satisfy Ward identities given by
\begin{align}\label{wids}
D_a  T^{ab} =F^{ba} J_a\,,\qquad D_aJ^a=0\,,
\end{align}
where $T^{ab}$ is the stress tensor and $J^a$ is the global $U(1)$ current current of the CFT, while
$D_a$ is the covariant derivative with respect to the background metric $g_{ab}$. 
Now for an arbitrary vector field $k^a$ on the boundary we have
\begin{align}
D_a[(T^a{}_b+J^a A_b)k^b]=\frac{1}{2}\mathcal{L}_kg_{ab} T^{ab}+ \mathcal{L}_kA_aJ^a\,.
\end{align}
Thus, when the background metric has a Killing vector $k$ that also preserves the background gauge field,
$\mathcal{L}_kA=0$, the
right hand side vanishes and there is an extra conserved current in the boundary theory given by
$T^a{}_bk^b+ (A\cdot k)J^a$. The non gauge-invariance of this current mirrors that of the bulk using
\eqref{expforG}--\eqref{capjaydef}.
Note also that if we have, more generally, $\mathcal{L}_kA=d\Lambda$, then the current is given by
$T^a{}_bk^b+J^a (A\cdot k-\Lambda)$.

\vskip 3mm

\resumetocwriting
\section{Killing spinor equations}\label{appeee}
We first recall the Killing spinors on a  $SE_7$. The metric is given by
\begin{align}
ds^2=ds^2(KE_6) +(\tfrac{1}{4}d\psip+\sigma)^2\,.
\end{align}
We take purely imaginary $D=7$ gamma matrices with
$\rho_{1234567}=-\ii$. 
The Killing spinor equation for the $SE_7$ is taken to be
\begin{align}\label{ksese7}
\D_A\chi=\frac{\ii}{2}\rho_A\chi\,.
\end{align}
After introducing the obvious orthonormal frame, we can solve this equation as in e.g. \cite{Gibbons:2002th}.
We impose the following projections on the Killing spinor:
\begin{align}
\rho_{12}\chi=\rho_{34}\chi=\rho_{56}\chi=\ii\chi,\qquad \Rightarrow\qquad \rho_7\chi=\chi\,,
\end{align}
and find that 
\begin{align}
\chi =\ex^{\ii  {\psip/2}}\chi_0\,,
\end{align}
where $\chi_0$ is a spinor on $KE_6$ satisfying \begin{align}\label{keeq}
D_m\chi_0=2\ii\sigma_m\chi_0\,,
\end{align}
where here $D_m$ is the covariant derivative on the $KE_6$, which always has a solution. Notice in particular,
that the spinor has the dependence $\ex^{\ii\psip/2}$ mentioned in section \ref{sec:gensetting}.

We turn now to the $D=11$ Killing spinor (KS) equation as given in \cite{Gauntlett:2002fz}:
\begin{align}\label{11kse}
\nabla_M\varepsilon+\frac{1}{12\times 4!}\Gamma_M{}^{N_1N_2N_3N_4}G_{N_1N_2N_3N_4}\varepsilon-
\frac{1}{6\times 3!}\Gamma^{N_1N_2N_3}G_{MN_1N_2N_3}\varepsilon=0\,.
\end{align}
We decompose the $D=11$ Clifford algebra via
\begin{align}\label{11dCliff}
\Gamma_{a} & = -\ii\gamma_{a} \gamma_5\otimes 1_8\, , \qquad  a=0,1,2,3\, ,\nn
\Gamma_{A+3} & = \gamma_5\otimes \rho_{ A}\, , \qquad\qquad  A=1,\ldots,7\, .
\end{align}
where $\gamma_5\equiv -\ii\gamma_0\gamma_1\gamma_2\gamma_3$. 
We then substitute the $D=11$ uplift of the $D=4$ solution given \eqref{lift} into \eqref{11kse}. The directions of the equation
tangent to the $SE_7$ are satisfied with $\chi$ solving \eqref{ksese7}, as above. For the remaining directions,
using a frame adapted to $ds^2_4$, we find
\begin{align}
\left(\nabla_a-\ii A_a+\frac{1}{2}\gamma_a +\frac{\ii}{4}F_{bc}\gamma^{bc}\gamma_a \right)\epsilon=0\,,
\end{align}
in agreement with e.g. (2.1) of \cite{Klemm:2013eca} (with a different sign choice in the definition of $\gamma_5$).

In order to construct spinor bilinears in section \ref{sec:Rsymmetry} it is helpful to identify various intertwiners. In $D=7$ we
take $A_7=C_7=1_8$ and hence $\rho_i=\rho_i^{\dagger}=-\rho_i^{T}$. For $D=4$, for the gamma matrices as in \eqref{4dgammas}
we can take
\begin{align}
A_4&=\sigma^1\otimes \sigma^3\,,\qquad \, A_4\gamma_a\,A_4^{-1}=\gamma_a^{\dagger}\,,\qquad \quad \, A_4^\dagger=A_4\,,\nn
C_4&=\sigma^2\otimes \sigma^1\,,\qquad
C_4^{-1}\,\gamma_a\,C_4=-\gamma_a^{T}\,,\qquad C_4^T=-C_4\, .
\end{align}
We can also define $D_4=C_4\,A_4^T=-\sigma^3\otimes \sigma^2$ and $\tilde D_4=\gamma_5 D_4=i 1\otimes \sigma^1$. Notice
that $D_4D_4^*=-1$, so cannot be used for a Majorana condition.
However, $\tilde D_4 \tilde D_4^*=+1$, so we can define a four-dimensional Majorana spinor as one satisfying  $\epsilon=\tilde D_4\,\epsilon^*$.
We can define the barred $D=4$ spinor as $\bar\epsilon=\epsilon^\dagger A_4$.

For the $D=11$ intertwiners we then have
\begin{align}
A_{11}&=A_4 \otimes C_7 \rightarrow A_{11}\,\Gamma_M\,A_{11}^{-1}=-\Gamma_M^{\dagger}\,,\qquad\quad A_{11}^\dagger=A_{11}\,,\nn
C_{11}&=\gamma_5 C_4 \otimes C_7 \rightarrow C_{11}^{-1}\,\Gamma_M\,C_{11}=-\Gamma_M^{T}\,,\qquad C_{11}^T=-C_{11}\,,
\end{align}
where $\gamma_5 C_4=\sigma^1\otimes \sigma^2$.
We also have $D_{11}=C_{11}\,A_{11}^T=\tilde D_4\otimes 1_8$ and we note that
$D_{11} D_{11}^*=+1$ so that the $D=11$ Majorana condition is $\varepsilon=D_{11}\,\varepsilon^*$. 
We can also define the barred $D=11$ spinor as $\bar\varepsilon=\varepsilon^\dagger A_{11}$. Thus if we have a $D=11$ spinor 
$\varepsilon=\epsilon\otimes\chi$ which is possibly complex (i.e. constructed from two $D=11$ Majorana spinors) we have
\begin{align}
\varepsilon=\epsilon\otimes \chi\,,\qquad\bar\varepsilon=\bar\epsilon\otimes \chi^\dagger\,.
\end{align}

\section{Bulk Killing spinor and boundary limit}\label{bulkks}

In this appendix we show how the conformal Killing spinor (CKS) \eqref{nonextremespinor}, arises as a limit of a bulk Killing spinor (KS). After finding the bulk Killing spinor of the BPS and non-extremal black hole, we derive the boundary metric with a suitable change of variables. Then we introduce a rotation that connects the bulk frame with the boundary frame, and rotate the spinor accordingly. Finally, we change our basis of gamma matrices to one that is suitable to interpret the boundary limit of the bulk KS as a tensor product of the CKS with some constant $2d$ spinor.

In the PD-type coordinates of section \ref{sec:nonextreme}, the full non-rotating solution is given in 
\eqref{metric1303app}:
\begin{align}\label{metric1303appendix}
\begin{split}
ds^2&=\frac{1}{(p+q)^2}\,\left(-\mathcal{Q}(q)\,d\tau^2+\frac{dq^2}{\mathcal{Q}(q)}+\frac{dp^2}{\mathcal{P}(p)}+\mathcal{P}(p)\,d\sigma^2\right)\,,\\
A&=\mathsf{Q}\,q\,d\tau-\mathsf{P}\,p\,d\sigma\,,
\end{split}
\end{align}
where
\begin{align}
\mathcal{P}(p)=\mathsf{C}^{-1}\,\mathcal{P}_1(p)\,\mathcal{P}_2(p)\,, \quad
\mathcal{Q}(q)=\mathsf{C}^{-1}\,\mathcal{Q}_1(q)\,\mathcal{Q}_2(q)\,,
\end{align}
and 
\begin{align}\nonumber
\mathcal{P}_1(p)&=-(1-p) (\mathsf{C} p+\mathsf{C}-\mathsf{P})\,,\quad
&& \mathcal{Q}_1(q)=\mathsf{C} q^2-\mathsf{C}+\mathsf{P} q+\ii \mathsf{Q}\,,\\
\mathcal{P}_2(p)&=-(1+p) (\mathsf{C} p-\mathsf{C}-\mathsf{P})
\,,\quad 
&&\mathcal{Q}_2(q)=\mathsf{C} q^2-\mathsf{C}+\mathsf{P} q-\ii \mathsf{Q}\,.
\end{align}

Using the frame
\begin{align}\label{framePD}
E^{\tau}=\frac{\sqrt{\mathcal{Q}(q)}}{p+q}d\tau\,, \quad
E^{q}=\frac{dq}{(p+q)\,\sqrt{\mathcal{Q}(q)}}\,, \quad
E^{p}=-\frac{dp}{(p+q)\,\sqrt{\mathcal{P}(p)}}\,, \quad
E^{\sigma}=\frac{\sqrt{\mathcal{P}(p)}}{p+q}d\sigma\,,
\end{align}
and $4d$ gamma matrices $\gamma_a$ introduced in \eqref{4dgammas}, we find the bulk Killing spinor (up to an overall normalization)
\begin{align}\label{bulkKS}
\psi=\frac{\ex^{-\ii\,(\kappa_1\,\tau+\kappa_2\,\sigma)}}{\sqrt{p+q}}
\begin{pmatrix}
\sqrt{\mathcal{Q}_1(q)\,\mathcal{P}_1(p)}\\-\frac{\mathsf{P}+\ii\,\mathsf{Q}}{\mathsf{P}^2+\mathsf{Q}^2}\sqrt{\mathcal{Q}_1(q)\,\mathcal{P}_2(p)}\\
\frac{\mathsf{P}+\ii\,\mathsf{Q}}{\mathsf{P}^2+\mathsf{Q}^2}\sqrt{\mathcal{Q}_2(q)\,\mathcal{P}_1(p)}\\\sqrt{\mathcal{Q}_2(q)\,\mathcal{P}_2(p)}
\end{pmatrix}\,.
\end{align}
Here $\kappa_1$ and $\kappa_2$ are the same constants introduced in \eqref{kappa12}, and we note that the dependence on $\tau$ and $\sigma$ could be simply removed with a gauge transformation.

Next, we derive the boundary metric \eqref{pdbmetric} starting from \eqref{metric1303appendix}. Recalling that the conformal boundary is given by $q=-p$, we introduce 
new coordinates ${\bar q},\,{\bar p}$ via
\begin{align}\label{PD_to_FG}
q=-{\bar p}+f_1({\bar p})\,{\bar q}+...\,\,, \quad
p={\bar p}+g_1({\bar p})\,{\bar q}+...\,\,,
\end{align}
where the boundary is approached in the limit ${\bar q}=0$. We find that for
\begin{align}
f_1({\bar p})=(1-\mathcal{P}({\bar p}))^{3/2}\,, \quad
g_1({\bar p})=\mathcal{P}({\bar p})\,(1-\mathcal{P}({\bar p}))^{1/2}\,,
\end{align}
at leading order in the small ${\bar q}$ expansion the metric reads
\begin{align}
ds^2\simeq \frac{1}{{\bar q}^2}\left(d{\bar q}^2+ds^2_{3d}\right)\,,
\end{align}
where $ds^2_{3d}$ is the boundary metric given in \eqref{pdbmetric} (dropping the bars).\footnote{In section \ref{sec:nonextreme} we use $p$ to denote the angular variable of the boundary metric. In this appendix, we reserve $q$ and $p$ for the bulk coordinates, while denoting with barred quantities the new
coordinates. In comparing quantities computed here for the boundary with those of section \ref{sec:nonextreme}, one should then simply set ${\bar p}_{\mathrm{here}}=p_{\mathrm{there}}$.} We notice that the CKS \eqref{nonextremespinor} was computed using the frame \eqref{boundaryframePD}, which is adapted to the boundary metric. This differs from the boundary limit of the frame \eqref{framePD}, and therefore in order to compare the CKS to the bulk KS of eq. \eqref{bulkKS}, one should first find the rotation that connects the two frames. We can define
\begin{align}
\tilde{E}^q=\cos\beta\, E^q-\sin\beta\,E^p\,, \quad
\tilde{E}^p=\sin\beta\, E^q+\cos\beta\,E^p\,,
\end{align}
with $E^{\tau}$ and $E^{\sigma}$ unchanged, where we have introduced 
\begin{align}
\cos \beta=\frac{\sqrt{\mathcal{Q}(q)}}{\sqrt{\mathcal{P}(p)+\mathcal{Q}(q)}}\,, \quad 
\sin \beta=\frac{\sqrt{\mathcal{P}(p)}}{\sqrt{\mathcal{P}(p)+\mathcal{Q}(q)}}\,.
\end{align}
Note that this is a suitable frame for our purposes because when $p+q=0$  
\begin{align}
E^{\tau}=\frac{(1-\mathcal{P}(\bar p))^{1/2}}{\bar q}\,e^0\,, \quad 
\tilde{E}^q=0\,, \quad
\tilde{E}^p=\frac{(1-\mathcal{P}(\bar p))^{1/2}}{\bar q}\,e^1\,, \quad
E^{\sigma}=\frac{(1-\mathcal{P}(\bar p))^{1/2}}{\bar q}\,e^2\,, 
\end{align}
where $e^i$ ($i=0,1,2$) is precisely the frame introduced in \eqref{boundaryframePD} to study the CKSE. The overall conformal factor is irrelevant because the CKSE is invariant under rescalings by a conformal factor. We can then define the rotated spinor 
\begin{align}
\tilde{\psi}=\ex^{-\tfrac{1}{2}\beta\,\gamma_{12}}\,\psi\,,
\end{align}
where we have used that $\tfrac{1}{2}\gamma_{12}$, generates rotations in the $(q,p)$ plane, in the spin $1/2$ representation. We can now take the boundary limit of $\tilde{\psi}$, which is implemented by changing coordinates as in \eqref{PD_to_FG} and retaining only the leading order term in the small ${\bar q}$ expansion. We find: 
\begin{align}
\tilde{\psi}\simeq 
\sqrt{\frac{\mathsf{P}\,(\mathsf{P}+\ii\,\mathsf{Q})}{2}} \,\frac{\ex^{-\ii\,(\kappa_1\tau+\kappa_2\sigma)}}{\sqrt{{\bar q}}}\,\chi + \ldots\, ,
\end{align}
where $\kappa_{1,2}$ are those introduced in \eqref{kappa12}, and we have introduced 
\begin{align}
\chi=
\begin{pmatrix}
\zeta_1({\bar p})+\zeta_2({\bar p})\\
\zeta_1({\bar p})-\zeta_2({\bar p})\\
\zeta_1({\bar p})+\zeta_2({\bar p})\\
\zeta_1({\bar p})-\zeta_2({\bar p})
\end{pmatrix}\,\,.
\end{align}
Here $\zeta_1$ and $\zeta_2$ are precisely the functions introduced in \eqref{boundaryspinorcomponentsPD}. Ideally, we would like to interpret this as the tensor product between the CKS \eqref{nonextremespinor} and some constant $2d$ spinor. However, this is not quite the case, and the reason is that the set of gamma matrices that we have used is not suitable for the decomposition of the $4d$ spacetime in three boundary directions and a radial one. This can be easily fixed by introducing a new set of gamma matrices,
\begin{align}
\tilde{\gamma}_0=\gammathree_0\otimes \sigma^3\,, \quad
\tilde{\gamma}_1=1_2\otimes \sigma^1\,, \quad
\tilde{\gamma}_2=\gammathree_1\otimes \sigma^3\,, \quad
\tilde{\gamma}_3=\gammathree_2\otimes \sigma^3\,,
\end{align}
where $\gammathree_i$ are the $3d$ gamma matrices used in section \ref{sec:conboundary}. They are related to the $\gamma_a$ of \eqref{4dgammas} by the similarity transformation
\begin{align}
\tilde{\gamma}_a=M\,\gamma_a\,M^{\dagger}\,,
\end{align}
with $M$ the unitary matrix
\begin{align}
M=\frac{1+\ii}{2\sqrt{2}}\,
\begin{pmatrix}
1 & -\ii & -\ii & 1\\
-\ii & 1 & 1 & -\ii \\
1 & \ii & -\ii & -1\\
-\ii & -1 & 1 & \ii \\
\end{pmatrix}\,.
\end{align}
Expressing the spinor $\chi$ in this new basis, we find
\begin{align}
M\,\chi=
\frac{1}{\sqrt{2}}
\begin{pmatrix}
\zeta_1({\bar p})\\
\zeta_2({\bar p})
\end{pmatrix}
\otimes 
\begin{pmatrix}
1\\1
\end{pmatrix}\,,
\end{align}
which completes our derivation of the CKS \eqref{nonextremespinor} from the bulk KS \eqref{bulkKS}.

\providecommand{\href}[2]{#2}\begingroup\raggedright\endgroup

\end{document}